\documentclass[useAMS,usenatbib]{mn2e}
\usepackage{graphicx}
\usepackage{lscape}
\usepackage{subfig}
\usepackage{epstopdf}
 \usepackage{epsfig}
\usepackage{grffile}
\usepackage{lscape}
\usepackage{verbatim}

\title[Red subdwarfs in binary systems]{A spectroscopic and proper motion search of Sloan digital Sky Survey: red subdwarfs in binary systems}
\author[Z. H. Zhang et al.]{Z. H. Zhang,$^{1,2,3}$\thanks{E-mail:
zenghuazhang@gmail.com} D. J. Pinfield,$^{1}$ B. Burningham,$^{1}$ H. R. A. Jones,$^{1}$ 
\newauthor
  M. C. G\'{a}lvez-Ortiz,$^{1,4}$ S. Catal\'an,$^{1}$ R. L. Smart,$^{3}$ S. L\'epine,$^{5}$  J. R. A. Clarke,$^{1,6}$ 
\newauthor 
 Ya. V. Pavlenko,$^{1,7}$ D. N. Murray,$^{1}$  M. K. Kuznetsov,$^{7}$   A. C. Day-Jones,$^{1}$ 
\newauthor 
  J. Gomes,$^{1}$   F. Marocco$^{1}$ and B. Sip\H{o}cz$^{1}$  \\
$^{1}$Centre for Astrophysics Research, Science and Technology Research Institute, University of Hertfordshire, Hatfield AL10 9AB, UK \\
$^{2}$Shanghai Astronomical Observatory, Chinese Academy of Sciences, 80 Nandan Road, Shanghai 200030, China \\
$^{3}$Istituto Nazionale di Astrofisica, Osservatorio Astronomico di Torino, Strada Osservatrio 20, 10025 Pino Torinese, Italy \\
$^{4}$Centro de Astrobiolog\'{i}a (CSIC-INTA), Ctra. Ajalvir km 4, E-28850 Torrej\'{o}n de Ardoz, Madrid, Spain \\
$^{5}$Department of Astrophysics, Division of Physical Sciences, American Museum of Natural History, New York, NY 10024, USA \\
$^{6}$Departamento de F\'{i}sica y Astronom\'{i}a, Universidad de Valpara\'{i}so, Av. Gran Breta\~na 1111, Casilla 5030, Valpara\'{i}so, Chile \\
$^{7}$Main Astronomical Observatory, Academy of Sciences of Ukraine, Golosiiv Woods, Kyiv-127, 03680, Ukraine }

\begin{document}

\date{Accepted 2013 June 06. Received 2013 May 15; in original form 2012 August 13}

\pagerange{\pageref{firstpage}--\pageref{lastpage}} \pubyear{2013}

\maketitle

\label{firstpage}

\begin{abstract}

Red subdwarfs in binary systems are crucial for both model calibration and spectral classification. We search for red subdwarfs in binary systems from a sample of high proper motion objects with Sloan digital Sky Survey spectroscopy. We present here discoveries from this search, as well as highlight several additional objects of interest. We find 30 red subdwarfs in wide binary systems including: two with spectral type of esdM5.5, 6 companions to white dwarfs  and 3 carbon enhanced red subdwarfs with normal red subdwarf companions. 15 red subdwarfs in our sample are partially resolved close binary systems. With this binary sample, we estimate the low limit of the red subdwarf binary fraction of $\sim$ 10 \%. We find that the binary fraction goes down with  decreasing masses and metallicities of red subdwarfs. A spectroscopic esdK7 subdwarf + white dwarf binary candidate is also reported. 30 new M subdwarfs have spectral type of $\geqslant$M6  in our sample.  We also derive relationships between spectral types and absolute magnitudes in the optical and near-infrared for M and L subdwarfs, and we present an M subdwarf sample with measured $U, V, W$ space velocities.

\end{abstract}


\section{Introduction}

Dwarf stars with subsolar metallicity are bluer than solar abundance dwarfs or main-sequence stars of the equivalence mass. They lie below the main sequence in the Hertzsprung--Russell diagram and appear less luminous than main-sequence stars. These objects were thus called ``subdwarfs" by \citet{kui39}. Evolving subdwarfs are referred to as cool subdwarfs to provide distinction from hot subdwarfs, a different class of objects \citep[e.g.][]{han07}. Cool dwarfs and cool subdwarfs are observationally and kinematically distinct, and were separated into Population I and II categories, respectively by \citet{baa44}. Population I and II are associated with the Galactic disc and spheroid respectively. \citet{rom50,rom52,rom54} find that the old, high-velocity Population II stars were also metal deficient.  

Red dwarfs are low-mass and relatively cool stars on the main sequence with spectral types of  late-type K and M, masses between   $\sim$ 0.6  and $\sim$ 0.08 $M_{\sun}$,  and surface effective temperatures between  $\sim$ 4000  and $ \sim $ 2300 K \citep{kal09}. Red dwarfs are the most common type of star both in the Milky Way \citep{kir12} and other galaxies \citep{van10}. The spectra of red dwarfs are dominated by molecular absorption bands of metal oxides and hydrides. TiO and CaH near 7000 \AA~are the most prominent bands \citep{bes82,bes91}. Red subdwarfs (RSDs) are the subsolar counterparts of red dwarfs by chemical abundance. RSDs are significantly rarer than red dwarfs. The  Kapteyn's Star (sdM1 type) is the only cool subdwarf among the sample of 8 pc within the Sun which contains more than 244 known stars and brown dwarfs, including 157 M dwarfs \citep{kir12}. The known sample of late-type M subdwarfs is significantly smaller than that of early-type M subdwarfs because they are fainter and have lower space density according to the halo mass function \citep[e.g.][]{cha03}. Late-type M subdwarfs have the most complex stellar atmospheres because they are ultracool  and have large-scale variation of chemical abundance and gravity.

Although RSDs are less luminous than F, G and mid-K subdwarfs, they are numerous and have more notable spectral features caused by chemical abundance and gravity, thus they are better targets for observational and theoretical studies. It is difficult to distinguish F, G and mid-K subdwarfs from normal dwarf stars of the same spectral type using their optical spectra because they are featureless. Dwarfs and subdwarfs with spectral types of late-type K and M, however, have a number of very different spectral features. Model atmospheres also suggest that  optical spectra of M subdwarfs are dramatically affected by metallicity variations. It is thus possible to use low-resolution spectroscopy for exploring effective temperature, metallicity and gravity effects  on the spectra of these cool objects \citep{all95}.

Classification and characterization of M subdwarfs is a rapidly evolving field \citep{giz97,lep07,jao08,dhi12}. M subdwarfs are classified into three metal classes: subdwarf (sdM), extreme subdwarf (esdM) and ultra subdwarf (usdM) based on the ratio of TiO to CaH indices \citep{lep07}. CaH and TiO indices are easy to measure and sensitive to temperature and metallicity. However, \citet{jao08} found that the CaH and TiO indices are affected in complicated ways by combinations of temperatures, metallicities and gravities of RSDs. Model spectra show that  the TiO5 index is more sensitive to metallicity while the CaH2 and CaH3 indices are more sensitive to gravity. This suggests that  the effect of gravity, which was previously ignored, should be considered in the classification of M subdwarfs. An ideal testbed for the impact  of gravity on spectra of M subdwarfs is binary systems with two M subdwarfs which share  the same age and metallicity. With the same effect on spectra from metallicity,  it is possible to measure the difference of gravity on the broad indices of CaH2, CaH3 and TiO5. 

M subdwarfs in binary systems are the key for both model calibration, and spectral classification and characterization. Discovery of a sample of RSD binary systems is therefore crucial. In this paper we present the discovery of 45 RSD binary systems from the Sloan Digital Sky Survey \citep[SDSS,][]{yor00} and the UKIRT Infrared Deep Sky Survey \citep[UKIDSS,][]{law07}. At least one companion in each system is confirmed as an RSD with SDSS spectra. The selection and classification processes of our RSD sample are presented in Section~\ref{sample}. The identification of RSD binary systems is presented in Section \ref{sbinary}. Section~\ref{snote} presents further discussion of RSD binary systems of particular interests. Summary and conclusions is described in Section~\ref{ssummary}.

\section{The SDSS sample}
\label{sample}
%Objects were detected in the SDSS, and were classified and measured by the image analysis software. These imaging data are used to select targets for the SDSS spectroscopic survey. 
The eighth data release (DR8) of SDSS includes 14555 deg$^{2}$ of  imaging data, and 9274 deg$^{2}$ of spectroscopic data. There are over 1.84 million spectra in total, including 0.6 million stars, 0.13 million quasars and 0.95 million galaxies \citep{aih11}. The SDSS DR8 also includes proper motions (PMs) for objects derived by combining SDSS astrometry with USNO-B positions, re-calibrated against SDSS. The errors of PMs are typically less than 10 mas$\cdot$yr$^{-1}$ \citep{mun04}. 

\begin{figure}%  figure placement: here, top, bottom, or page
   \centering
   \includegraphics[width=\columnwidth]{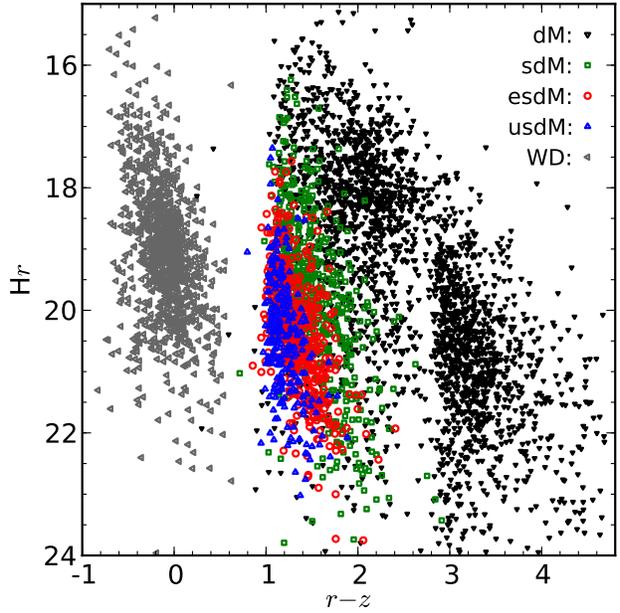}
   \caption{SDSS $r$ band reduced PMs $H_r$ and $r-z$ colour of our PM selected objects. Symbols in the figure are black down-pointing triangles: M dwarfs; green squares: sdMs; red circles: esdMs; blue up-pointing triangles: usdMs; grey left-pointing triangles: WDs. }
   \label{hr}
\end{figure}

\begin{figure*} %  figure placement: here, top, bottom, or page
   \centering
   \includegraphics[width=0.9\textwidth, angle=0]{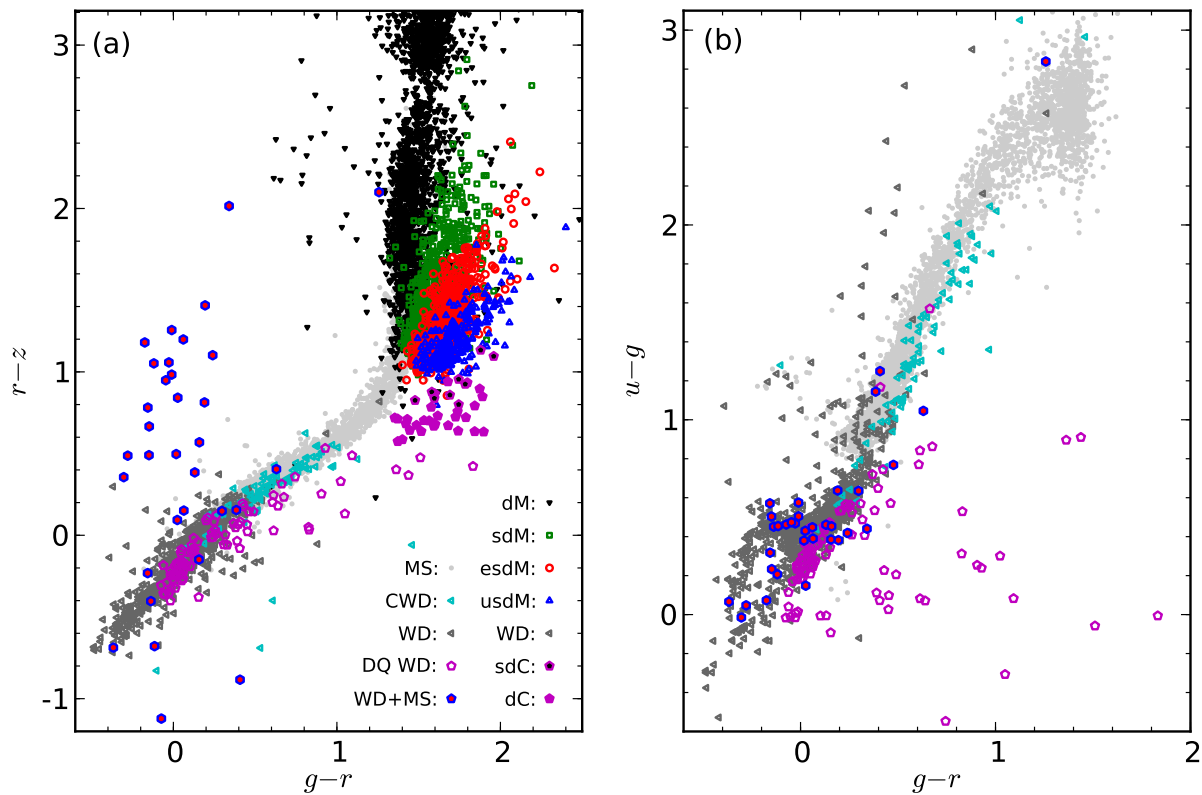}
   \caption{$g-r$ versus $r-z$ (left) and $g-r$ versus $r-z$ (right) colours of our PM selected sample. Symbols in the figure are black down-pointing triangles: dMs; green squares: sdMs; red circles: esdMs; blue up-pointing triangles: usdMs; filled magenta pentagons: carbon dwarfs (dC); magenta pentagons filled with black: carbon subdwarfs (sdC); open magenta pentagons: DQ WDs; blue hexagon filled with red: WD + MS binaries; dark grey left-pointing triangles: WDs; cyan left-pointing: cool WDs (CWD); light grey dots: 3028 point sources (MS) with $17 < r < 18$ selected from 10 square degrees of SDSS.}
   \label{ugrz}
\end{figure*}

\begin{figure*} %  figure placement: here, top, bottom, or page
   \centering
   \includegraphics[width=\textwidth, angle=0]{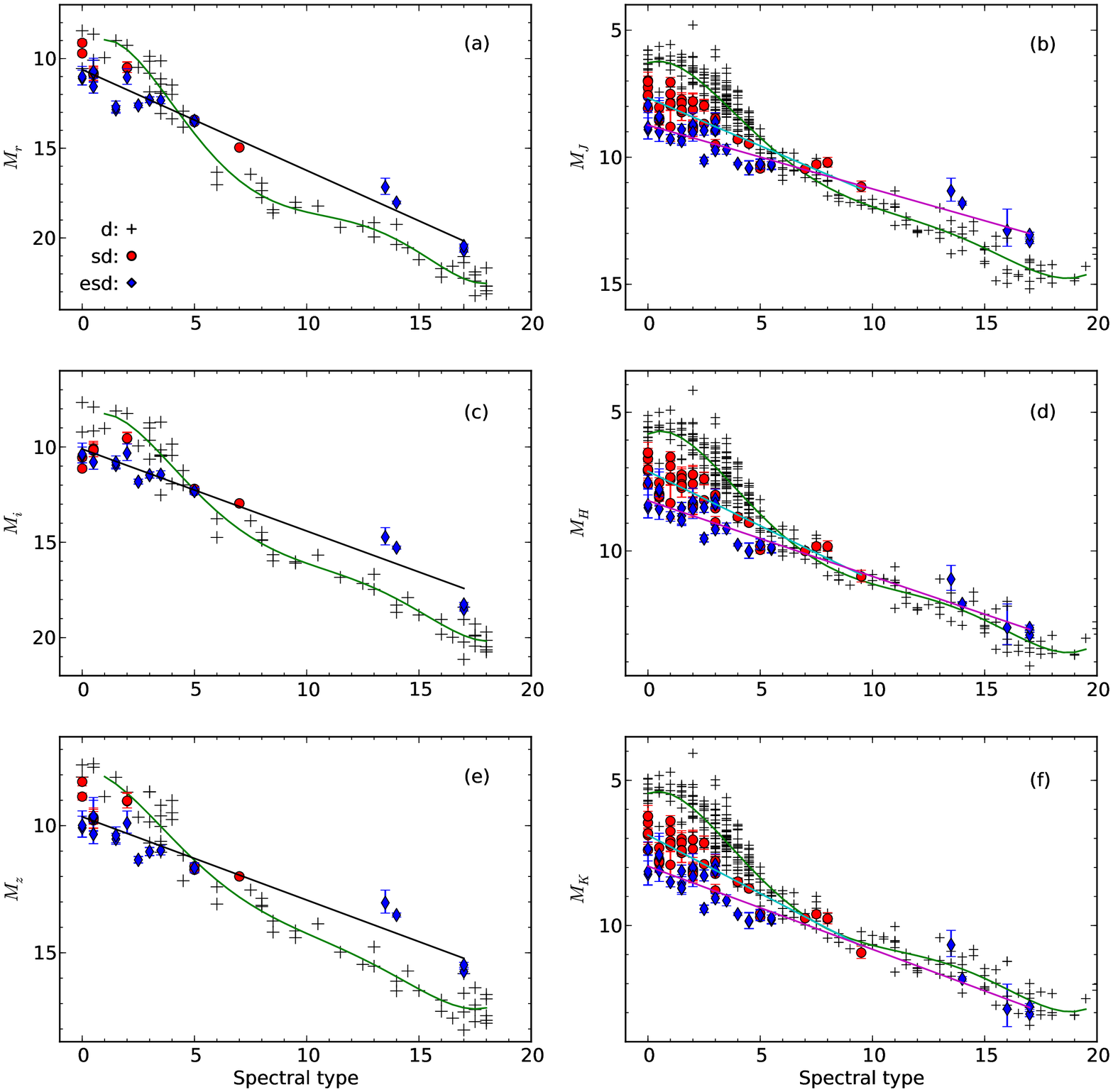}
   \caption{Spectral type and absolute magnitude relationships of dwarfs (black pluses), subdwarfs (red dots) and extreme subdwarfs (blue diamonds). M0 = 0, M5 = 5, L0 = 10, L5 = 15. Green and black lines on the left-hand panels are best fits of dwarfs and all subdwarfs respectively.  Green, cyan and magenta lines on the right-hand panels are best fits of dwarfs, subdwarfs and extreme subdwarfs, respectively. M and L subdwarfs with two independent parallax measurements are plotted twice.}
   \label{sptm}
\end{figure*}

\subsection{Selection}
We selected our candidates using SDSS CasJobs by combining the spectroscopic and PM catalogues. We required PM greater than 100 mas$\cdot$yr$^{-1}$. No photometric criteria were applied but spectral observations for red dwarfs in SDSS is limited to $r \sim 21.0, i \sim 20$ and $z \sim 19.5$.  Most of RSDs in SDSS spectroscopic data base are at distances of 200 $\sim$ 400 pc. Objects with tangential velocity of 200 km s$^{-1}$ at a distance of 400 pc (or 100 km s$^{-1}$ and  200 pc) will have PMs higher than 100 mas$\cdot$yr$^{-1}$. A PM cut of 100 mas$\cdot$yr$^{-1}$ allows a good balance in order to select most late-type K, early-type M and almost all mid-late type M   subdwarfs in the combined SDSS PM + spectroscopic catalogues while minimizing the contamination by dwarf stars.  Since not all spectra in DR7 \citep{aba09} are reproduced in DR8 we applied our search to both data releases finding 7499 and 8445 spectra respectively. There were 9146 spectra in total for 8236 objects (some objects had more than one spectrum). 

\subsection{Spectral Classification}
%The current classification system of M dwarfs was refined to four metallicity classes (dwarfs: dM; subdwarfs: sdM; and extreme subdwarfs: esdM; usdM: ultra subdwarfs) from three metallicity classes system \citep[sdM; sdM; esdM;][]{giz97} by \citet{lep07}. 
%The spectroscopic definition of these classes is based on the relative strength of molecular absorption bands (CaH2, CaH3, and TiO5) near 7000 \AA. 
%Gravity has not been considered in this classification system, thought it will impact the classification results in some cases \citep{jao08}. 

SDSS spectra were reduced and classified with the {\rm idlspec2d} code by SDSS \citep{aih11}. 
For 8236 objects in our sample, 5687 were classified as stars, 1558 as galaxies and 348 as QSOs, 643 of them were not classified. Not all objects in the sample are classified by SDSS. RSDs are generally classified as stars or galaxies in some cases.

To pick out and classify RSDs in our sample properly we used a K and M subdwarf classification code developed by \citet{lep07}. We ran the code on the spectra of all 8236 objects to identify subdwarfs and assign their spectral types. The code classified objects into nine groups: dM (2455), dK (80), sdM (689), sdK (326), esdM (483), esdK (189), usdM (256), usdK (189), and unclassified (3442). We found many objects, originally classified as galaxies by SDSS, that were re-classified as sdK, esdK or usdK. 

We inspected the spectra by eye in each group to ensure that the correct classification was applied in each case. We found 2004 galaxies with false PMs were selected into our sample: 383 of them were classified as stars (mostly late-type K subdwarfs). We found that 463 late-type K subdwarfs and 1363 M  subdwarfs survived the eyeball check. 54 objects were classified as sdM but removed from our subdwarf sample because 
%their CaH and TiO ratio are too high to be subdwarfs and 
they do not have typical halo kinematics. Fig. \ref{hr} shows reduced PMs and $r-z$ colour of these objects. Three sequences from left to right show the location of white dwarfs (WDs), M subdwarfs and M dwarfs. Fig. \ref{ugrz} shows two colour diagrams of $g-r$ versus $r-z$ and $g-r$ versus $r-z$ of the sample. The WD sample will be discussed in a future paper.

\subsection{The M subdwarf sample}

\begin{table*}
 \centering
 %\begin{minipage}{140mm}
  \caption{K7, M and L subdwarfs with parallax measurements}
  \begin{tabular}{l c l c c c c c}
\hline\hline
    Name &  2MASS &  SpT & Ref1 & $\pi$ (mas) & Ref2 & $\pi$ (mas) & Ref3 \\ 
\hline
LP 406$-$47 & J01002474+1711272 & sdM5 & 13 & 15.7$\pm$1.2 & 8 & 15.7$\pm$1.1 & 15 \\
G 003$-$036 & J02025226+0542205 & sdM0 & 8 & 35.9$\pm$3.1 & 8 & 32.38$\pm$5.09 & 16 \\
LHS 164 & J03014052$-$3457548 & sdK7 & 11 & 19.16$\pm$1.55 & 11 &    &  \\
G 038$-$001 & J03285302+3722579 & sdK7 & 8 & 35.3$\pm$3.1 & 8 & 39.32$\pm$2.24 & 16 \\
GJ 1062 & J03381558$-$1129102 & sdM2.5 & 8 & 64.8$\pm$2.5 & 8 &    &  \\
G 079$-$059 & J03422933+1231368 & sdM1.5 & 8 & 45.1$\pm$6.6 & 8 & 45.1$\pm$7.8 & 9 \\
WT0135 & J04112712$-$4418097 & sdM3 & 11 & 39.04$\pm$2.42 & 11 &    &  \\
Kapteyn's & J05114046$-$4501051 & sdM1 & 8 & 258.3$\pm$6.5 & 8 & 255.66$\pm$0.91 & 16 \\
G 099$-$033 & J05480018+0822142 & sdM0 & 8 & 18.8$\pm$2.9 & 8 & 19.3$\pm$3.1 & 15 \\
G 105$-$023 & J06140146+1509570 & sdM2 & 8 & 30.6$\pm$3.0 & 8 & 30.1$\pm$3.1 & 15 \\
G 251$-$044 & J07432434+7248500 & sdK7 & 8 & 18.7$\pm$1.8 & 8 & 18.2$\pm$2.8 & 15 \\
LHS 272 & J09434633$-$1747066 & sdM3 & 8 & 73.95$\pm$1.18 & 11 &    &  \\
SSSPM 1013$-$1356 & J10130734$-$1356204 & sdM9.5 & 6 & 20.28$\pm$1.96 & 14 &    &  \\
SCR1107$-$4135 & J11075597$-$4135529 & sdM0.5 & 11 & 14.79$\pm$1.18 & 11 &    &  \\
LHS 300AB & J11111376$-$4105326 & sdM0: & 11 & 33.03$\pm$1.36 & 11 &    &  \\
G 176$-$040 & J11324528+4359444 & sdM0.5 & 8 & 18.3$\pm$2.7 & 8 & 17.9$\pm$3.3 & 15 \\
Ross 451 & J11402025+6715349 & sdM0 & 8 & 32.7$\pm$2.5 & 8 & 42.79$\pm$2.70 & 16 \\
LHS 318 & J11565479+2639586 & sdM2: & 11 & 18.76$\pm$2.32 & 11 &    &  \\
G 011$-$035 & J12023365+0825505 & sdM2 & 8 & 26.0$\pm$3.4 & 8 & 25.8$\pm$3.6 & 12 \\
LHS 326 & J12242688$-$0443361 & sdM3 & 11 & 20.39$\pm$1.94 & 11 &    &  \\
LHS 2734A & J13251422$-$2127120 & sdK7 & 11 & 3.94$\pm$1.19 & 11 &    &  \\
LHS 2734B & J13251572$-$2128176 & sdM1 & 11 & 3.94$\pm$1.19 & 11 &    &  \\
LSR 1425+7102 & J14250510+7102097 & sdM8 & 3 & 12.19$\pm$1.07 & 14 & 12.27$\pm$0.45 &  \\
LP 440$-$52 & J14390030+1839385 & sdM7 & 8 & 28.4$\pm$0.8 & 8 & 28.4$\pm$0.7 & 15 \\
G 137$-$008 & J15281403+1643109 & sdK7 & 8 & 18.7$\pm$3.4 & 8 & 10$\pm$5 & 10 \\
LHS 406 & J15431836$-$2015310 & sdM2 & 11 & 46.73$\pm$1.17 & 11 &    &  \\
LP 803$-$27 & J15454034$-$2036157 & sdM5 & 8 & 31.5$\pm$2 & 8 & 33.5$\pm$2.0 & 15 \\
G 138$-$059 & J16420431+1025583 & sdM2 & 8 & 26.1$\pm$4.7 & 8 & 26$\pm$6 & 10 \\
LHS 440 & J17182561$-$4326373 & sdM1 & 11 & 36.40$\pm$1.22 & 11 &    &  \\
G 021$-$023 & J18413636+0055145 & sdK7 & 8 & 36.2$\pm$2 & 8 & 30.63$\pm$3.76 & 16 \\
LP 141$-$1 & J18455236+5227400 & sdM4.5 & 8 & 50.1$\pm$1.3 & 8 & 50.1$\pm$2.5 & 9 \\
SCR1916$-$3638 & J19164658$-$3638040 & sdM3 & 11 & 14.78$\pm$1.37 & 11 &    &  \\
LP 869$-$24 & J19442199$-$2230534 & sdM4 & 8 & 17.7$\pm$0.8 & 8 & 17.7$\pm$0.8 & 15 \\
LP 753$-$21 & J19451476$-$0917581 & sdM2.5 & 8 & 6.7$\pm$0.7 & 8 & 6.7$\pm$0.7 & 15 \\
G 142$-$052 & J19464860+1204580 & sdM1 & 8 & 22.4$\pm$2.3 & 8 & 21.7$\pm$2.7 & 15 \\
Gl 781 & J20050227+5426037 & sdM1.5 & 8 & 60.3$\pm$1.7 & 8 & 63.17$\pm$3.82 & 16 \\
G 210$-$019 & J20272905+3559245 & sdM1.5 & 8 & 21.1$\pm$3.6 & 8 & 20.4$\pm$4.2 & 15 \\
LSR 2036+5059 & J20362165+5100051 & sdM7.5 & 3 & 21.60$\pm$1.26 & 14 &    &  \\
LHS 3620 & J21042537$-$2752453 & sdM2 & 11 & 12.88$\pm$1.40 & 11 &    &  \\
Wo 9722 & J21075543+5943198 & sdM1.5 & 8 & 41.9$\pm$2.2 & 8 & 40$\pm$3 & 10 \\
LHS 518 & J22202698$-$2421500 & sdM1 & 11 & 15.92$\pm$1.31 & 11 &    &  \\
LHS 521 & J22275918$-$3009324 & sdM0.5 & 11 & 18.46$\pm$1.07 & 11 &    &  \\
G 128$-$034 & J23082608+3140240 & sdM0.5 & 8 & 22.7$\pm$2.5 & 8 & 22.4$\pm$3.0 & 15 \\
\hline
\end{tabular}
%\begin{list}{}{}
%\item[]Note. 
%\item[$^{a}$] 
%%\end{minipage}
\label{tplx}
\end{table*}

\addtocounter{table}{-1}
\begin{table*}
 \centering
 %\begin{minipage}{140mm}
  \caption{continued.}
  \begin{tabular}{l c l c c c c c}
\hline\hline
    Name &  2MASS &  SpT & Ref1 & $\pi$ (mas) & Ref2 & $\pi$ (mas) & Ref3 \\ 
\hline
G 004$-$029 & J02341234+1745527 & esdM3 & 17 & 27.3$\pm$4.2 & 15 &    &  \\
G 075$-$047 & J02524557+0155501 & esdM2 & 17 & 25.3$\pm$5 & 15 &    &  \\
G 005$-$022 & J03132412+1849390 & esdK7 & 8 & 30.9$\pm$2.3 & 8 &    &  \\
G 095$-$059 & J03501388+4325407 & esdM0 & 8 & 23.4$\pm$2.5 & 8 &    &  \\
G 007$-$017 & J04013654+1843423 & esdM0.5 & 8 & 16.7$\pm$4.6 & 8 & 16.2$\pm$4.7 & 15 \\
LP 302$-$31 & J04305244+2812001 & esdM1 & 8 & 10.2$\pm$0.8 & 8 & 10.2$\pm$0.8 & 15 \\
LP 417$-$42 & J05103896+1924078 & esdM5.5 & 8 & 13.4$\pm$1.2 & 8 & 13.8$\pm$1.0 & 15 \\
LP 417$-$44 & J05195663+2010545 & esdM4.5 & 8 & 10.4$\pm$1.3 & 8 & 10.3$\pm$1.3 & 15 \\
2MASS 0532+8246 & J05325346+8246465 & esdL7 & 2,1 & 42.28$\pm$1.76 & 14 & 37.5$\pm$1.7 & 2 \\
2MASS 0616$-$6407 & J06164006$-$6407194 & esdL6 & 4,1  & 19.85$\pm$6.45 & 7 &    &  \\
LSR0627+0616 & J06273330+0616591 & esdM1.5 & 11 & 16.43$\pm$1.25 & 11 &    &  \\
LP 484$-$6 & J08012900+1043041 & esdM2.5 & 8 & 12.9$\pm$0.8 & 8 & 12.9$\pm$0.7 & 15 \\
SDSS J1256$-$0224 & J12563716$-$0224522 & esdL3.5 & 3,1 & 11.10$\pm$2.28 & 14 &    &  \\
G 165$-$047 & J14065553+3836577 & esdM1.5 & 8 & 37.4$\pm$3.7 & 8 &    &  \\
LP 857$-$48 & J14313832$-$2525328 & esdM4 & 8 & 41.7$\pm$1 & 15 &    &  \\
LP 502$-$32 & J15202946+1434391 & esdM5 & 8 & 8.9$\pm$0.8 & 8 & 8.9$\pm$0.7 & 15 \\
2MASS 1626+3925 & J16262034+3925190 & esdL4 & 3,1 & 29.85$\pm$1.08 & 14 &    &  \\
LP 686$-$36 & J16595712$-$0333136 & esdM0.5 & 8 & 6.4$\pm$1.2 & 8 & 6.4$\pm$1.2 & 15 \\
LP 139$-$14 & J17395137+5127176 & esdM3.5 & 8 & 10.3$\pm$0.9 & 8 &    &  \\
LP 24$-$219 & J18215294+7709303 & esdM2.5 & 8 & 10.4$\pm$0.9 & 8 & 10.4$\pm$0.9 & 15 \\
LP 515$-$3 & J20190458+1235056 & esdM0 & 8 & 18.9$\pm$3.6 & 8 & 18.1$\pm$4.2 & 12 \\
LP 695$-$96 & J20253705$-$0612366 & esdM3 & 8 & 8.3$\pm$0.6 & 8 & 8.4$\pm$0.6 & 15 \\
LP 757$-$13 & J21073416$-$1326557 & esdM1.5 & 8 & 8.8$\pm$0.8 & 8 & 8.8$\pm$0.8 & 15 \\
G 018$-$051 & J22284904+0548128 & esdK7 & 8 & 26.8$\pm$2.1 & 8 & 21$\pm$3 & 10 \\
LSPM J2321+4704 & J23212321+4704382 &  esdM2 & 18 &  10.34$\pm$1.77 & 19 & & \\
\hline
\end{tabular}
\begin{list}{}{}
\item[]References. 1: \cite{zha13}; 2: \cite{bur08}; 3: \cite{bur09}; 4: \cite{cus09}; 5: \cite{dah08}; 6: \cite{fah09}; 7: \cite{fah12}; 8: \cite{giz97}; 9: \cite{gli91}; 10: \cite{har80}; 11: \cite{jao11}; 12: \cite{jen52}; 13: \cite{kir10}; 14: \cite{sch09}; 15: \cite{van95}; 16: \cite{van07}; 17: \cite{woo09}; 18: \cite{lep03a}; 19: \cite{gat09}.
%\item[$^{a}$] Spectral types are modified from original papers according to \citet{zha13}. 
%\item[$^{a}$] 
\end{list}
%\end{minipage}
\label{tplxe}
\end{table*}

\subsubsection{Spectroscopic distances }
\label{specd}
 M, L and T dwarfs   are known in relatively large numbers in the solar neighbourhood and have recently improved absolute magnitude versus spectral type relationships \cite[e.g.][]{fah12,dup12}. M and L subdwarfs are much less numerous in nearby space and their absolute magnitude and spectral type relationships have not been well constrained. To estimate distances of our M subdwarfs we determined relationships between spectral types and absolute magnitudes ($M_{r, i, z, J, H, Ks}$) based on SDSS and 2MASS filters. We collected a sample of M and L dwarfs\footnote[1]{References of parallax distances for M and L dwarfs: \cite{gli91,per97,giz97,dah02,vrb04,cos05,cos06,jao05,jao11,hen06,lep09,sma10,rie10,and11,fah12}. } and subdwarfs\footnote[2]{See table \ref{tplx} for references of parallax distances of M and L subdwarfs.} with parallax distances from the literature.

Fig. \ref{sptm} shows spectral type - absolute magnitude relationships of M and L types of dwarfs and subdwarfs. The parallax sample of available M subdwarfs is classified under the system of \citet{giz97} which has three metal classes: dM, sdM and esdM. M subdwarfs with metal class of usdM \citep[see Fig. 3 of ][]{lep07} are included in the esdM metal class of \citet{giz97}.  Table \ref{tplx} shows parallax measurements of M and L subdwarfs used in Fig. \ref{sptm}. We fitted $M_{r, i, z}$ of  M and L subdwarfs and extreme subdwarfs together with straight lines. We fitted $M_{J, H, Ks}$ of M subdwarfs and M0--L7 extreme subdwarfs  with straight lines separately. 
The differences in absolute magnitudes between esdM and usdM subdwarfs of the same subtype are smaller than the fitting errors.
% The distribution of data and sample size do not warrant higher order polynomial fits and are reasonably well represented by these straight lines. 

Tables \ref{tpolyfitsd}  \& \ref{tpolyfitd} show the coefficients of polynomial fits of the SDSS and 2MASS magnitudes as a function of spectral type for the M and L  subdwarfs and dwarfs plotted in Fig. \ref{sptm}. Table \ref{sdk7} shows average absolute magnitudes in 2MASS $ J, H, Ks $ bands for K7 subdwarfs in Tables \ref{tpolyfitd} \& \ref{tpolyfitsd}. 
Early-type M subdwarfs are fainter than the same subtype dwarfs in optical bands and even more so in near-infrared bands. It is clear that esdMs are fainter than sdMs, and sdMs are fainter than dMs for $\leq$M5 types.  However, late-type M and L subdwarfs appear to be brighter than normal dwarfs with the same subtypes for $M_{r, i, z, J}$, and similar to that of dwarfs for $M_{H, Ks}$. The H$_2$ collision-induced absorption \citep{sau94} becomes stronger as metallicity goes down and suppresses the $H$ and $Ks$ band flux, thus the near-infrared spectra become bluer. While the dust cloud delays the suppressing of  spectra below 1 $\mu$m \citep{wit09}. 
%Red and ultracool dwarfs with sub-solar abundance are bluer than these with solar abundance, thus could be called purple dwarfs. 
From Fig. \ref{sptm} we can see ultracool subdwarfs (UCSDs) may not be a suitable name for metal-deficient ultracool dwarfs (UCDs) because they are not less luminous than the same subtype UCDs. `Purple dwarfs' might be a sensible name for such bluish and very-red UCDs with subsolar abundance.

\begin{table}
 \centering
 %\begin{minipage}{140mm}
  \caption{Coefficients of polynomial fits to magnitudes versus spectral types of M and L subdwarfs}
  \begin{tabular}{c c c c c}
\hline\hline
    $M_{\rm abs}$ &  $c_0$ &  $c_1$ & rms (mag) & SpT range \\ 
\hline
$M_{r}$ & 10.6168 & 0.5619  & 0.66 &  e/sdM0--e/sdL7 \\
$M_{i}$ &  10.1092 & 0.4300  & 0.62 &  e/sdM0--e/sdL7 \\
$M_{z}$ & 9.6501 & 0.3275 & 0.60 & e/sdM0--e/sdL7\\
$M_{J}$ & 7.6626 & 0.3773  &0.45 & sdM0--sdM9.5\\
$M_{J}$ & 8.7385 & 0.2507  &0.42 & esdM0--esdL7 \\
$M_{H}$ & 7.1279 & 0.3928  &0.46 & sdM0--sdM9.5 \\
$M_{H}$ & 8.1861 & 0.2731   &0.43 & esdM0--esdL7 \\
$M_{Ks}$ & 6.8886 & 0.4013  &0.45 & sdM0--sdM9.5 \\
$M_{Ks}$ & 7.9591 & 0.2860   & 0.47& esdM0--esdL7 \\
\hline
\end{tabular}
\begin{list}{}{}
\item[]Note. Coefficients of first-order polynomial fits of the SDSS, 2MASS absolute magnitudes ($M_{\rm abs}$) as a function of spectral types (SpT) for M0--L7 subdwarfs in Fig. \ref{sptm}. The fits are defined as  $M_{\rm abs} = c_0 + c_1 \times {\rm SpT}$ where SpT = 0 for M0, SpT = 5 for M5, SpT = 10 for L0, SpT = 15 for L5. The rms errors and applicable ranges of spectral types are indicated in the last two columns.  
%\item[$^{a}$] 
%%\end{minipage}
\end{list}
\label{tpolyfitsd}
\end{table}

\begin{table*}
 \centering
 %\begin{minipage}{140mm}
  \caption{Coefficients of polynomial fits to magnitudes versus spectral types of M and L dwarfs}
  \begin{tabular}{c c c c c c c c}
\hline\hline
    $M_{\rm abs}$ &  $c_0$ &  $c_1$ &  $c_2$ &  $c_3$&  $c_4$ &  $c_5$ & rms (mag)  \\ 
\hline
$M_{r}$ &  9.8326 & -1.8452 & 1.1111 &   -1.5169E--1 & 8.4503E--3 & -1.6751E--4 & 0.71 \\
$M_{i}$ & 8.6772 &  -1.0400 & 7.1114E--1  & -9.6139E--2 & 5.3482E--3 & -1.0634E--4 &  0.70 \\
$M_{z}$ & 7.8303  & -4.7257E--2 & 3.3379E--1 & -4.9888e--2 & 2.9652E--3 & -6.2600E--5 &  0.60 \\
$M_{J}$ &  6.2826 &  -2.9785E--1 & 3.6508E--1 & -4.9840E--2 & 2.7283E--3 &   -5.3057E--5 & 0.52 \\ 
$M_{H}$ &  5.7798 & -3.9566E--1 & 4.0397E--1 & -5.5512E--2 & 3.0343E--3 & -5.8608E--5 &  0.54 \\
$M_{Ks}$ & 5.4490  & -3.0303E--1 & 3.7299E--1 & -5.2136E--2 &  2.8626E--3 & -5.5258E--5 &  0.53 \\
\hline
\end{tabular}
\begin{list}{}{}
\item[]Note. Coefficients of fifth-order polynomial fits of the SDSS, 2MASS absolute magnitudes  as a function of spectral types for M0-L8 dwarfs in Fig. \ref{sptm}. The fits are defined as $$M_{\rm abs} = \sum_{i=0}^5 c_i \times ({\rm SpT})^{i}$$ where SpT = 1 for M1, SpT = 5 for M5, SpT = 10 for L0, SpT = 15 for L5. Optical spectral types are applied to fits of SDSS $r, i, z$ magnitudes. Near-infrared spectral types are applied to fits of 2MASS $J, H, Ks$ magnitudes. The rms errors are indicated in the last column. The fits  are applicable from M1 to L8 for $M_{r, i, z}$ and from M1 to L9 for $M_{J, H, Ks}$. 
%\item[$^{a}$] 
%%\end{minipage}
\end{list}
\label{tpolyfitd}
\end{table*}

\begin{table}
 \centering
 %\begin{minipage}{140mm}
  \caption{Absolute magnitudes of K7 type subdwarfs}
  \begin{tabular}{c c c}
\hline\hline
 $M_{\rm abs}$ &  sdK7 & esdK7 \\ 
\hline
$M_{J}$ & 6.99$\pm$0.47 & 8.62$\pm$0.38 \\
$M_{H}$ & 6.41$\pm$0.48 &  8.14$\pm$0.38 \\
$M_{Ks}$ & 6.24$\pm$0.49 & 7.96$\pm$0.38 \\
\hline
\end{tabular}
%\begin{list}{}{}
%%\end{minipage}
%\end{list}
\label{sdk7}
\end{table}

\subsubsection{Space motions}
We estimated distances of our M dwarf and subdwarf samples based on spectral type -- absolute magnitude relationships derived in Section \ref{specd}. For objects detected in 2MASS, we used the mean value of distances estimated with spectral type versus $M_{J, H, Ks}$ relationships, since they have smaller root-mean-square errors than the optical bands (usdMs are treated as esdMs). For objects not detected in 2MASS or which have no errors for $J, H, Ks$ band magnitudes, we estimated their distances with spectral type versus $M_{r, i, z}$ relationships, and the mean value of three distances for each object is adopted as a final distance.

\begin{figure} %  figure placement: here, top, bottom, or page
   \centering
   \includegraphics[width=\columnwidth, angle=0]{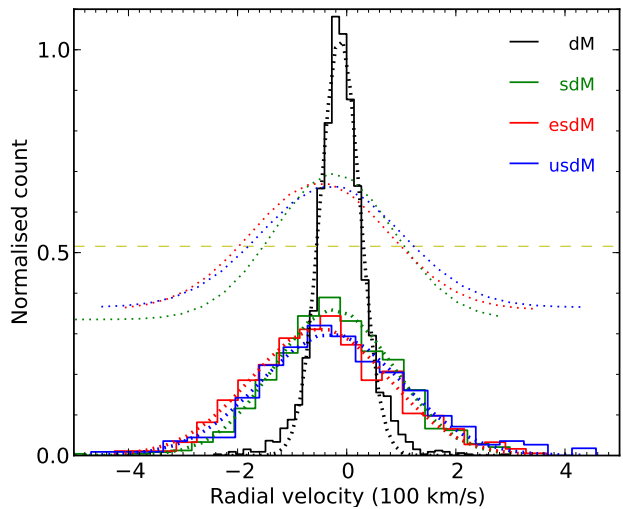}
   \caption{Radial velocity and error distribution of dMs (black), sdMs (green), esdMs (red) and usdMs (blue). All distributions are normalized so that the area, e.g. the sample for each class, is  equal to one. Dotted lines are best Gaussian fits. A yellow dashed line shows the half-maximum of dMs fit. Additional fit lines of sdMs, esdMs and usdMs are also plotted and shifted to the yellow line by their half-maximum.  Typical errors of radial velocities are 3-10 km$\cdot$s$ ^{-1} $. }
   \label{rv}
\end{figure}

Stellar Doppler shifts are computed using the ELODIE library \citep{pru01} spectra as templates with the SDSS pipeline \citep{ade08,aih11}. These Doppler shifts represent the best estimate of the radial velocity of the star.
 Fig. \ref{rv} shows the normalized radial velocity and error distribution of dMs, sdMs, esdMs and usdMs. The full width at half-maximum (FWHM) of the best Gaussian fits for dMs, sdMs, esdMs and usdMs are 85.71, 257.54, 296.54 and 303.11 km s$^{-1}$ respectively. The M subdwarfs velocities are larger than that of M dwarfs, which is consistent with these M subdwarfs being members of the Galactic halo; while the M dwarfs derive from the Galactic disc. 
 
\begin{figure*} %  figure placement: here, top, bottom, or page
   \centering
   \includegraphics[width=\textwidth, angle=0]{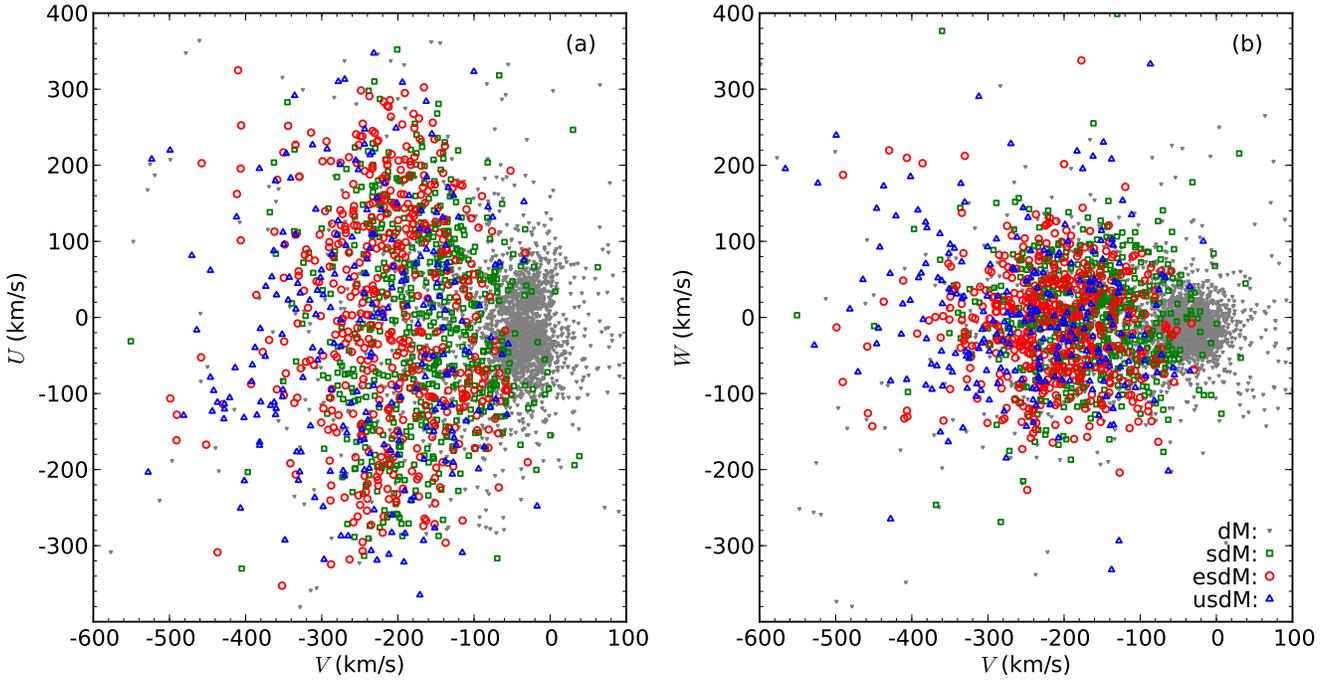}
   \caption{ $ U, V, W $ Galactic velocities of dM (grey down-pointing triangles), sdM (green squares), esdM (red circles) and usdM (blue up-pointing triangles) dwarfs. Note that $ U $ is positive towards the Galactic anti centre.}
   \label{uvwv}
\end{figure*}

\begin{figure*} %  figure placement: here, top, bottom, or page
   \centering
   \includegraphics[width=\textwidth, angle=0]{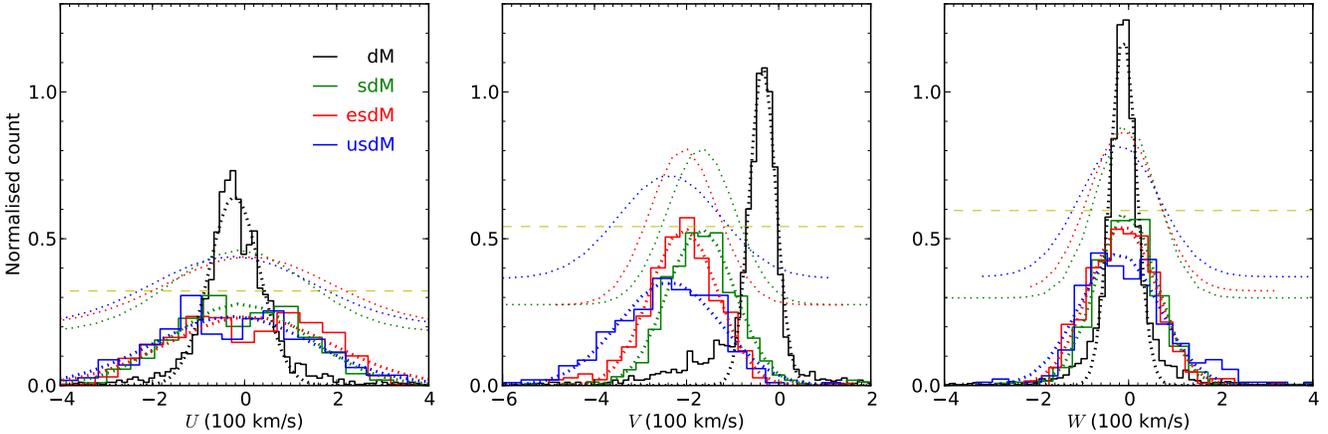}
   \caption{Histograms of $ U, V, W $ Galactic velocities for dM, sdM, esdM and usdM dwarfs. All distributions are normalized so that the area, e.g. the sample for each class, is  equal to one. Dotted lines are best Gaussian fits. A yellow dashed line show the half-maximum of dMs fit. Additional fit lines of sdMs, esdMs and usdMs are also plotted and shifted to the yellow line by their half-maximum.}
   \label{uvwhis}
\end{figure*}

\begin{figure} %  figure placement: here, top, bottom, or page
   \centering
   \includegraphics[width=\columnwidth, angle=0]{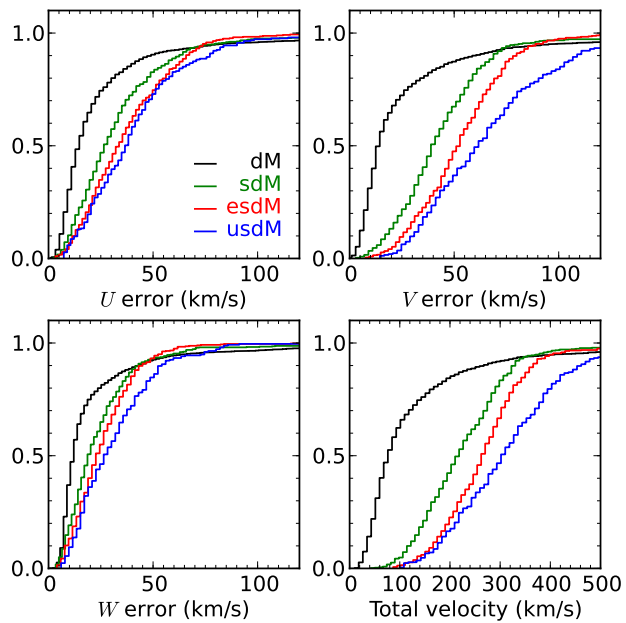}
   \caption{Cumulative histograms of errors of $ U, V, W $ Galactic velocities and total space velocity for dM, sdM, esdM and usdM dwarfs (from top left to bottom right). }
   \label{uvwverr}
\end{figure}

With  PMs and radial velocities from SDSS and spectroscopic distances estimated from Section \ref{specd}, we calculated the \emph{U, V, W}  space velocities of our M subdwarfs. Fig. \ref{uvwv} shows the space velocities in \emph{V - U}  and  \emph{V - W} spaces. Fig. \ref{uvwhis} shows distributions of $U, V, W $ Galactic velocities for dMs, sdMs, esdMs and usdMs. Fig. \ref{uvwverr} shows cumulative histograms of errors of $U, V, W$ and total space velocities. The lack of objects around $ U \sim 0$ km s$^{-1}$  reflects the fact that only objects with PMs higher than 100 mas$\cdot$yr$^{-1}$ are selected in our sample, thus some distant early-type M subdwarfs are missed. 
The $V$ space velocity distributions of dMs, sdMs, esdMs and usdMs have their maxima at  $ -34.64, -168.09, -206.08$ and $ -237.19 $ km s$^{-1}$, and we find FWHM of 67.95, 172.43, 171.65 and 257.59 km s$^{-1}$ respectively, assuming a Gaussian distribution. The $V$ velocity distribution of M dwarfs can not be fitted well with a single Gaussian line (Fig. \ref{uvwhis}). It appears that some of the M dwarfs ($\sim$18\%) have halo-like velocities ($ V < -100$ km s$^{-1}$). A single Gaussian also can not fit the distribution of $ W $ velocity (17\% have $ W > 50$ km s$^{-1}$ or $ W < -80$ km s$^{-1}$). This means a fraction of M dwarfs  have halo like kinematics. There are also some M subdwarfs which have disc kinematics in the original sample. The study by \citet{spa10} based on FGK stars shows that the metallicity tail of the thick disc population goes down to [$M$/H]$\sim -$1.2. Halo and disc populations have overlaps in metallicity and kinematics. Thus if a star has either metallicity as low as [$M$/H] $\sim -$1.2 or halo-like kinematics, this does not always mean it is belongs to the Population II.

\begin{figure*} %  figure placement: here, top, bottom, or page
   \centering
   \includegraphics[width=\textwidth, angle=0]{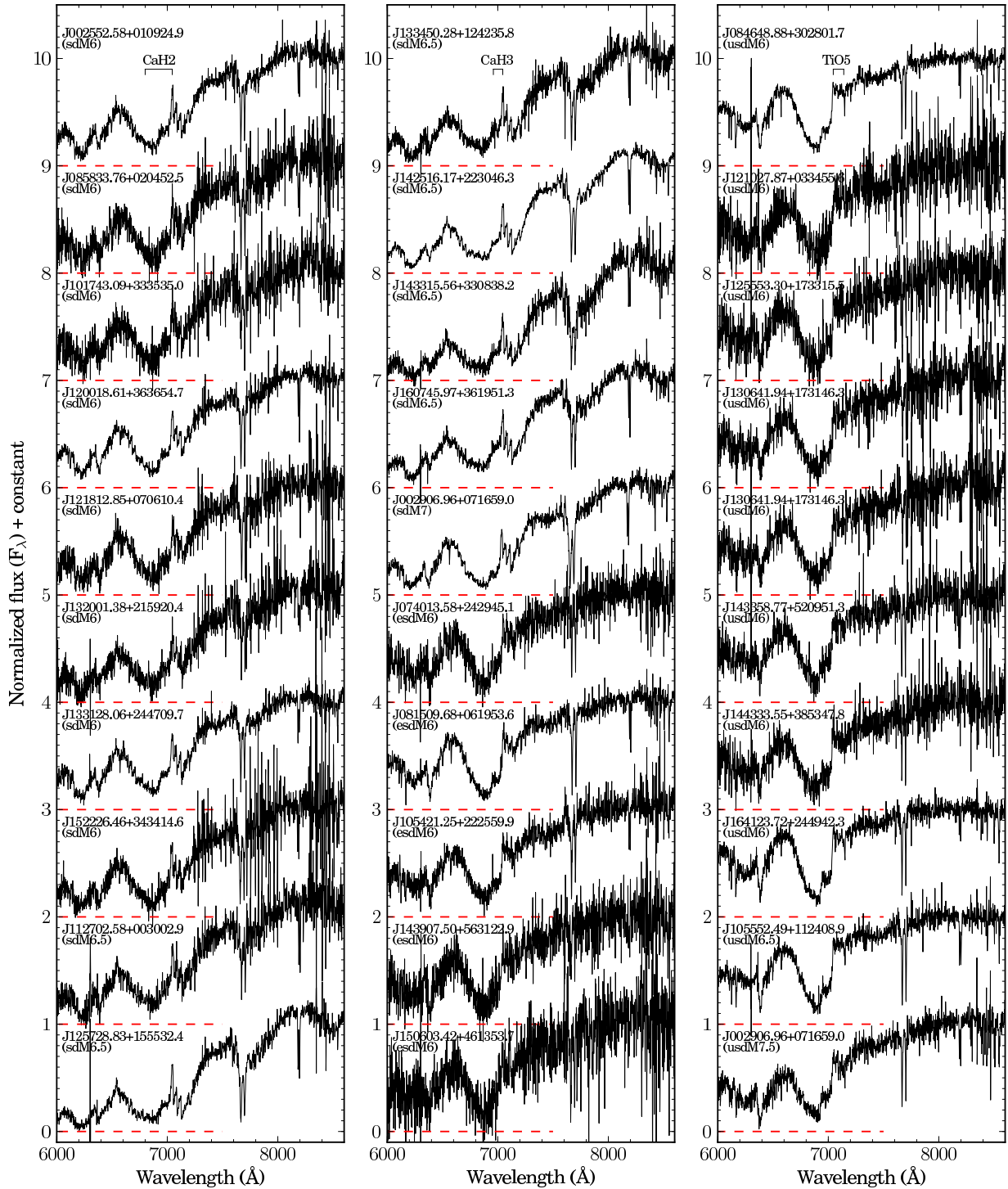}
   \caption{ SDSS optical spectra of 30 RSDs with spectral types of  $\geqslant$sdM6, $\geqslant$esdM6 and $\geqslant$usdM6. The SDSS name and spectral type are indicated above each spectrum. Absorption bands of CaH2, CaH3 and TiO5 are also indicated above top spectra.  All spectra are normalized at  8000 \AA.}
   \label{ucsd}
\end{figure*}

\begin{figure} %  figure placement: here, top, bottom, or page
   \centering
   \includegraphics[width=\columnwidth]{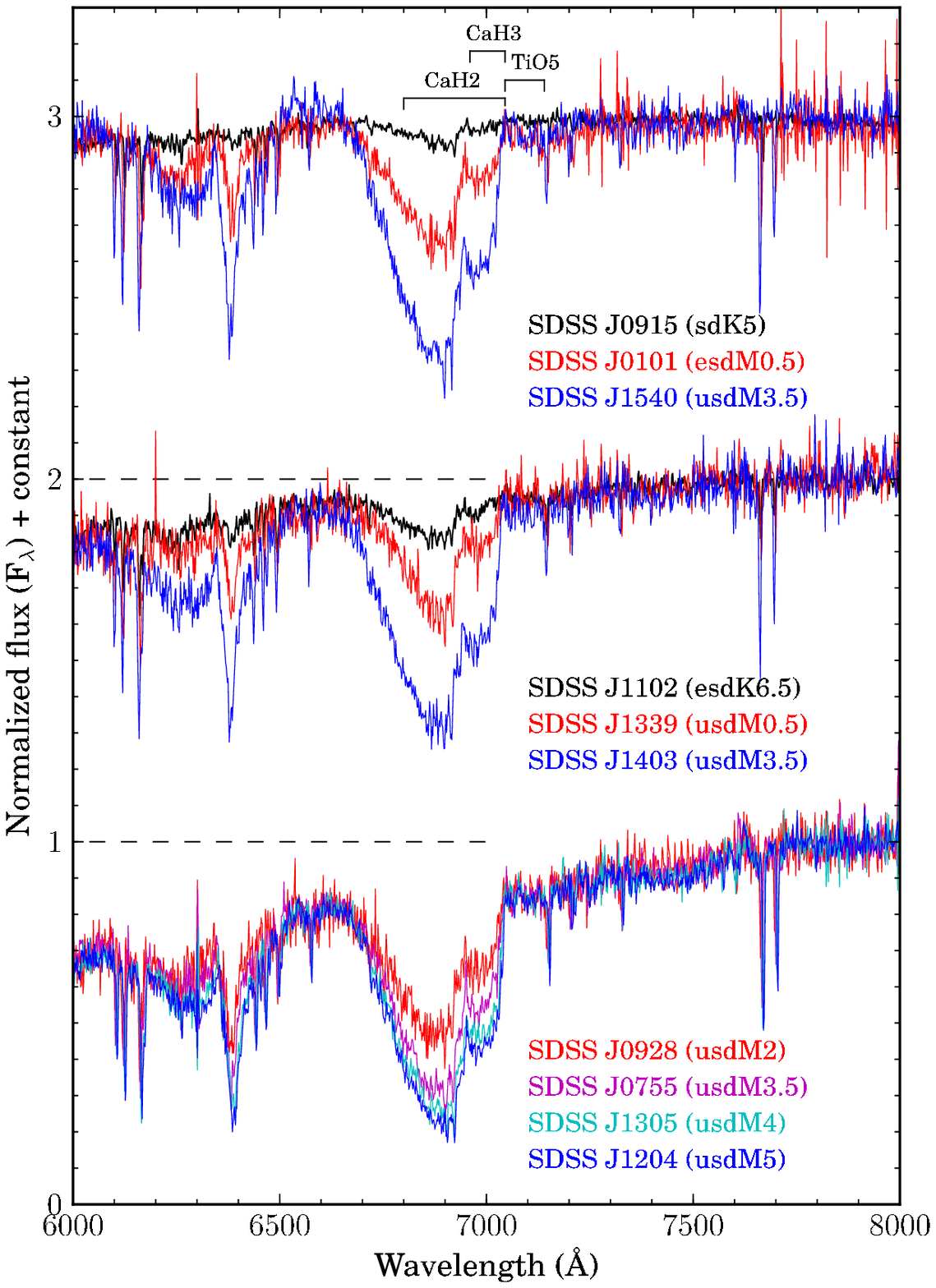}
  \caption{SDSS spectra of RSDs with different gravities. Subtypes indicated in the plot are flawed for high gravity subdwarfs. Small shifts between lines of different spectra along wavelength are due to different high radial velocities. All spectra are normalized at  8000 \AA.}   
   \label{highg}
\end{figure}

\subsubsection{New late-type M subdwarfs}
Cool subdwarfs with spectral types of late-type M and L are referred to as UCSDs \citep[e.g.][]{bur09b} following the definition of UCDs \citep[e.g.][]{kir97}.  UCSDs are important for our understanding of metal-poor ultracool atmospheres. UCSDs exhibit complex spectra dominated by molecular absorption bands and metal lines. Spectra of UCSDs are affected by their low effective temperature, subsolar abundance and gravity in a complicated way. Current atmospheric models do not reproduce observed spectra of UCSDs \citep{bur07}. UCSDs with different properties ($T_{\rm eff}$, [$M$/H], gravity, multiplicity) are very useful to test and calibrate models of ultracool atmospheres \citep{bur01,mar02,hel08,wit09} and low-mass stellar evolution scenarios \citep{bar97,bar03,mon00}. 

To date, there have been only about 80 UCSDs  discovered \cite{giz97,giz06,sch99,lep03b,lep03c,lep04,lep07,lep08,bur03,bur04,bur06,bur07,sch04a,sch04b,mar08,jao08,siv09,cus09,lod10,lod12,kir10}. 42 subdwarfs in our sample have spectral types of $\geqslant$M6 , 12 of which are known $\geqslant$M7 \citep{lep08}, 30 of which are new $\geqslant$M6  including 9 M6.5--M7.5. Table \ref{m6} shows photometry and PMs of these 30 new $\geqslant$M6  subdwarfs. Fig. \ref{ucsd} shows spectral sequences of these  $\geqslant$sdM6, $\geqslant$esdM6 and $\geqslant$usdM6 subdwarfs. Spectral types are assigned  according to a metallicity index $\zeta_{\rm TiO/CaH}$ defined by absorption bands of CaH2, CaH3 and TiO5 \citep{lep07}. From these spectra we can see that both CaH and TiO bands are sensitive to effective temperature, but that the TiO bands are more sensitive to metallicity compared to CaH bands. Spectral types of these late-type M subdwarfs have uncertainties of 0.5-1.0 because some spectra do not have a very high signal-to-noise ratio. The actual uncertainty of the spectral type classification could be larger because the effects of gravity  are not included in the $\zeta_{\rm TiO/CaH}$ index. In some extreme cases, gravity  could changes the $\zeta_{\rm TiO/CaH}$ index by an equivalent  of three subtypes. We will discuss the impact of gravity on the spectra of M subdwarfs in Section \ref{sgravity}.   

\begin{table*}
 \centering
 %\begin{minipage}{140mm}
  \caption{30 new subdwarfs with spectral types of M6 and later.}
  \begin{tabular}{c c c c c r r r}
\hline\hline
    SDSS name &  SDSS \emph{g} &  SDSS \emph{r}  & SDSS \emph{i} & SDSS \emph{z} & $\mu_{\rm RA}$(mas/yr)  & $\mu_{\rm Dec}$(mas/yr) &  SpT~~  \\
\hline
J002552.58+010924.9 & 21.06$\pm$0.03 & 19.30$\pm$0.02 & 17.83$\pm$0.02 & 17.06$\pm$0.02 & 159.34$\pm$5.13 & -195.61$\pm$5.13 & sdM6~~ \\
J002906.96+071659.0 & 21.31$\pm$0.04 & 19.24$\pm$0.02 & 17.62$\pm$0.02 & 16.85$\pm$0.02 & 185.85$\pm$4.60 & -176.32$\pm$4.60  & sdM7~~ \\
J013346.24+132822.4 & 20.82$\pm$0.04 & 18.96$\pm$0.02 & 17.82$\pm$0.02 & 17.19$\pm$0.02 & 80.02$\pm$4.70 & -296.67$\pm$4.70 & usdM7.5 \\
J074013.58+242945.1 & 22.15$\pm$0.08 & 20.18$\pm$0.03 & 19.09$\pm$0.02 & 18.55$\pm$0.05 & -119.28$\pm$5.29 & -115.95$\pm$5.29  & esdM6~~ \\
J081509.68+061953.6 & 21.42$\pm$0.04 & 19.34$\pm$0.02 & 18.10$\pm$0.02 & 17.43$\pm$0.02 & 41.59$\pm$3.09 & -251.88$\pm$3.09 &  esdM6~~ \\
J084648.88+302801.7 & 20.52$\pm$0.03 & 18.51$\pm$0.02 & 17.48$\pm$0.01 & 16.83$\pm$0.02 & -26.22$\pm$4.34 & -375.02$\pm$4.34  & usdM6~~ \\
J085833.76+020452.5 & 22.02$\pm$0.09 & 20.40$\pm$0.03 & 19.03$\pm$0.02 & 18.20$\pm$0.03 & 96.82$\pm$3.90 & -226.60$\pm$3.90 &  sdM6~~ \\
J101743.09+333535.0 & 22.44$\pm$0.10 & 20.45$\pm$0.03 & 19.15$\pm$0.02 & 18.47$\pm$0.04 & -109.52$\pm$5.29 & -243.27$\pm$5.29 & sdM6~~ \\
J105421.25+222559.9 & 22.17$\pm$0.07 & 20.16$\pm$0.03 & 19.03$\pm$0.02 & 18.40$\pm$0.03 & -201.90$\pm$17.37 & -310.00$\pm$17.37  & esdM6~~ \\
J105552.49+112408.9 & 21.08$\pm$0.04 & 19.01$\pm$0.02 & 17.93$\pm$0.02 & 17.33$\pm$0.02 & -232.08$\pm$4.53 & -20.84$\pm$4.53 &  usdM6.5 \\
J112702.58+003002.9 & 21.71$\pm$0.06 & 19.91$\pm$0.02 & 18.48$\pm$0.02 & 17.67$\pm$0.02 & -67.31$\pm$8.09 & -174.90$\pm$8.09 &  sdM6.5 \\
J120018.61+363654.7 & 20.94$\pm$0.04 & 18.98$\pm$0.02 & 17.61$\pm$0.01 & 16.83$\pm$0.02 & -314.66$\pm$3.28 & -197.36$\pm$3.28 & sdM6~~ \\
J121027.87+033455.6 & 22.27$\pm$0.10 & 20.30$\pm$0.03 & 19.36$\pm$0.02 & 18.89$\pm$0.05 & 137.05$\pm$6.09 & 132.79$\pm$6.09 &  usdM6~~ \\
J121812.85+070610.4 & 22.07$\pm$0.06 & 20.24$\pm$0.03 & 18.93$\pm$0.02 & 18.18$\pm$0.02 & -343.08$\pm$5.20 & -44.55$\pm$5.20 &  sdM6~~ \\
J125553.30+173315.5 & 22.62$\pm$0.09 & 20.53$\pm$0.03 & 19.56$\pm$0.02 & 19.03$\pm$0.04 & 4.21$\pm$3.66 & -153.31$\pm$3.66 & usdM6~~ \\
J125728.83+155532.4 & 21.45$\pm$0.05 & 19.65$\pm$0.02 & 17.75$\pm$0.01 & 16.74$\pm$0.02 & 157.97$\pm$13.22 & -546.98$\pm$13.22 &  sdM6.5 \\
J130641.94+173146.3 & 22.33$\pm$0.07 & 20.32$\pm$0.03 & 19.37$\pm$0.02 & 18.78$\pm$0.04 & -177.61$\pm$4.51 & -55.11$\pm$4.51 &  usdM6~~ \\
J132001.38+215920.4 & 22.19$\pm$0.08 & 20.45$\pm$0.03 & 18.97$\pm$0.03 & 18.35$\pm$0.03 & -157.81$\pm$5.39 & -100.18$\pm$5.39 & sdM6~~ \\
J133128.06+244709.7 & 20.70$\pm$0.03 & 18.93$\pm$0.02 & 17.56$\pm$0.02 & 16.90$\pm$0.02 & -489.51$\pm$13.24 & -231.33$\pm$13.24  & sdM6~~ \\
J133450.28+124235.8 & 21.65$\pm$0.05 & 19.85$\pm$0.03 & 18.25$\pm$0.02 & 17.41$\pm$0.02 & -79.76$\pm$4.86 & -79.06$\pm$4.86 &  sdM6.5 \\
J142516.17+223046.3 & 20.60$\pm$0.03 & 18.81$\pm$0.02 & 17.07$\pm$0.01 & 16.19$\pm$0.02 & -243.34$\pm$3.14 & -88.49$\pm$3.14 & sdM6.5 \\
J143315.56+330838.2 & 22.20$\pm$0.08 & 20.46$\pm$0.03 & 18.55$\pm$0.01 & 17.62$\pm$0.02 & -73.11$\pm$4.96 & -327.09$\pm$4.96 &  sdM6.5 \\
J143358.77+520951.3 & 22.12$\pm$0.08 & 19.94$\pm$0.02 & 18.94$\pm$0.02 & 18.36$\pm$0.03 & -157.79$\pm$5.31 & -128.54$\pm$5.31 & usdM6~~ \\
J143907.50+563122.9 & 22.83$\pm$0.23 & 20.50$\pm$0.04 & 19.37$\pm$0.03 & 18.86$\pm$0.05 & -97.47$\pm$5.77 & -115.19$\pm$5.77 &  esdM6~~ \\
J144333.55+385347.8 & 22.13$\pm$0.08 & 20.10$\pm$0.03 & 19.05$\pm$0.02 & 18.49$\pm$0.03 & -153.15$\pm$4.65 & -191.93$\pm$4.65 & usdM6~~ \\
J150603.42+461353.7 & 22.21$\pm$0.09 & 20.33$\pm$0.03 & 19.30$\pm$0.02 & 18.67$\pm$0.04 & -120.98$\pm$5.68 & -49.45$\pm$5.68 &  esdM6~~ \\
J152226.46+343414.6 & 21.85$\pm$0.05 & 20.05$\pm$0.02 & 18.56$\pm$0.02 & 17.78$\pm$0.02 & -166.36$\pm$4.76 & -117.88$\pm$4.76 &  sdM6~~ \\
J153647.08+025501.5 & 22.09$\pm$0.09 & 20.12$\pm$0.03 & 19.07$\pm$0.02 & 18.48$\pm$0.04 & -192.88$\pm$5.33 & -42.51$\pm$5.33 &  esdM6~~ \\
J160745.97+361951.3 & 21.11$\pm$0.04 & 19.40$\pm$0.02 & 17.80$\pm$0.01 & 17.00$\pm$0.02 & 44.95$\pm$4.23 & -136.49$\pm$4.23 &  sdM6.5 \\
J164123.72+244942.3 & 20.33$\pm$0.02 & 18.31$\pm$0.01 & 17.23$\pm$0.02 & 16.61$\pm$0.02 & -630.32$\pm$15.91 & -177.37$\pm$15.91 &  usdM6~~ \\
\hline
\end{tabular}
%\begin{list}{}{}
%\item[]Note. 
%\item[$^{a}$] 
%\end{list}
%\end{minipage}
\label{m6}
\end{table*}

\begin{figure} %  figure placement: here, top, bottom, or page. 
   \centering
   \includegraphics[width= \columnwidth]{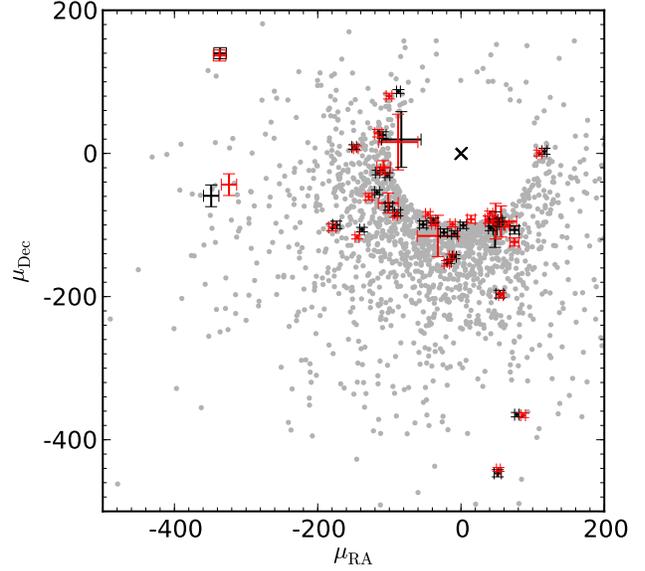}
   \caption{30 subdwarf common PM pairs. Companions in   each binary are plotted in black and red error bars. A multiplication   symbol indicates the location of (0, 0).  Light grey dots are RSDs in our sample.  }
   \label{pm}
\end{figure}

\subsubsection{High-gravity M ultra subdwarfs}
\label{sgravity}
During our visual inspection of M subdwarf spectra we found some M subdwarfs that have very strong CaH bands, and appear up to three subtypes later than normal M subdwarfs with the same overall profile. The CaH and TiO indices are used to assign spectral types and metal class for M subdwarfs \citep{giz97,lep07}. Gravity is not considered in the classification of M subdwarfs. The study by \citet{jao08} based on Gaia model grids \citep{bro05} suggests that CaH and TiO absorption bands are both good indicators of effective temperature. The TiO is more sensitive to metallicity changes compared to CaH.  The CaH is very sensitive to gravity changes but the TiO does not appear to be sensitive to gravity at all.  \citet{jao08} also suggests that overall spectral profiles could be used as a major indicator of effective temperature. In this system, spectra with similar overall profiles and TiO indices would have similar effective temperatures and metallicities, and the variation of CaH indices would represents  gravity changes.

We inspected all  subdwarf spectra for unusual, relative CaH strength and found a large variation of the depth of CaH bands among M subdwarfs with the same overall profile and depth of TiO band. Fig. \ref{highg} shows SDSS spectra of M subdwarfs with different gravity features.  Spectra of M subdwarfs with normal gravity are over-plotted for comparison: SDSS J091559.72+290817.4 (SDSS J0915; sdK5), SDSS J010131.32$-$002325.6 (SDSS J0101; esdM0.5), SDSS J110252.67+274203.7 (SDSS J1102; esdK6.5), SDSS J133945.66+134747.3 (SDSS J1339; usdM0.5), SDSS J092833.78+425428.8 (SDSS J0928; usdM2). Each set of spectra in Fig. \ref{highg} have similar overall profile and depth of TiO band but have very different depths of CaH bands which are indicators of gravity.  They are classified as different spectral types and metallicity classes according to the classification system of \citet{lep07} as indicated in Fig. \ref{highg}. We note, however, that spectral types of these unusual M subdwarfs as they stand are flawed since they do not take account of the high-gravity of these sources. They should have similar subtypes as normal M subdwarf spectra (SDSS J0101, SDSS J1339, SDSS J0928) over plotted with them in Fig. \ref{highg}, e.g. spectral type of SDSS J1204 should be an usdM2 rather than usdM5.  Table \ref{5hg} shows SDSS photometry, PMs and spectral types of these five high gravity M ultra subdwarfs. 

We only found such large variation of CaH band in our usdM subdwarf sample. The TiO band in spectra of usdM subdwarfs is very weak and barely visible, thus has large measurement errors. If  strengthening CaH bands in these five usdM subdwarfs does not represent high gravity, it probably  indicates low metallicity beyond normal usdM subdwarfs.

\begin{table*}
 \centering
 %\begin{minipage}{140mm}
  \caption{Five M ultra subdwarfs with high gravity}
  \begin{tabular}{c c c c c r r c}
\hline\hline
    SDSS name &  SDSS \emph{g} &  SDSS \emph{r}  & SDSS \emph{i} & SDSS \emph{z} & $\mu_{\rm RA}$ (mas/yr)  & $\mu_{\rm Dec}$ (mas/yr)  & SpT$^{a}$  \\
\hline
J075526.13+482837.3 & 19.71$\pm$0.02 & 17.86$\pm$0.02 & 17.11$\pm$0.02 & 16.62$\pm$0.02 & 244.13$\pm$3.25 & $-$284.59$\pm$3.25 &  usdM3.5 (2.0)\\
J120426.90+132923.3 & 19.34$\pm$0.03 & 17.40$\pm$0.02 & 16.55$\pm$0.02 & 16.06$\pm$0.02 & 109.01$\pm$3.05 & $-$384.54$\pm$3.05 &  usdM5.0 (2.0)\\
J130509.05+641753.2 & 19.75$\pm$0.02 & 17.83$\pm$0.02 & 16.95$\pm$0.01 & 16.48$\pm$0.02 & $-$202.49$\pm$3.33 & $-$251.65$\pm$3.33 &  usdM4.0 (2.0)\\
J140305.50+282424.4 & 19.48$\pm$0.02 & 17.63$\pm$0.02 & 16.77$\pm$0.01 & 16.33$\pm$0.02 & $-$235.77$\pm$2.82 & $-$328.94$\pm$2.82  & usdM3.5 (0.5)\\
J154041.61+265812.0 & 18.60$\pm$0.02 & 16.78$\pm$0.01 & 15.95$\pm$0.01 & 15.48$\pm$0.01 & $-$784.14$\pm$2.63 & $-$120.55$\pm$2.63  & usdM3.5 (0.0)\\
\hline
\end{tabular}
\begin{list}{}{}
\item[$^{a}$] Spectral types listed in the table are based on the classification system of \citet{lep07} which has no consideration for gravity effects, and thus are flawed for these high-gravity objects.  More practicable subtypes of these high-gravity M subdwarfs are indicated in parentheses after their spectral types. 
\end{list}
%\end{minipage}
\label{5hg}
\end{table*}

\subsubsection{Carbon subdwarfs}
\citet{dah77} identified the first dwarf carbon star G 77--61 (LHS 1555), and hypothesized that this object was in fact a double star. The primary ejected carbon material on to the surface of its lower mass companion during its giant branch phase, and then evolves to become a cool WD and is much fainter than the carbon dwarf secondary \citep[e.g.][]{ste05}. Radial velocity variations proved this hypothesis of an unseen component in G 77-61 \citep{dea86}. The $U, V, W$ space motion \citep{dah77} and spectrum fits \citep{gas88,ple05} of G 77--61 indicate that it is a low metallicity object of the Galactic halo. 

Carbon dwarfs are rare objects compared to normal red dwarfs, with only about 120 published \citep[e.g.][]{marg02,low03,dow04}. The metal-poor carbon dwarfs have features of both carbon dwarfs and RSDs, and have not  been distinguished from carbon dwarfs as a new population. 

Five cool carbon dwarfs with strong CaH indices were noticed by \citet{marg02}. They argue that CaH indices present in these stars may be an effective low-resolution luminosity indicator. However, it may be more natural to explain the presence of CaH indices with low metallicity. Strong CaH and weak TiO indices are main features of late-type K and M subdwarfs. These carbon-enriched metal-deficient objects could be called 'carbon subdwarfs' because they have features of both cool subdwarfs and carbon dwarfs (see Section \ref{3sdc} for further discussion). Nine late-type K and M type carbon subdwarfs have been identified in our sample. Five mid-K-type carbon subdwarf candidates are also identified. Table \ref{37dc} shows photometry and PMs of these carbon subdwarfs and 22 cool carbon dwarfs..

\begin{figure*} %  figure placement: here, top, bottom, or page
   \centering
   \includegraphics[width=0.9\textwidth, angle=0]{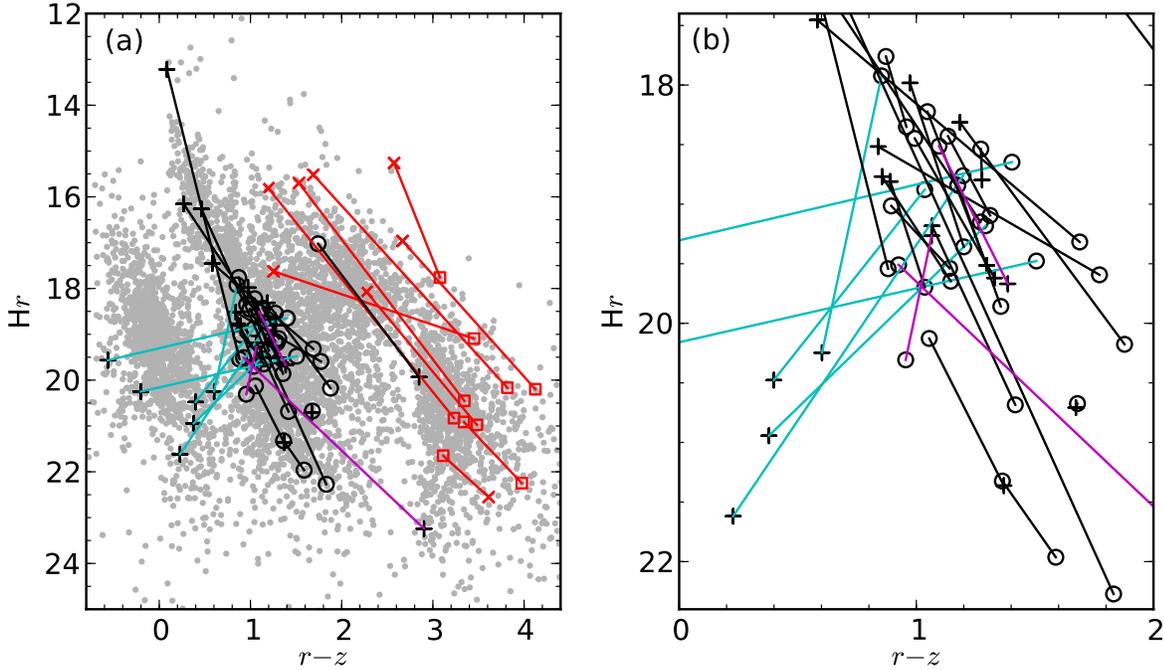}
   \caption{Reduced PMs and $r-z$ colour of our 30 RSDs
     binaries and 9 binaries containing $\geqslant$M6  dwarfs. Panel (b) on the right is a zoom in plot for panel (a) on the left. Our PM
     selected sample (known galaxies excluded) are also plotted on the left-hand
     panel for reference. Black circles represent spectroscopically confirmed RSDs,
     black pluses represent their companions without spectra. RSD + RSD binaries are
     joined with black lines; RSD + WD binaries are joined with cyan lines. Red
     squares represent spectroscopically confirmed $\geqslant$M6  companions, red crosses
     \textbf{x} are companions without spectra, they are joined with red
     lines. Binaries with carbon
     subdwarf companions are joined with magenta lines.  }
   \label{hrzb2}
\end{figure*}

\begin{figure*} %  figure placement: here, top, bottom, or page
   \centering
   \includegraphics[width=0.9\textwidth, angle=0]{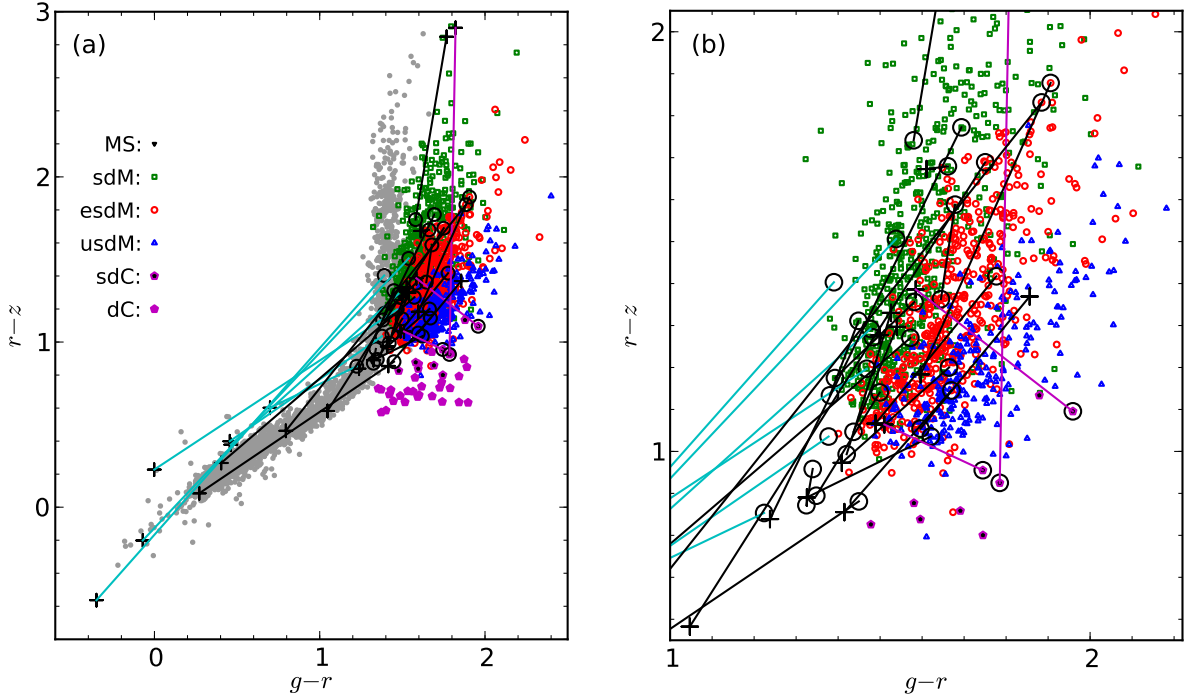}
   \caption{$g-r$ versus $r-z$ colours of our 30 RSD binaries. Green
     squares: sdMs; red circles: esdMs; blue up-pointing triangles: usdMs;
     filled magenta pentagons: carbon dwarfs; magenta pentagons filled with
     black: carbon subdwarfs; grey dots: 3028 point sources with $17<r<18$
     selected from 10 square degrees of SDSS. Black circles represent spectra
     confirmed RSDs, black pluses represent their companions without
     spectra. RSD + RSD binaries are joined with black lines; RSD + WD binaries
     are joined with cyan lines. RSD + carbon subdwarf binaries are joined with
     magenta lines.} 
   \label{grzb2}
\end{figure*}

\begin{figure*} %  figure placement: here, top, bottom, or page
   \centering
   \includegraphics[width=\textwidth, angle=0]{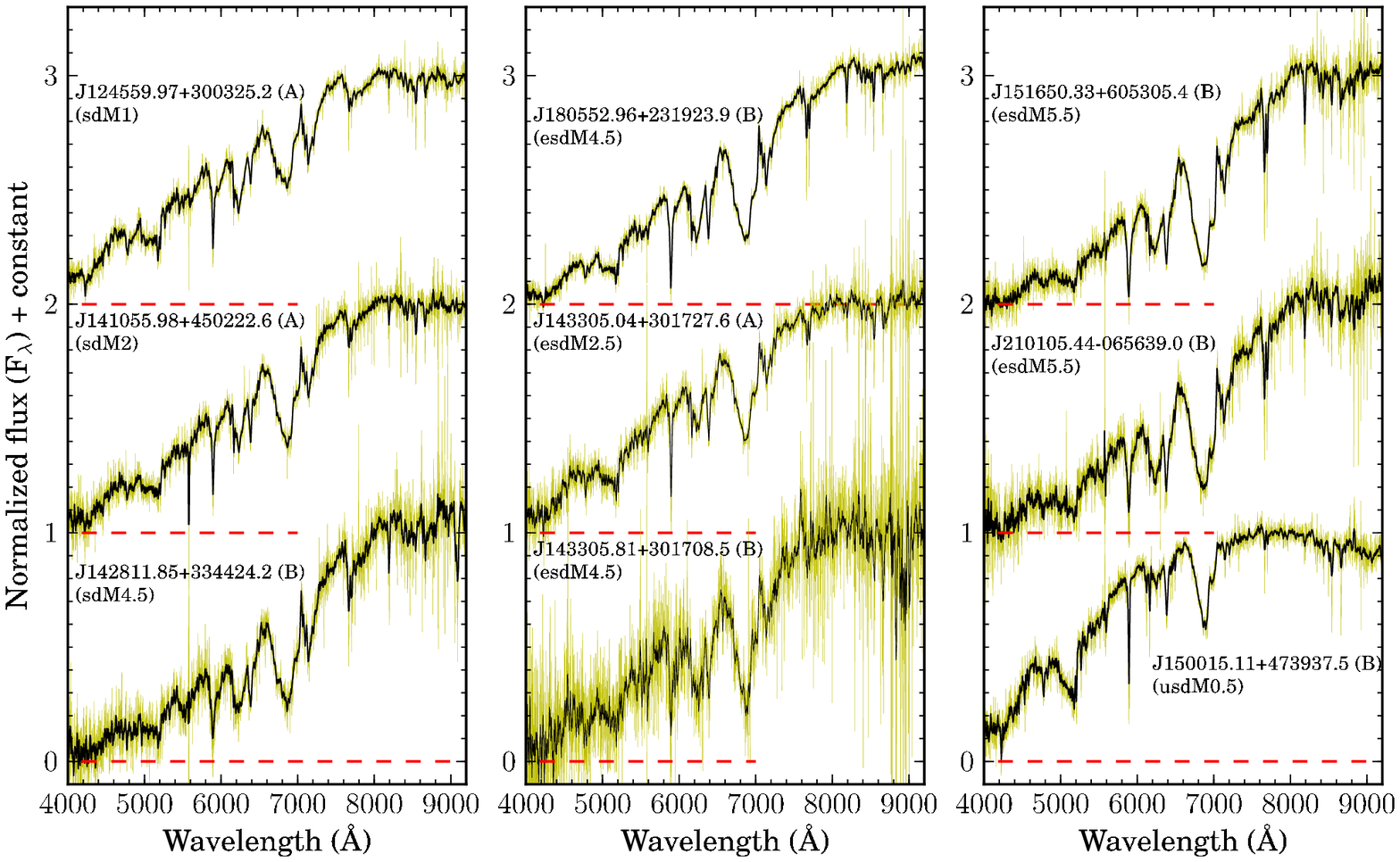}   
      \caption{SDSS spectra of nine subdwarf companions in eight wide binary systems. All spectra are normalized at  8000 \AA. Spectra are binned by 11 pixels, original spectra are plotted in yellow.}
   \label{wbspec}
\end{figure*}

\section{New binary systems}
\label{sbinary}
We used three different methods to identify subdwarf binary systems with different
separations. 
\subsection{Wide common PM binaries} 
Common PM is one of the most useful indicators of wide binary
systems ($>$ 100 au). Many ultracool dwarf binary systems have been successfully identified by this method (e.g. \citealt{fah10,zha10,bur10,day11,pin12}).  

\subsubsection{Cross match}
\label{crossm}
The statistical probability that two objects with common PM higher than 100 mas$\cdot$yr$^{-1}$
and errors less than $ \sim $ 15 mas$\cdot$yr$^{-1}$ within a few arcminutes could occur by random is
usually very small \citep[e.g. $\ll$ 1\%,][]{zha10}. We obtained a sample of 1.81 million objects
with PMs larger than 80 mas$\cdot$yr$^{-1}$ from SDSS DR8 and cross matched this
sample with our PM and spectroscopy sample. To include some possible very wide
binaries we used a separation limit of 9 arcmin, and PM difference of 
15 mas$\cdot$yr$^{-1}$ during this cross match.  Objects in the SDSS PM catalogue generally have errors better than 15 mas$\cdot$yr$^{-1}$ for $ r < 20 $ \citep[Figure 4 of][]{mun04} . Objects with errors larger than 15 mas$\cdot$yr$^{-1}$ are not reliable. 

We first estimate the expected number of random common PM pairs within
our RSD sample and SDSS PM catalogue. The SDSS PM catalogue has 1.81 million
objects with PM $>$ 80 mas$\cdot$yr$^{-1}$. For each of these 1.81 million objects, we
counted the number of common PM pairs with PM differences of less than 15 mas$\cdot$yr$^{-1}$
in the PM sample of 1.81 million objects, without separation constraints. We
then divided the total number of common PM pairs (1.642 billion) by the total number
of objects (1.808 million) and the total coverage of the PM catalogue (14555
 arcmin$^2$) to get the average random common PM density of the whole
sample. The possibility of finding common PM companions within 15 mas$\cdot$yr$^{-1}$ by 
random  within a small area of radius of 9 arcmin 
%is times the area 254.5.1 square arcmins with the density, which 
is  $4.41\times10^{-3}$. Thus, we would
expect to find $1880\times4.41\times10^{-3} = 8.3$ random common PM pairs 
between our RSD sample and the PM catalogue with
1.81 million objects.

50 M subdwarf common PM pairs were found in our PM pair search. Fig. \ref{pm}
(a) shows PMs of 30 RSD binaries. Their PMs are listed in Table
\ref{tcsdb}. To confirm the binary status of our common PM pairs, we did a colour
consistency check of our common PM pairs according to three rules: (I) fainter
companion should have redder colours;  (II) companions
of a binary should associate and line up on the same ridge in the 
reduced PM versus $r-z$ colour plot (Fig. \ref{hrzb2}); (III) companions of a binary should
associate and line up on the same metallicity sequences (Figure
\ref{grzb2}). Rule (I) is also applied when we use rule (II) or (III) for
binarity checking. These three rules do not apply on binaries with WD companions. 32 survived rule (II), and 24 of them also
survived rule (III). We thus believe these 24 common PM pairs are genuine binary
systems. 

Table \ref{tcsdb} shows properties  of  the RSD binaries identified with
common PMs. Of all our wide binaries identified with common PMs, at least one of the
companions is a confirmed RSD with SDSS spectra. Colours and relative brightnesses
of these RSD are consistent with their common PM companions. Fig. \ref{wbspec} shows SDSS spectra of some of these companions. In five systems spectra 
of both components were taken by SDSS.  Fig. \ref{hrzb2} shows the SDSS $r$ band reduced PM and $r-z$ colours of
 these 30 RSD binaries and 9 $\geqslant$M6  dwarf binaries (discovered as a by-product, see Table \ref{tucdb} in the appendix). Our sample (grey dots)
 are separated into three sequences: WD, RSD and M dwarfs from left
 to right.    Fig. \ref{grzb2} shows the $g-r$ and $r-z$ colours of 30 RSD binary
  systems. Four metal sequences of dwarfs, M subdwarfs, M extreme subdwarfs,
  and M ultra subdwarfs are plotted for comparison. A number of carbon
  dwarfs/subdwarfs are also over plotted.   
  Three carbon subdwarfs look inconsistent in Figures \ref{hrzb2} and \ref{grzb2}, suggesting that only one companion in each system is a carbon subdwarf (see Section \ref{3sdc} for further discussion).

\subsubsection{Visual inspection} 
We conducted a systematic search for companions to our M subdwarfs. This
search was conducted by visual inspection of the region of sky around each of
our M subdwarfs using the SDSS Navigate Tool. We inspected images covering
separations out to 1 arcmin on the sky, and looked for objects that could be K
or M subdwarf companions according to rule (I)  (Section
\ref{crossm}). Then we measured PMs of all such selected candidates following the method described in section 5 of \citet{zha09} to test their companionship.  Images available from online data bases of the SDSS, UKIDSS and POSS are used for our PM measurements. Five common PM pairs were found by this method. 

We applied this method to a larger SDSS RSD sample without PM measurements to search for fainter companions. We select binary candidates by their colours, then measure their common PMs to confirm their binary status.  SDSS J150015.11+473937.5 (SDSS J1500; usdM0.5) is found to have a fainter companion with spectral type of usdM3 according to Fig. \ref{sptm}. We are following up more binary candidates from a colour + spectroscopy selected RSD sample.

\begin{figure*}
  \centering
  
  \subfloat[]{\label{j0935ir}\includegraphics[width=0.166\textwidth]{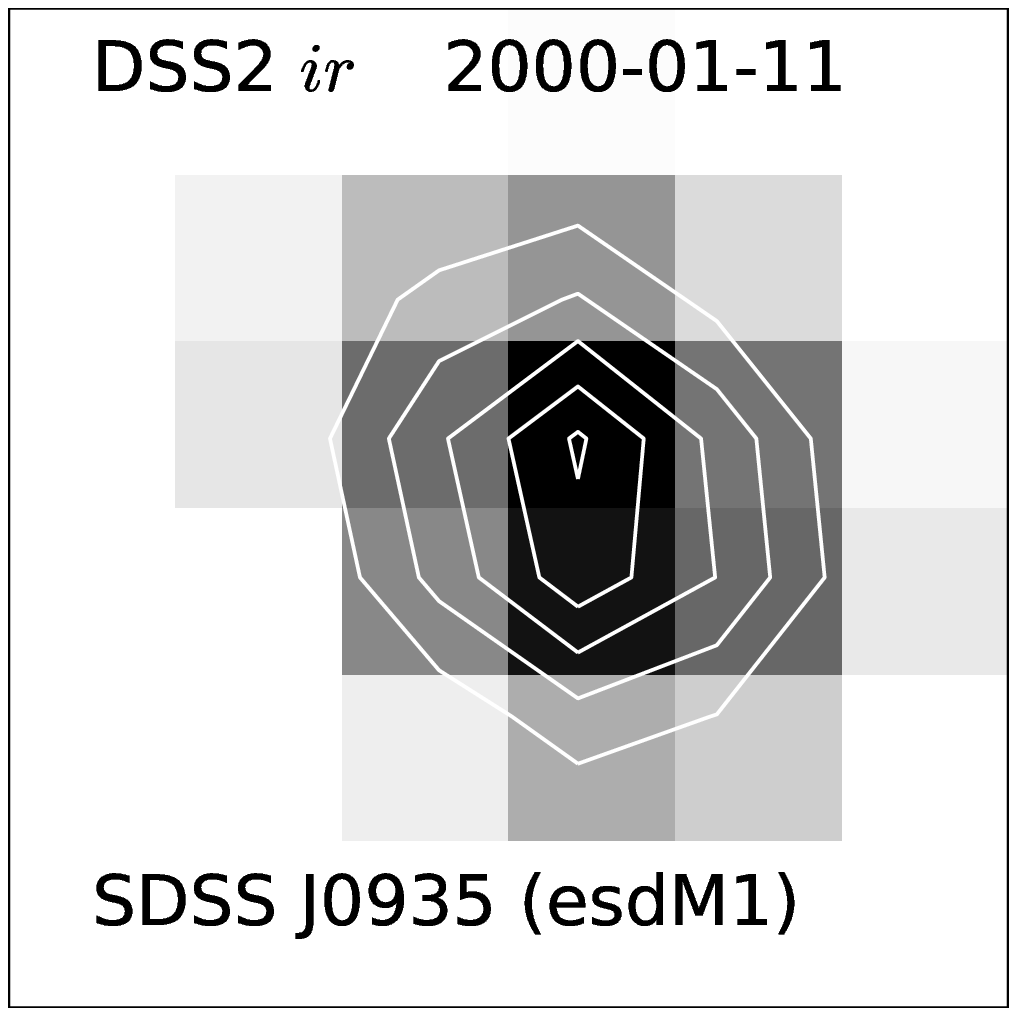}}                
  \subfloat[]{\label{j0935i}\includegraphics[width=0.166\textwidth]{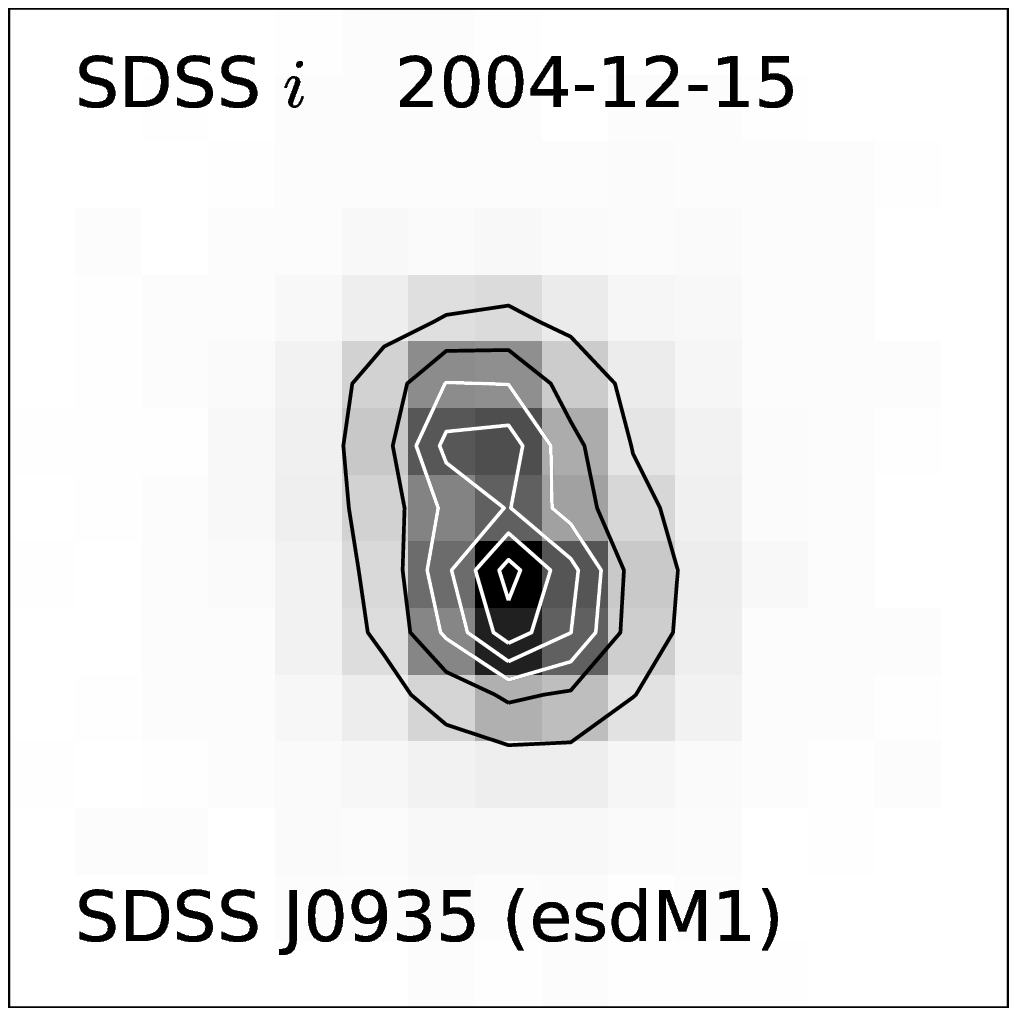}}
  \subfloat[]{\label{j0935z}\includegraphics[width=0.166\textwidth]{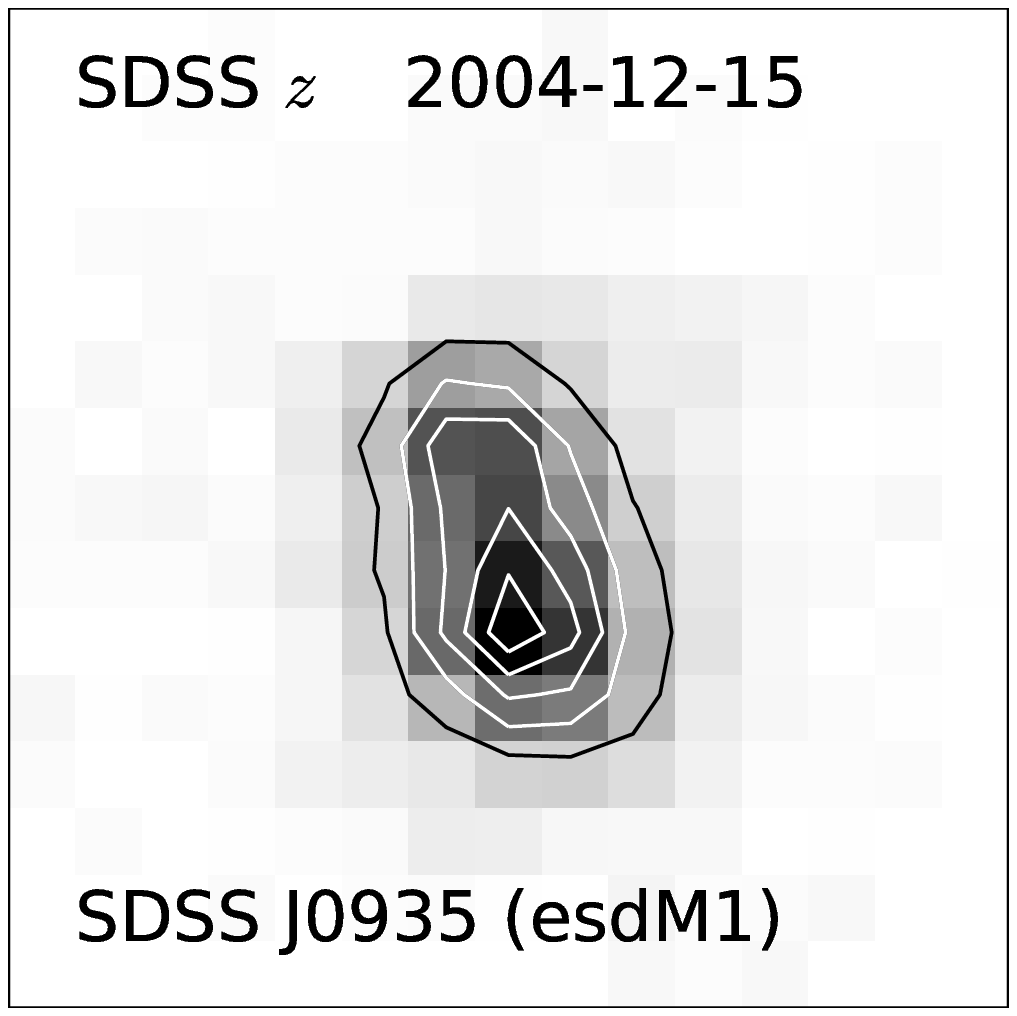}}     
  \subfloat[]{\label{j1422ir}\includegraphics[width=0.166\textwidth]{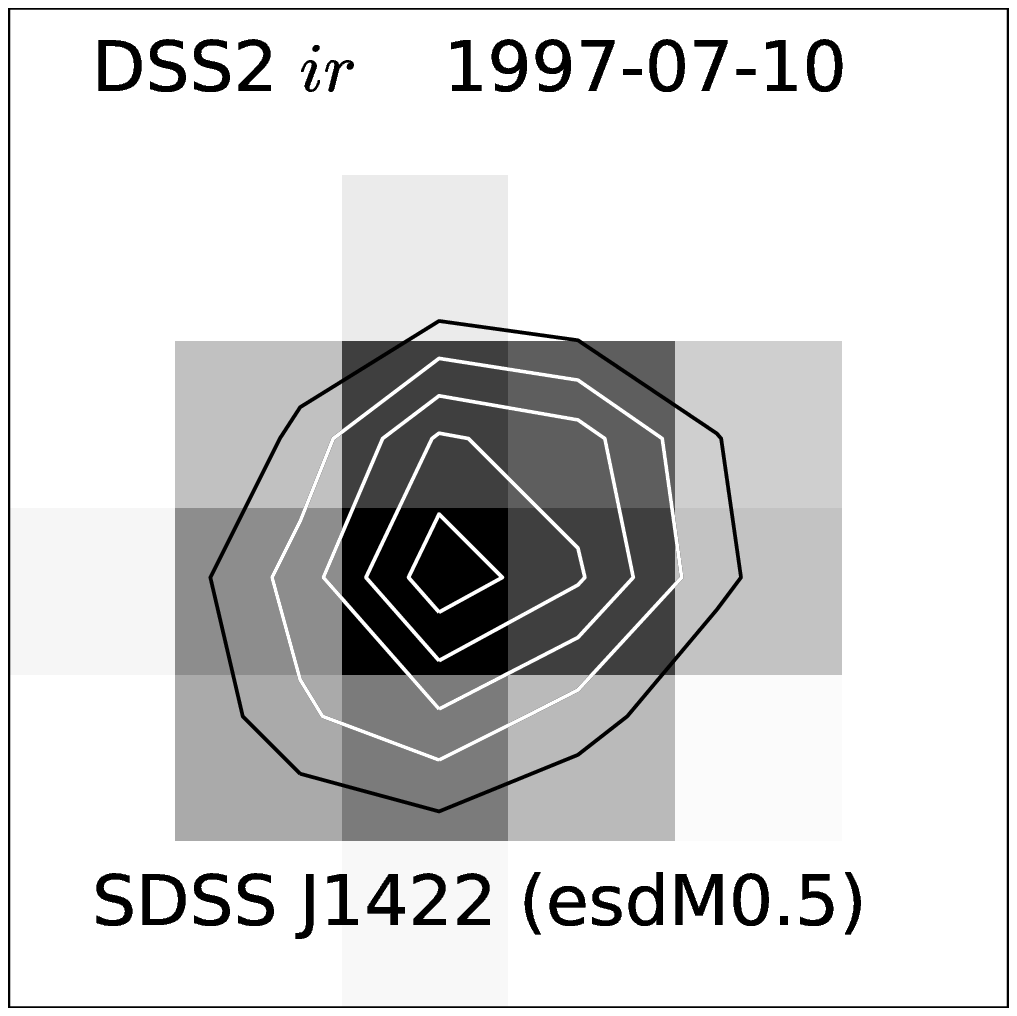}}                
  \subfloat[]{\label{j1422i}\includegraphics[width=0.166\textwidth]{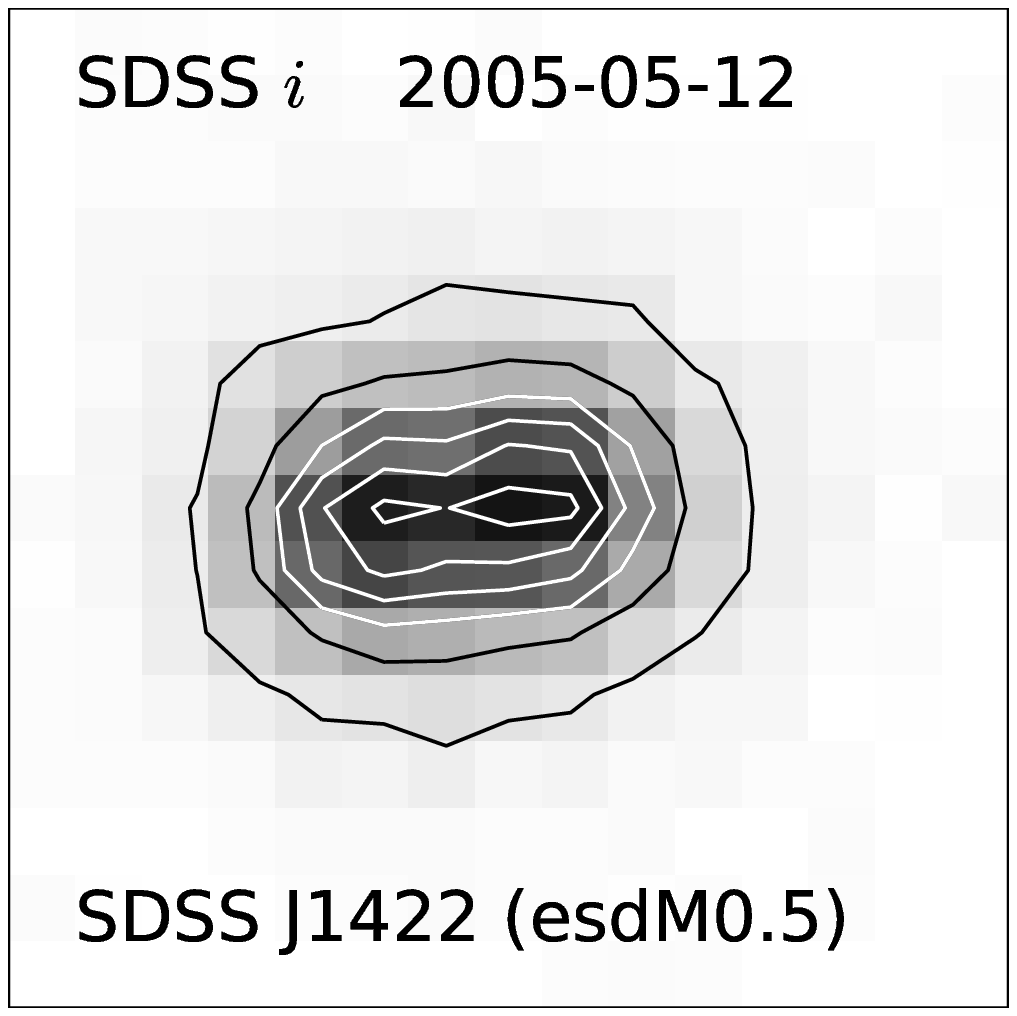}}
  \subfloat[]{\label{j1422z}\includegraphics[width=0.166\textwidth]{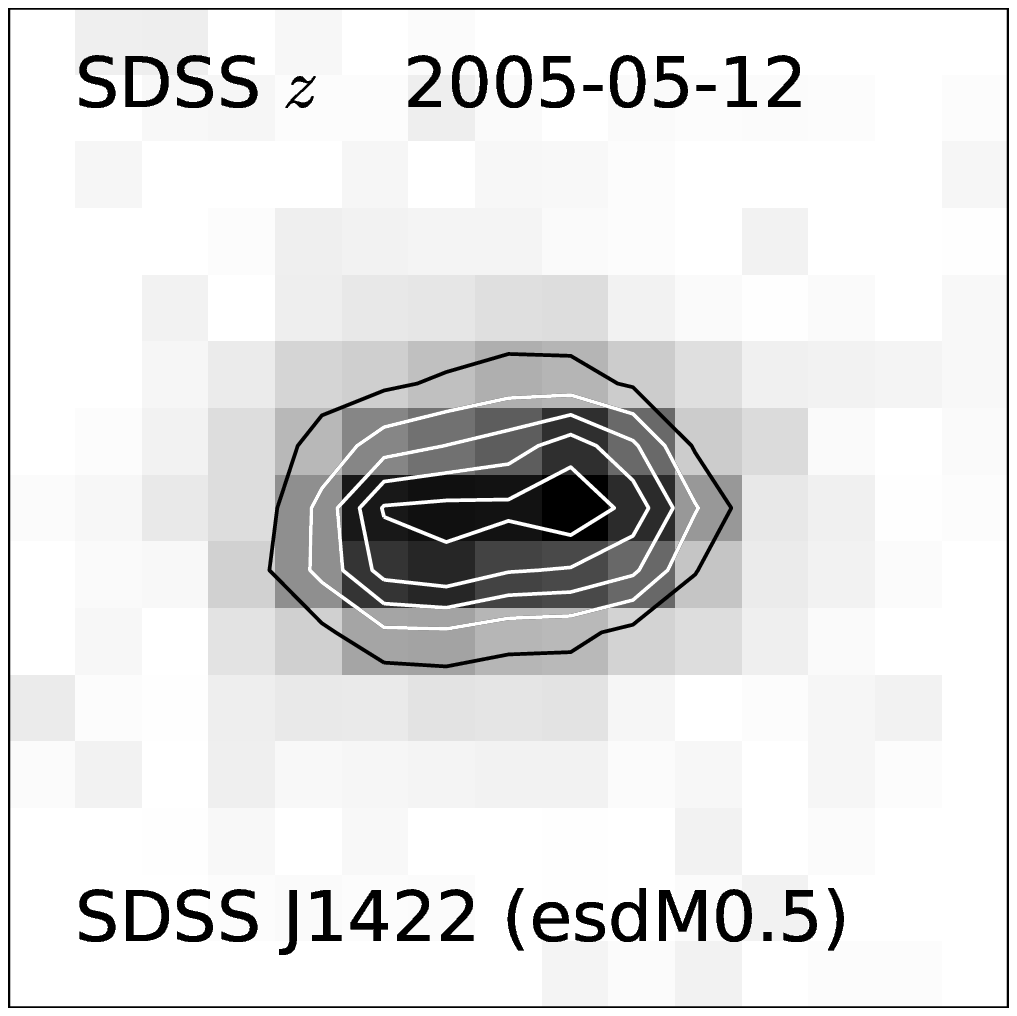}} \\
 
 \subfloat[]{\label{j1215ir}\includegraphics[width=0.166\textwidth]{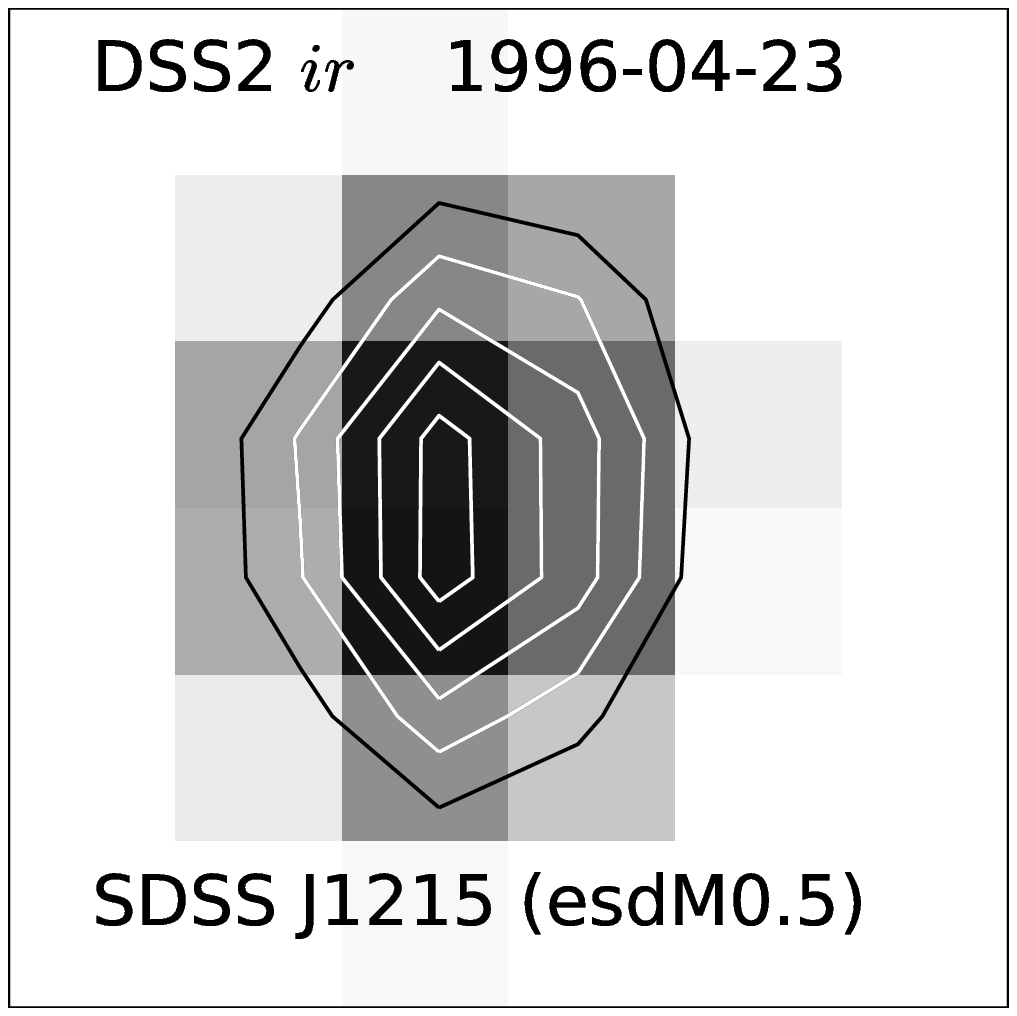}}                
  \subfloat[]{\label{j1215i}\includegraphics[width=0.166\textwidth]{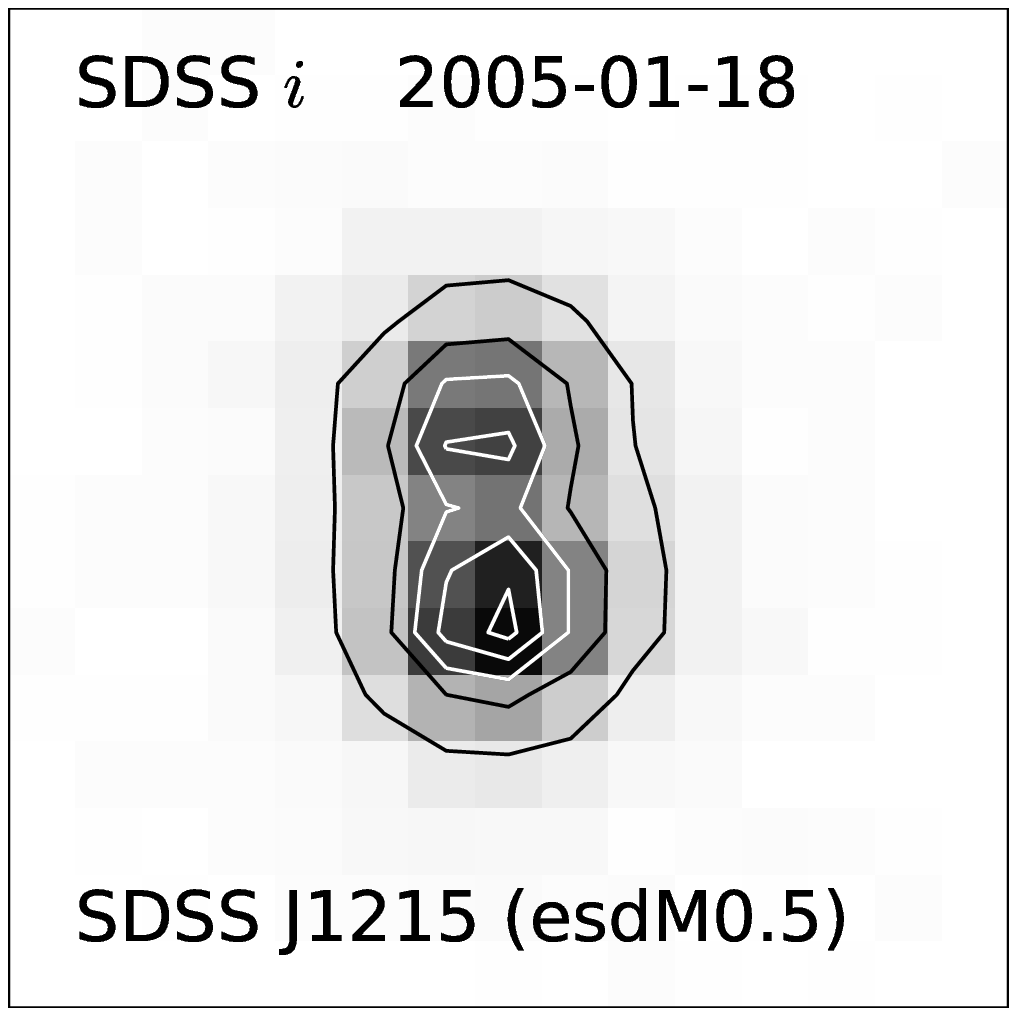}}
  \subfloat[]{\label{j1215z}\includegraphics[width=0.166\textwidth]{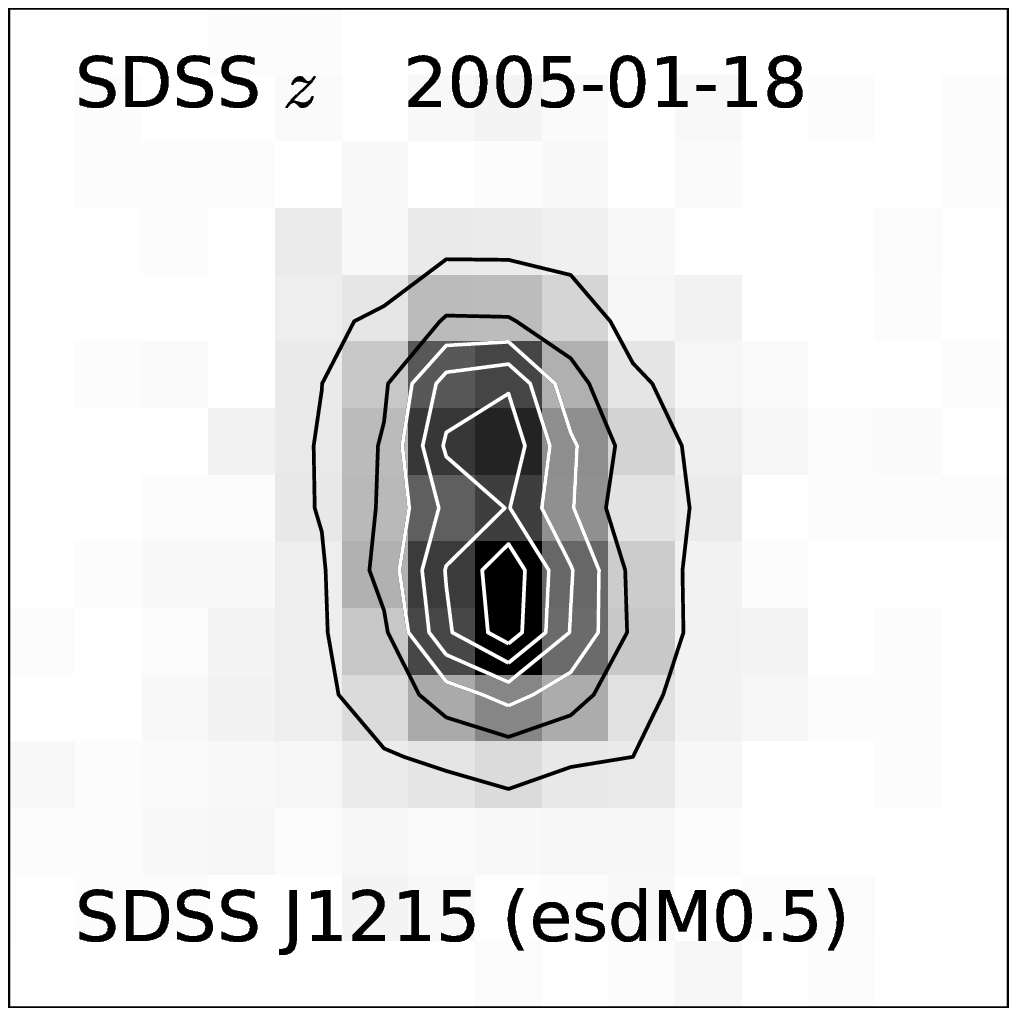}}
  \subfloat[]{\label{j1215j}\includegraphics[width=0.166\textwidth]{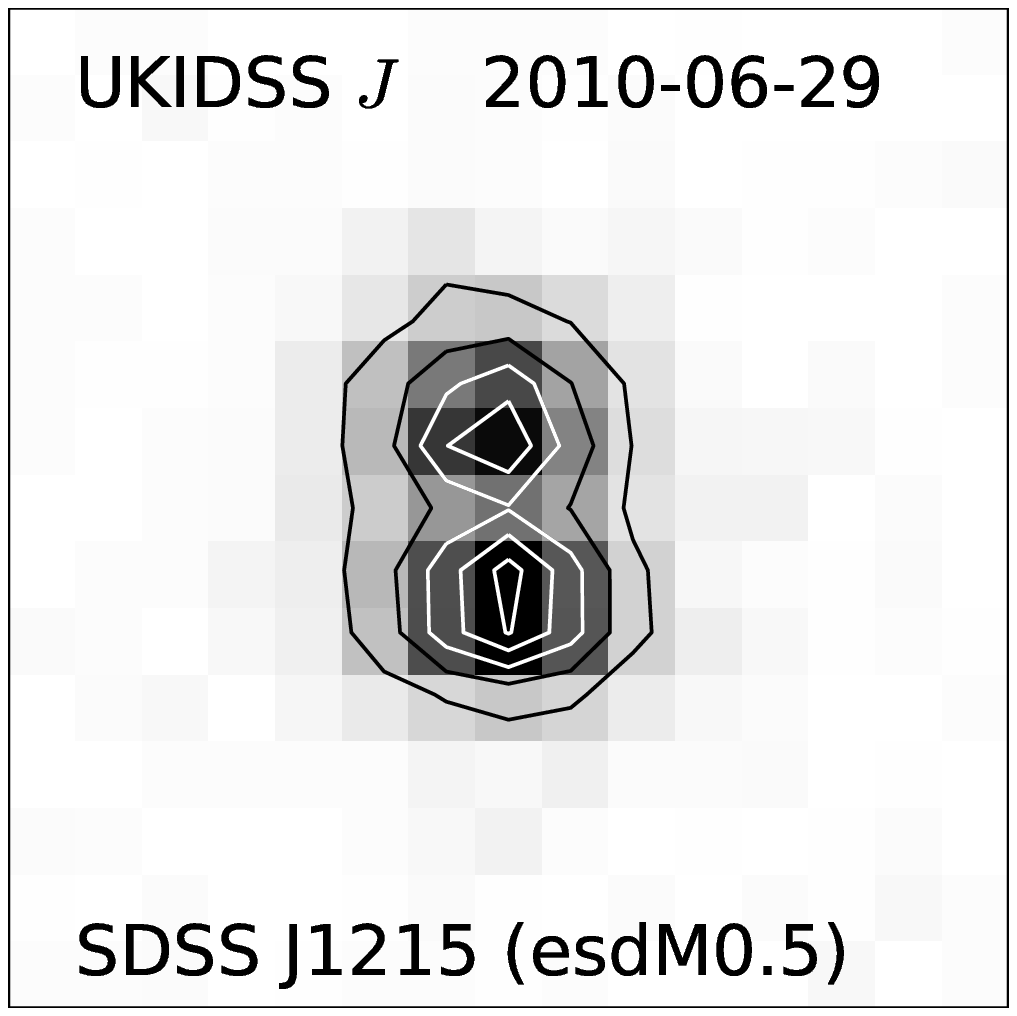}}
  \subfloat[]{\label{j1215h}\includegraphics[width=0.166\textwidth]{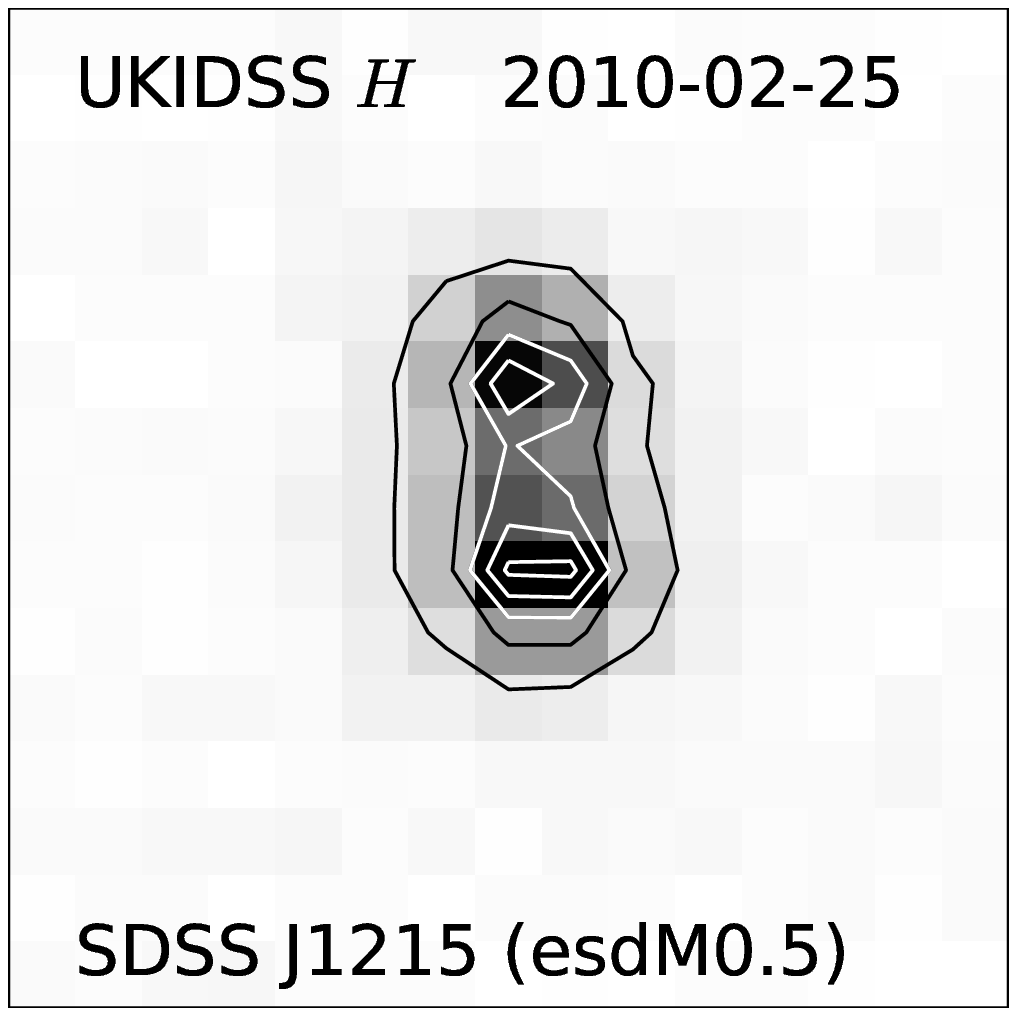}}
  \subfloat[]{\label{j1215k}\includegraphics[width=0.166\textwidth]{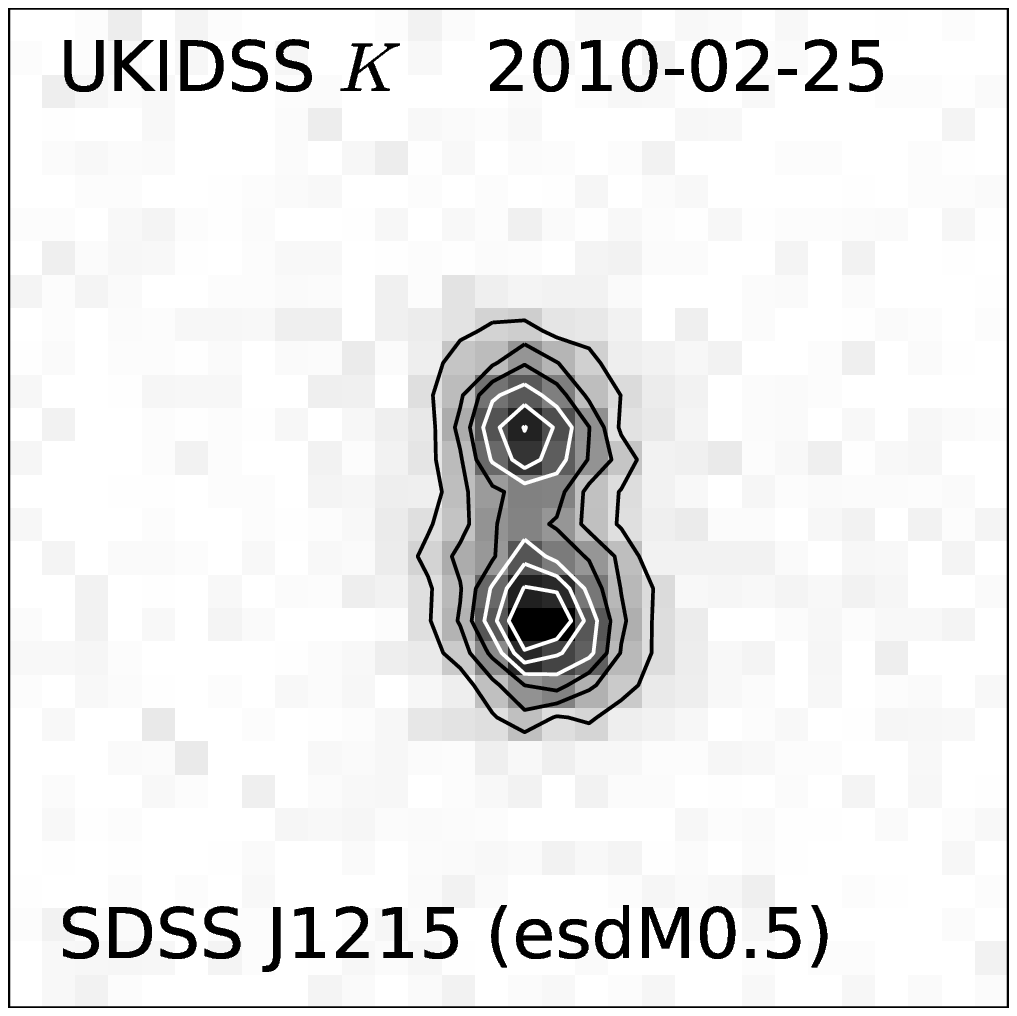}}\\

  \subfloat[]{\label{j1313ir}\includegraphics[width=0.166\textwidth]{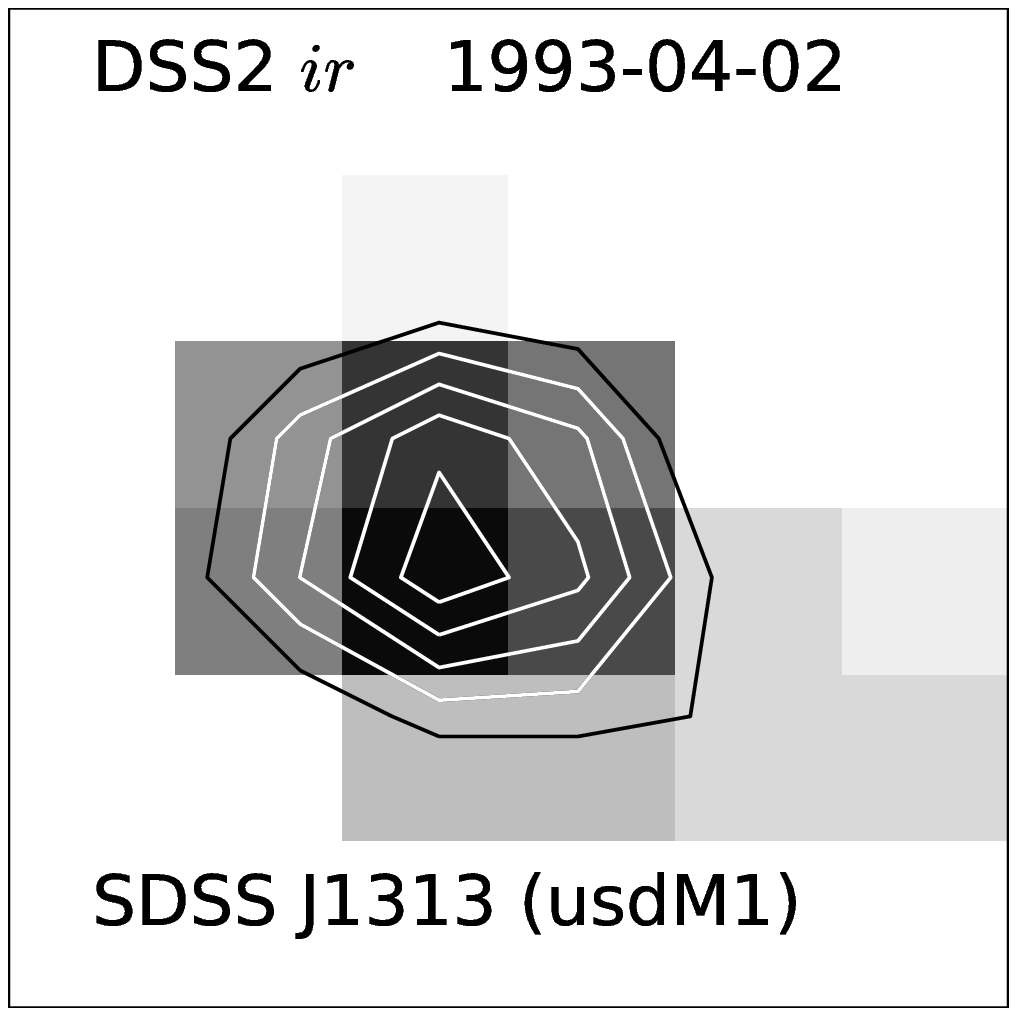}}                
  \subfloat[]{\label{j1313i}\includegraphics[width=0.166\textwidth]{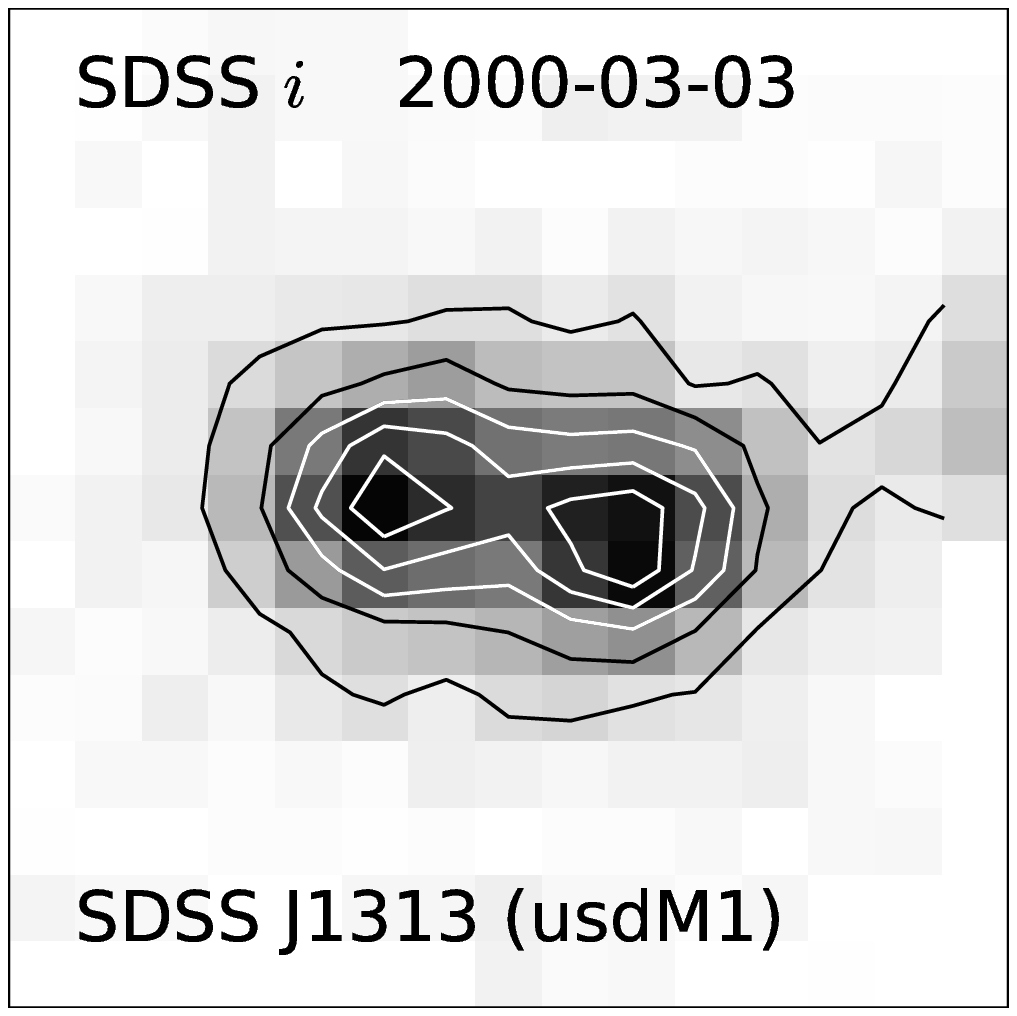}}
  \subfloat[]{\label{j1313z}\includegraphics[width=0.166\textwidth]{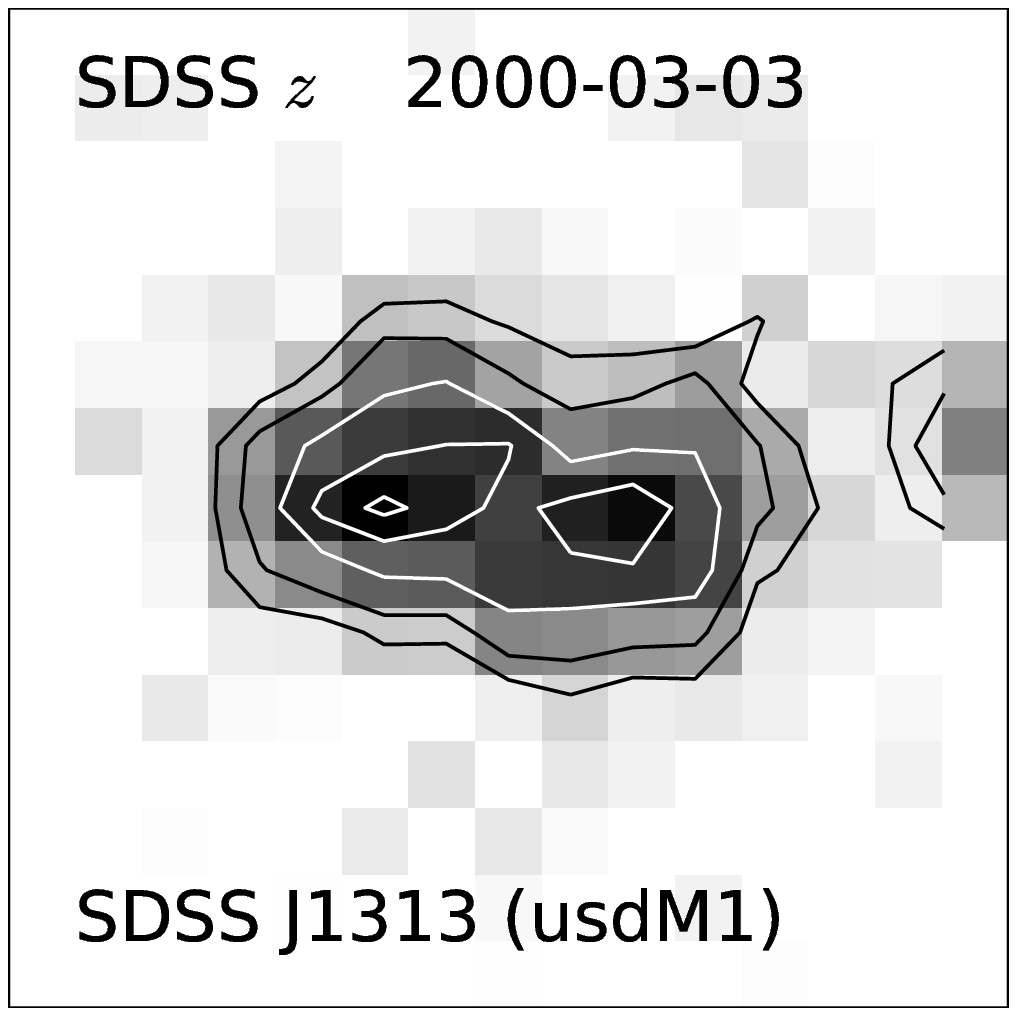}} 
  \subfloat[]{\label{j1313h}\includegraphics[width=0.166\textwidth]{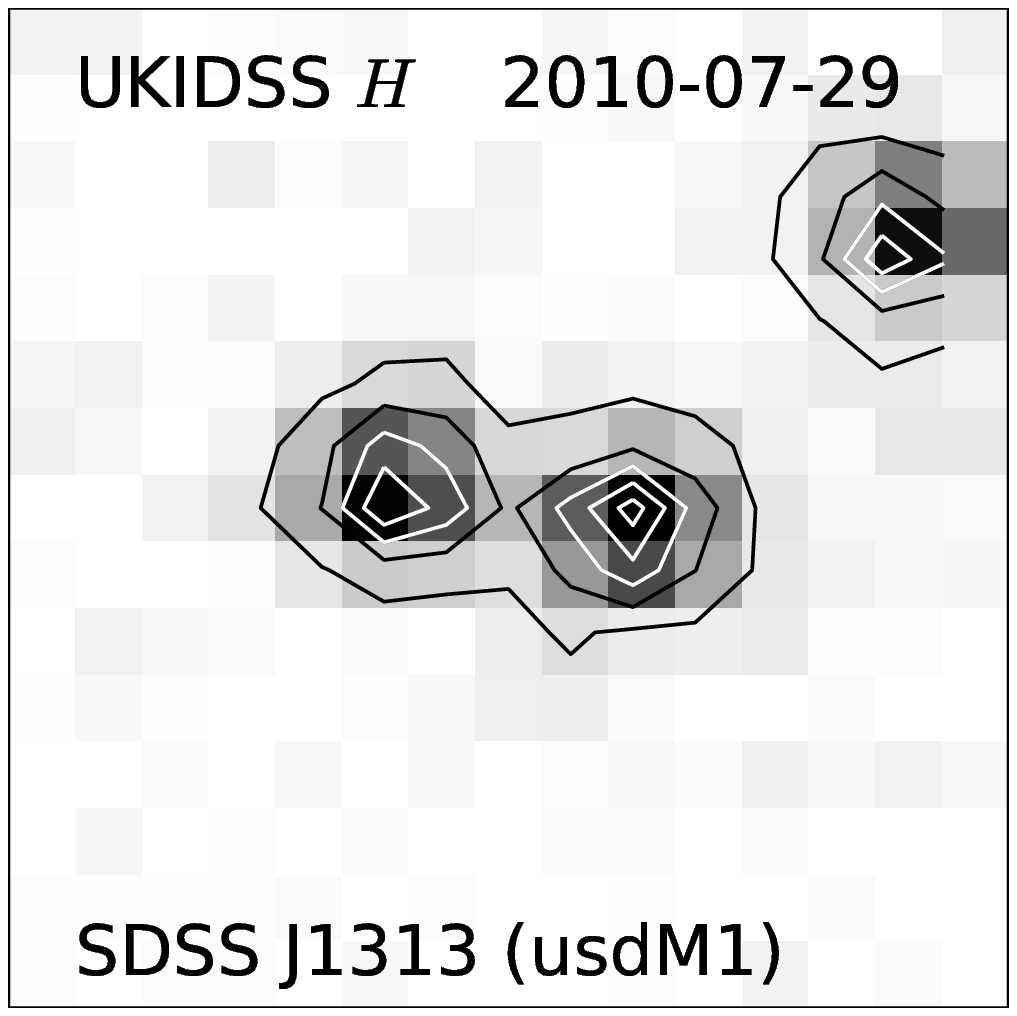}} 
  \subfloat[]{\label{j1313k}\includegraphics[width=0.166\textwidth]{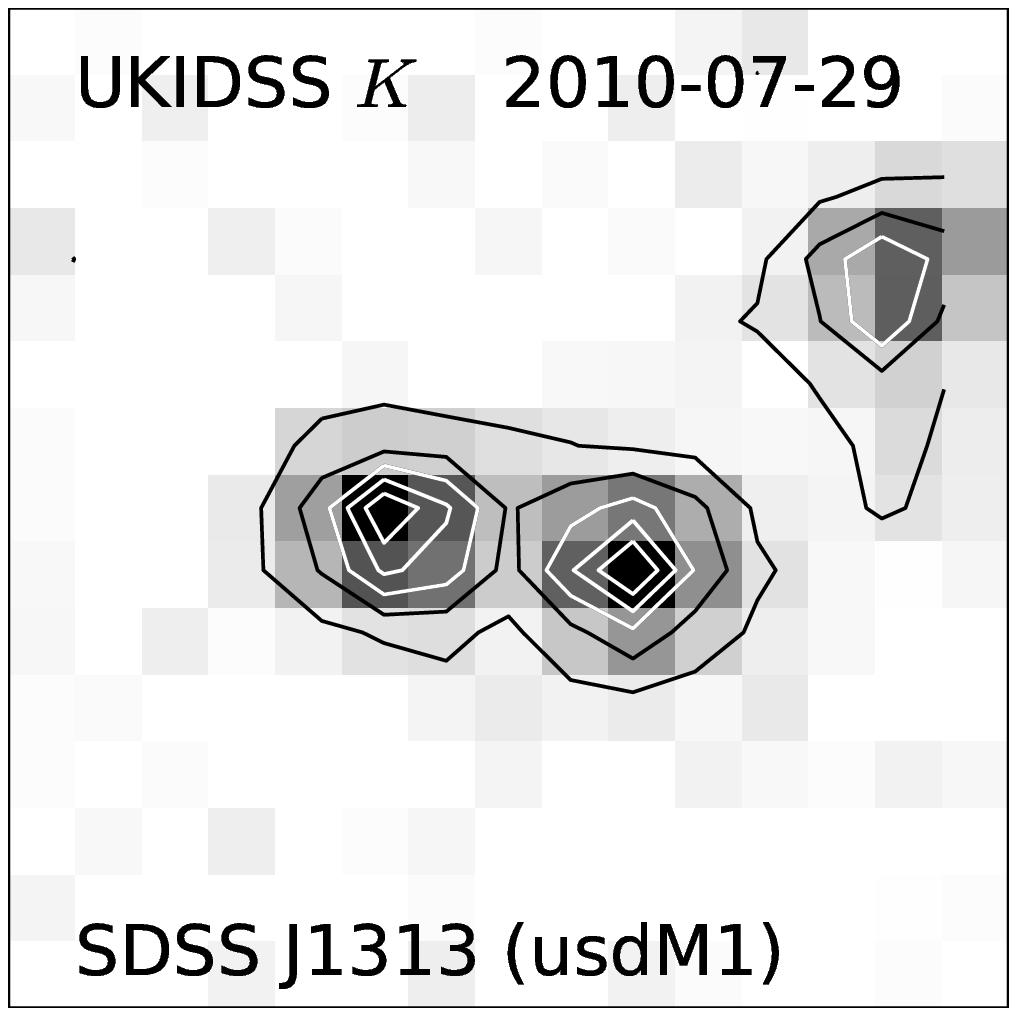}} 

\caption{DSS2, SDSS and UKIDSS images of close binaries. All images have a size of 6 arcsec $\times$ 6 arcsec with north up and east left. Level differences between contour lines in images are equal and on a linear scale.}
  \label{closeim1}
\end{figure*}

\subsection{Partially resolved binaries}

Objects classified as galaxies by imaging data, but as stars by spectroscopy or PMs,  are often in fact partially resolved  binary systems. A binary system with relative small separation (e.g. 0.5 arcsec -- 3 arcsec for SDSS images) and similar luminosity for each companion (e.g. dM+dM, dM+WD or WD+WD) will be classified as an extended source, e.g. a galaxy. 

Four spectroscopically confirmed M subdwarfs in our sample were classified as galaxies by the SDSS pipelines based on their imaging data. We found that they have a peanut-like configuration and double peaks in their images (SDSS J093517.25+242139.4, esdM1; SDSS J121502.52+271706.7, esdM0.5; SDSS J131304.72$-$033102.4, usdM1; SDSS J142259.37+144335.9, esdM0.5). These objects all have high PMs. 
To confirm that they are binaries, we checked their POSS1, POSS2 and UKIDSS
images (Fig. \ref{closeim1}). These systems are all detected but not resolved in POSS1 $r$ band images, and have substantially moved their positions in the SDSS images. 
They have elliptical shapes in POSS2 $ir$ images
(1.0 arcsec pixel$ ^{-1} $), and are consistent with their SDSS $i $ and $z$ band images
(0.4 arcsec pixel$ ^{-1}$). SDSS J1215 and SDSS J1313 are also detected in the UKIDSS
images (0.2 arcsec pixel$ ^{-1}$, 0.4 arcsec pixel$ ^{-1}$). These peaks in each pair
generally the same separations and position angles in SDSS and UKIDSS
images. As an example, Fig. \ref{closeim1} (o) and 
(p) show that SDSS J1313 is passing by a background object
to its western side from SDSS to UKIDSS epochs. Thus we conclude that these four objects are common PM binary systems.  

Four late-type K subdwarf binary systems (SDSS J091956.86+324844.2, sdK7; SDSS
J111523.79+270216.3, sdK7; SDSS J124951.09+324521.4, sdK6.5 and SDSS
J152733.23+113853.2, sdK7.5) were found in the same way. 
Companions in these  eight binary systems generally have similar
magnitudes. To find close binary systems which contain fainter companions (and
may not be classified as galaxies by SDSS) in our sample, we visually
inspected all M subdwarfs in our sample in SDSS and UKIDSS, and found another
seven close RSD binary systems. They all have  double flux peaks and
common PMs. In total we found 15 partially resolved binary systems from our RSD sample. Nearby stars around these binaries do not have double peaks.   

 Five M and two late-type K subdwarfs have faint companions detected nearby but
there are not good enough second epoch images to confirm their common PMs. Table \ref{close} shows photometry and PMs of 15 partially resolved RSD
binaries and 7 candidate systems. 
SDSS spectra of these RSD binaries can not be distinguished visually from that of single RSDs. A combined spectrum of two equal spectral type companions in a close RSD binary would looks similar to spectra of each companion. While a combined spectrum of two companions with more than 2-3 subtypes different would be dominated by the brighter companions.

\begin{landscape}
\begin{table}
 \centering
 %\begin{minipage}{140mm}
  \caption{30 common PM  confirmed RSD binaries.}
  \begin{tabular}{c l c c c c c r r r c}
\hline\hline
  Comp & SDSS name  & SDSS \emph{u}  & SDSS \emph{g} & SDSS \emph{r} & SDSS \emph{i} & SDSS \emph{z} &$\mu_{\rm RA}$ (mas/yr)  & $\mu_{\rm Dec}$ (mas/yr)  & Sep (arcsec)  & SpT$^{a}$  \\
\hline
01A & J000633.94$-$043511.2 & 21.46$\pm$0.14 & 19.01$\pm$0.03 & 17.60$\pm$0.02 & 17.01$\pm$0.02 & 16.63$\pm$0.02 & 64.24$\pm$2.66 & $-$100.23$\pm$2.66 & 20.13 &   (esdK6) \\
01B & J000633.68$-$043531.0 & 22.90$\pm$0.47 & 20.74$\pm$0.04 & 19.07$\pm$0.02 & 18.37$\pm$0.02 & 17.87$\pm$0.03 & 56.14$\pm$3.35 & $-$99.29$\pm$3.35 &   & esdM1.5 \\
02A & J000744.29$-$045418.1 & 22.67$\pm$0.43 & 19.89$\pm$0.02 & 18.43$\pm$0.01 & 17.64$\pm$0.02 & 17.23$\pm$0.02 & 116.61$\pm$3.54 & 3.06$\pm$4.15 & 5.23 &  sdM0 \\
02B & J000744.56$-$045414.7 & 21.80$\pm$0.17 & 21.43$\pm$0.15 & 21.43$\pm$0.75 & 21.31$\pm$1.32 & 21.20$\pm$2.39 & 109.07$\pm$3.54 & 0.54$\pm$4.15 &  &    WD \\
03A & J041008.68$-$041234.7 & 22.36$\pm$0.24 & 19.11$\pm$0.02 & 17.62$\pm$0.02 & 16.93$\pm$0.01 & 16.55$\pm$0.02 & 55.72$\pm$3.19 & $-$197.95$\pm$3.19 & 22.06   & (esdM0) \\
03B & J041007.30$-$041242.6 & 22.99$\pm$0.46 & 20.91$\pm$0.04 & 19.14$\pm$0.02 & 18.21$\pm$0.02 & 17.72$\pm$0.03 & 54.08$\pm$5.39 & $-$196.63$\pm$5.39 &  &  esdM3 \\
04A & J092128.09+233235.5 & 22.38$\pm$0.18 & 20.07$\pm$0.02 & 18.68$\pm$0.02 & 17.96$\pm$0.01 & 17.50$\pm$0.02 & 54.65$\pm$3.16 & $-$92.86$\pm$3.16 & 372.81  & sdM0 \\
04B & J092127.99+232622.7 & 24.36$\pm$0.82 & 21.23$\pm$0.04 & 19.65$\pm$0.02 & 18.76$\pm$0.02 & 18.29$\pm$0.03 & 55.88$\pm$21.71 & $-$95.09$\pm$21.71 &  & sdM2 \\
05A & J093325.29+374940.5 & 22.54$\pm$0.26 & 20.14$\pm$0.03 & 18.82$\pm$0.02 & 18.28$\pm$0.02 & 17.93$\pm$0.03 & $-$11.82$\pm$3.67 & $-$98.87$\pm$3.67 & 152.75 &   (usdK7.5)  \\
05B & J093320.38+374719.3 & 23.99$\pm$0.95 & 21.06$\pm$0.04 & 19.44$\pm$0.02 & 18.78$\pm$0.02 & 18.40$\pm$0.03 & $-$9.28$\pm$3.95 & $-$112.16$\pm$3.95 &   & usdM0.5 \\
06A & J095419.65+545136.9 & 22.70$\pm$0.29 & 20.33$\pm$0.03 & 18.91$\pm$0.02 & 18.36$\pm$0.02 & 18.06$\pm$0.02 & 40.65$\pm$3.67 & $-$84.26$\pm$3.67 & 98.27 &   (esdM1) \\
06B & J095417.41+545000.6 & 24.18$\pm$0.78 & 21.18$\pm$0.04 & 19.51$\pm$0.02 & 18.75$\pm$0.02 & 18.37$\pm$0.03 & 45.15$\pm$5.32 & $-$96.19$\pm$5.32 &  & esdM2 \\
07A & J101235.45+181945.5 & 21.95$\pm$0.27 & 20.01$\pm$0.03 & 18.53$\pm$0.02 & 17.75$\pm$0.02 & 17.25$\pm$0.04 & 3.15$\pm$4.52 & $-$100.46$\pm$4.52 & 435.09  & sdM0 \\
07B & J101303.57+182236.0 & 23.02$\pm$0.36 & 20.45$\pm$0.03 & 18.95$\pm$0.02 & 18.14$\pm$0.02 & 17.67$\pm$0.04 & 14.10$\pm$5.56 & $-$92.04$\pm$5.56 &   &  (sdM1) \\
08A & J104321.03+010436.5 & 23.85$\pm$0.58 & 20.74$\pm$0.03 & 19.12$\pm$0.02 & 18.06$\pm$0.02 & 17.45$\pm$0.02 & $-$179.79$\pm$4.91 & $-$103.20$\pm$4.91 & 8.89 &  (sdM3) \\
08B & J104320.47+010439.4 & 23.48$\pm$0.47 & 20.83$\pm$0.03 & 19.16$\pm$0.02 & 18.09$\pm$0.02 & 17.48$\pm$0.02 & $-$173.83$\pm$5.23 & $-$99.79$\pm$5.23 &   & sdM3 \\
09A & J104921.61+012044.4 & 21.65$\pm$0.12 & 19.31$\pm$0.02 & 17.89$\pm$0.01 & 17.18$\pm$0.01 & 16.90$\pm$0.02 & $-$117.25$\pm$3.41 & $-$54.22$\pm$3.41 & 12.91  & sdK7.5 \\
09B & J104922.06+012055.3 & 23.00$\pm$0.38 & 20.38$\pm$0.03 & 18.85$\pm$0.02 & 18.02$\pm$0.02 & 17.52$\pm$0.02 & $-$128.73$\pm$4.73 & $-$60.52$\pm$4.73 & &    (sdM2.5) \\
10A & J110134.92+385614.5 & 21.87$\pm$0.18 & 19.16$\pm$0.02 & 17.84$\pm$0.02 & 17.25$\pm$0.02 & 16.97$\pm$0.03 & $-$46.28$\pm$3.32 & $-$84.65$\pm$3.32 & 234.96 &  sdK6.5 \\
10B & J110140.89+385230.2 & 22.40$\pm$0.27 & 19.66$\pm$0.02 & 18.32$\pm$0.02 & 17.68$\pm$0.02 & 17.36$\pm$0.03 & $-$38.63$\pm$3.44 & $-$93.76$\pm$3.44 &  & sdK7 \\
11A & J113454.19+125242.7 & 17.18$\pm$0.02 & 16.11$\pm$0.03 & 15.70$\pm$0.03 & 15.50$\pm$0.02 & 15.44$\pm$0.02 & $-$101.54$\pm$13.88 & $-$69.38$\pm$13.88 & 531.17 &   (sdK4) \\
11B & J113433.88+124522.4 & 22.97$\pm$0.57 & 20.24$\pm$0.04 & 18.67$\pm$0.03 & 17.87$\pm$0.02 & 17.40$\pm$0.02 & $-$100.53$\pm$5.32 & $-$74.18$\pm$5.32 & & sdM1.5 \\
12A & J120537.70+005747.5 & 21.80$\pm$0.17 & 18.94$\pm$0.02 & 17.36$\pm$0.01 & 16.25$\pm$0.01 & 15.62$\pm$0.02 & $-$83.32$\pm$27.58 & 19.44$\pm$38.91 & 4.42 &  sdM3 \\
12B & J120537.99+005748.2 & 23.13$\pm$0.53 & 21.95$\pm$0.08 & 20.18$\pm$0.03 & 18.31$\pm$0.02 & 17.33$\pm$0.02 & $-$87.74$\pm$27.58 & 15.87$\pm$38.91 &  &    (sdM8) \\
13A & J121158.45+001450.9 & 22.61$\pm$0.21 & 20.39$\pm$0.02 & 18.92$\pm$0.01 & 18.08$\pm$0.02 & 17.62$\pm$0.02 & $-$23.83$\pm$5.05 & $-$110.46$\pm$5.05 & 245.48 &  sdM1 \\
13B & J121155.81+001853.2 & 22.30$\pm$0.15 & 21.01$\pm$0.03 & 20.55$\pm$0.03 & 20.37$\pm$0.04 & 20.17$\pm$0.10 & $-$32.33$\pm$28.81 & $-$115.25$\pm$28.81 &  &   WD \\
14A & J124559.97+300325.2 & 22.08$\pm$0.18 & 19.71$\pm$0.06 & 18.32$\pm$0.03 & 17.40$\pm$0.02 & 16.91$\pm$0.02 & 47.57$\pm$2.83 & $-$106.36$\pm$24.87 & 4.95 & sdM1 \\
14B & J124559.97+300330.1 & 19.42$\pm$0.03 & 19.08$\pm$0.03 & 19.43$\pm$0.08 & 19.70$\pm$0.25 & 20.00$\pm$0.10 & 48.33$\pm$2.83 & $-$94.72$\pm$24.87 &  &   WD \\
15A & J124819.77+610930.9 & 14.23$\pm$0.02 & 13.38$\pm$0.00 & 13.10$\pm$0.00 & 13.96$\pm$0.00 & 13.02$\pm$0.03 & $-$38.26$\pm$2.79 & $-$98.25$\pm$2.79 & 221.30 &   (esdK1) \\
15B & J124758.27+610653.6 & 22.76$\pm$0.40 & 20.74$\pm$0.03 & 19.29$\pm$0.02 & 18.73$\pm$0.02 & 18.40$\pm$0.04 & $-$52.83$\pm$4.99 & $-$99.33$\pm$4.99 &  &  esdK6.5 \\
16A & J124841.64$-$021538.7 & 22.45$\pm$0.31 & 19.87$\pm$0.02 & 18.49$\pm$0.02 & 17.92$\pm$0.03 & 17.46$\pm$0.02 & $-$116.30$\pm$3.15 & $-$26.78$\pm$3.15 & 429.59 & sdK7 \\
16B & J124823.55$-$022111.9 & 21.63$\pm$0.17 & 20.73$\pm$0.04 & 20.27$\pm$0.03 & 20.06$\pm$0.04 & 19.88$\pm$0.09 & $-$108.02$\pm$9.39 & $-$19.50$\pm$9.39 &  &   WD \\
17A & J141055.98+450222.6 & 23.23$\pm$0.57 & 20.14$\pm$0.03 & 18.61$\pm$0.02 & 17.70$\pm$0.02 & 17.10$\pm$0.02 & $-$149.09$\pm$3.25 & 8.93$\pm$3.25 & 130.73  & sdM2 \\
17B & J141055.71+450011.9 & 19.80$\pm$0.04 & 19.34$\pm$0.02 & 19.41$\pm$0.02 & 19.53$\pm$0.03 & 19.62$\pm$0.09 & $-$146.75$\pm$3.32 & 7.32$\pm$3.32 & &   WD \\
18A & J142540.14+584045.6 & 22.88$\pm$0.31 & 20.11$\pm$0.03 & 18.76$\pm$0.01 & 18.18$\pm$0.02 & 17.87$\pm$0.03 & $-$109.29$\pm$4.71 & 25.63$\pm$4.71 & 174.45  & sdK6.5 \\
18B & J142600.85+583939.8 & 22.93$\pm$0.33 & 20.66$\pm$0.03 & 19.16$\pm$0.01 & 18.42$\pm$0.02 & 18.02$\pm$0.03 & $-$115.59$\pm$5.23 & 28.28$\pm$5.23 &   & sdM0 \\
19A & J142617.53+073749.2 & 21.44$\pm$0.11 & 18.89$\pm$0.03 & 17.66$\pm$0.01 & 17.13$\pm$0.01 & 16.81$\pm$0.02 & 41.44$\pm$3.11 & $-$104.77$\pm$3.11 & 321.59  & sdK6.5 \\
19B & J142639.13+073731.8 & 22.15$\pm$0.20 & 20.94$\pm$0.04 & 20.24$\pm$0.03 & 19.89$\pm$0.03 & 19.64$\pm$0.08 & 38.62$\pm$5.21 & $-$92.34$\pm$5.21 &  &    WD \\
20A & J142754.07+334736.7 & 22.09$\pm$0.14 & 19.51$\pm$0.03 & 18.28$\pm$0.01 & 17.73$\pm$0.02 & 17.44$\pm$0.03 & $-$109.21$\pm$3.00 & $-$23.18$\pm$3.00 & 293.65 &    (sdM3) \\
20B & J142811.85+334424.2 & 24.29$\pm$1.67 & 21.14$\pm$0.04 & 19.44$\pm$0.02 & 18.24$\pm$0.02 & 17.67$\pm$0.03 & $-$102.66$\pm$3.29 & $-$30.36$\pm$3.29 &   & sdM4.5 \\
21A & J143305.04+301727.6 & 22.65$\pm$0.30 & 20.11$\pm$0.02 & 18.46$\pm$0.02 & 17.58$\pm$0.01 & 17.10$\pm$0.02 & 77.93$\pm$3.11 & $-$364.94$\pm$3.11 & 21.53  & esdM2.5 \\
21B & J143305.81+301708.5 & 23.12$\pm$0.45 & 20.77$\pm$0.03 & 19.09$\pm$0.02 & 18.06$\pm$0.01 & 17.50$\pm$0.02 & 86.68$\pm$3.28 & $-$365.68$\pm$3.28 &  & esdM4.5 \\
\hline
\end{tabular}
%\begin{list}{}{}
%\item[]Note. 
%\item[$^{a}$] 
%\end{list}
%\end{minipage}
\label{tcsdb}
\end{table}
\end{landscape}

\addtocounter{table}{-1}
\begin{landscape}
\begin{table}
 \centering
 %\begin{minipage}{140mm}
  \caption{continued.}
  \begin{tabular}{c l c c c c c r r r c}
\hline\hline
 Comp & SDSS name  & SDSS \emph{u}  & SDSS \emph{g} & SDSS \emph{r} & SDSS \emph{i} & SDSS \emph{z} &$\mu_{\rm RA}$ (mas/yr)  & $\mu_{\rm Dec}$ (mas/yr)  & Sep (arcsec)  & SpT$^{a}$  \\
\hline 
22A & J145725.85+234125.4$^{c}$ & 21.38$\pm$0.09 & 18.55$\pm$0.02 & 16.77$\pm$0.02 & 16.12$\pm$0.02 & 15.84$\pm$0.02 & $-$348.71$\pm$10.73 & $-$59.22$\pm$15.17 & 3.96  & sdK7 \\
22B & J145726.02+234122.2 & 24.54$\pm$0.97 & 22.49$\pm$0.12 & 20.67$\pm$0.05 & 18.75$\pm$0.02 & 17.77$\pm$0.03 & $-$323.85$\pm$10.73 & $-$43.61$\pm$15.17 &  &   (esdM6.5) \\
23A & J150015.11+473937.5$^{b}$ & 21.99$\pm$0.20 & 18.92$\pm$0.02 & 17.32$\pm$0.02 & 16.69$\pm$0.02 & 16.27$\pm$0.02 & $-$336.22$\pm$8.72 & 139.96$\pm$7.77 & 4.36 & usdM0.5 \\
23B & J150014.95+473933.5$^{b}$ & 23.69$\pm$0.94 & 20.42$\pm$0.05 & 18.56$\pm$0.09 & 17.69$\pm$0.08 & 17.19$\pm$0.07 & $-$336.66$\pm$8.72 & 136.87$\pm$7.77 &  &   (usdM3) \\
24A & J151649.87+605437.6 & 15.56$\pm$0.02 & 13.82$\pm$0.02 & 13.02$\pm$0.00 & 12.72$\pm$0.00 & 12.56$\pm$0.02 & 52.25$\pm$2.73 & $-$441.52$\pm$2.73 & 92.25 &   (esdK2) \\
24B & J151650.33+605305.4 & 24.57$\pm$0.34 & 20.89$\pm$0.03 & 19.01$\pm$0.02 & 17.84$\pm$0.02 & 17.18$\pm$0.02 & 51.33$\pm$5.02 & $-$446.83$\pm$5.02 &  & esdM5.5 \\
25A & J153554.51+105328.0 & 24.32$\pm$0.85 & 20.27$\pm$0.02 & 18.76$\pm$0.01 & 18.11$\pm$0.01 & 17.70$\pm$0.02 & $-$91.92$\pm$3.05 & $-$86.20$\pm$3.05 & 6.17 &   (esdK5.5) \\
25B & J153554.81+105323.7$^{c}$ & 25.20$\pm$0.77 & 21.63$\pm$0.05 & 19.89$\pm$0.02 & 19.27$\pm$0.02 & 18.93$\pm$0.04 & $-$88.59$\pm$4.32 & $-$82.95$\pm$4.32 &  &  esdK7 \\
26A & J161454.33+145314.7$^{c}$ & 22.56$\pm$0.32 & 19.56$\pm$0.02 & 17.60$\pm$0.01 & 16.92$\pm$0.01 & 16.50$\pm$0.02 & $-$16.60$\pm$2.79 & $-$151.41$\pm$2.79 & 437.85 &  esdK7.5 \\
26B & J161519.36+145719.9 & 22.71$\pm$0.27 & 20.30$\pm$0.02 & 18.72$\pm$0.01 & 17.81$\pm$0.02 & 17.33$\pm$0.02 & $-$19.71$\pm$3.78 & $-$153.64$\pm$3.78 &  &   (esdM2.5) \\
27A & J173049.54+320123.9 & 21.68$\pm$0.13 & 19.21$\pm$0.02 & 17.78$\pm$0.01 & 17.13$\pm$0.01 & 16.73$\pm$0.02 & $-$87.28$\pm$2.74 & 86.44$\pm$2.74 & 7.07  & sdK7.5 \\
27B & J173049.00+320122.5 & 23.11$\pm$0.43 & 20.51$\pm$0.11 & 18.97$\pm$0.08 & 18.12$\pm$0.06 & 17.67$\pm$0.05 & $-$100.44$\pm$3.97 & 80.06$\pm$3.97 &  &    (sdM3) \\
28A & J180527.97+231935.9 & 19.69$\pm$0.04 & 17.15$\pm$0.03 & 16.11$\pm$0.03 & 15.73$\pm$0.02 & 15.52$\pm$0.02 & $-$145.01$\pm$2.82 & $-$116.68$\pm$2.82 & 344.34   & (esdM1.5) \\
28B & J180552.96+231923.9 & 23.33$\pm$0.56 & 19.86$\pm$0.06 & 18.11$\pm$0.01 & 17.04$\pm$0.01 & 16.42$\pm$0.02 & $-$138.20$\pm$3.17 & $-$106.46$\pm$3.17 &  &  esdM4.5 \\
29A & J210105.37$-$065633.0 & 21.87$\pm$0.18 & 19.11$\pm$0.01 & 17.51$\pm$0.01 & 16.80$\pm$0.01 & 16.33$\pm$0.01 & 74.49$\pm$6.21 & $-$123.73$\pm$5.74 & 6.13 &   (esdM1.5) \\
29B & J210105.44$-$065639.0 & 22.77$\pm$0.40 & 21.50$\pm$0.06 & 19.60$\pm$0.02 & 18.40$\pm$0.01 & 17.72$\pm$0.02 & 74.78$\pm$6.21 & $-$106.96$\pm$5.74 &  &  esdM5.5 \\
30A & J221549.63+005732.0 & 21.88$\pm$0.14 & 19.01$\pm$0.02 & 17.63$\pm$0.01 & 16.91$\pm$0.01 & 16.50$\pm$0.01 & $-$13.40$\pm$3.13 & $-$143.78$\pm$3.13 & 7.36  & sdK7.5 \\
30B & J221549.22+005736.1 & 22.43$\pm$0.22 & 19.74$\pm$0.04 & 18.29$\pm$0.03 & 17.47$\pm$0.04 & 16.98$\pm$0.03 & $-$9.26$\pm$2.78 & $-$144.38$\pm$2.78 &   & sdM0.5 \\
\hline
\end{tabular}
\begin{list}{}{}
%\item[]Note. 
\item[$^{a}$]Spectral types in brackets are estimated from relationships of spectral types and $r-, i-, z-$ band absolute magnitudes (Fig. \ref{sptm}). \\
\item[$^{b}$]This binary system is not in the original PM search. $^{c}$ Objects have carbon lines and are classified as carbon subdwarfs.
\end{list}
%\end{minipage}
\label{tcsdb1}
\end{table}
\end{landscape}

%\clearpage

%\begin{landscape}
\begin{table*}
 \centering
 %\begin{minipage}{140mm}
  \caption{15 partially resolved RSD binaries and seven candidate systems.}
  \begin{tabular}{c c r r c c c c c}
\hline\hline
  SDSS Name  &   SDSS \emph{i} & $\mu_{\rm RA}$ (mas/yr)  & $\mu_{\rm Dec}$ (mas/yr)   &  SpT  & Dis (pc) & Sep (arcsec) & Sep (au) & Binarity \\
\hline
J012958.44+073745.0 &  15.48$\pm$0.02 &  104.69$\pm$2.55 & $-$1.54$\pm$2.55 &  sdM2~~ & 86-121  & 1.20 & 103-145 & Yes \\
J091956.86+324844.2 &  16.12$\pm$0.01 &  $-$56.01$\pm$2.77 & $-$125.24$\pm$2.77 &  sdK7~~ & 201-285 & 0.85 & 171-242 & Yes \\
J093339.18+303908.5 &  17.58$\pm$0.03 &  53.40$\pm$2.88 & $-$109.90$\pm$2.88  & esdM0.5 &  294-416 & 1.69 & 497-703 & Yes \\
J093517.25+242139.4 &  17.72$\pm$0.03 &  8.51$\pm$2.78 & $-$114.72$\pm$2.78 & esdM1~~ & 276-390 & 1.06 & 292-414 & Yes \\
J111523.79+270216.3 &  17.78$\pm$0.02 &  22.28$\pm$2.86 & $-$100.29$\pm$2.86  & sdK7~~ & 435-615 &  0.90 & 391-553 & Yes \\
J120640.46+391543.8 &  16.70$\pm$0.02 &  16.74$\pm$2.83 & $-$100.19$\pm$2.83 &  sdK6.5 & 290-410 & 1.84 & 533-754 & Yes \\
J120936.28+040806.6 &  15.81$\pm$0.02 &  $-$228.60$\pm$3.12 & $-$127.11$\pm$3.12  & sdM1.5 & 109-154 & 1.38 & 150-212 & Yes \\
J121502.52+271706.7 &  17.52$\pm$0.02 &  $-$133.15$\pm$3.03 & 10.20$\pm$3.03 &  esdM0.5 & 277-392 & 1.07 & 297-420 & Yes \\
J122322.55$-$010349.5 &  17.48$\pm$0.02 &  $-$265.08$\pm$3.72 & 43.70$\pm$3.72 & esdM1.5 & 228-322 & 1.34 & 305-432 & Yes \\
J124951.09+324521.4 &  16.45$\pm$0.03 &  $-$99.36$\pm$2.46 & $-$60.54$\pm$2.46  & sdK6.5 & 260-367 & 1.39 & 361-510 & Yes \\
J131304.72$-$033102.4 &  18.43$\pm$0.03 &  $-$95.13$\pm$4.49 & $-$108.82$\pm$4.49  & usdM1~~ & 378-535 & 1.61 & 609-861 & Yes \\
J142259.37+144335.9 &  17.45$\pm$0.02 &  $-$33.32$\pm$3.04 & $-$142.30$\pm$3.04  & esdM0.5 & 277-392 & 0.84 & 233-329 & Yes \\
J142417.10+075530.6 &  18.60$\pm$0.02 &  $-$48.37$\pm$4.46 & $-$93.16$\pm$4.46 &  sdM0.5 & 468-662 & 1.53 & 716-1013 & Yes \\
J152733.23+113853.2 &  18.01$\pm$0.21 &  $-$157.85$\pm$3.52 & 61.44$\pm$3.52  & sdK7.5 & 435-615 & 0.50 & 218-308 & Yes \\
J214607.39+741129.0 &  17.61$\pm$0.02 &  49.51$\pm$3.97 & 152.23$\pm$3.97 &  sdM1~~ & 272-384 & 2.35 & 638-902 & Yes \\
%\hline
J032805.96+001928.2 &  17.18$\pm$0.01 &  85.74$\pm$2.80 & $-$106.72$\pm$2.80  & esdK6.5 & 364-514 & 1.86 & 676-957 & Cand \\
J083345.49+233515.0 &  18.60$\pm$0.02 &  $-$424.25$\pm$4.74 & $-$102.77$\pm$4.74  & usdM1~~ & 425-601 & 1.96 & 832-1177 & Cand \\
J121850.48+053530.2 &  18.66$\pm$0.02 &  $-$131.25$\pm$5.49 & $-$58.18$\pm$5.49  & usdM2~~ & 362-512 & 2.00 & 724-1024 & Cand \\
J123754.58+185229.9 &  19.21$\pm$0.02 &  $-$150.13$\pm$4.62 & $-$140.24$\pm$4.62 &  usdM1.5 &  505-715 & 1.60 & 808-1143 &  Cand \\
J132237.29+665826.0 &  17.67$\pm$0.02 &  42.50$\pm$3.64 & $-$110.26$\pm$3.64 &  esdM0.5 & 310-438 & 3.25 & 1006-1423 & Cand \\
J134040.52+190217.8 &  16.03$\pm$0.02 &  $-$454.75$\pm$2.79 & $-$792.05$\pm$2.79 &  esdM4~~ & 72-101 & 1.40 & 100-142 & Cand \\
J234212.45+091037.2 &  14.81$\pm$0.01 &  107.27$\pm$2.35 & 1.13$\pm$2.35 &  sdK7~~ & 110-155 & 2.40 & 263-373 & Cand \\
\hline
\end{tabular}
%\begin{list}{}{}
%\item[]Note. 
%\item[$^{a}$] 
%\end{list}
%\end{minipage}
\label{close}
\end{table*}
%\end{landscape}

\subsection{Binary fraction of RSDs}
The multiplicity fraction of dwarf stars decreases with mass, from 57\% for nearby solar-type main-sequence stars \citep{duq91} to 42\% for M dwarfs \citep{fis92}. The M subdwarf multiplicity fraction is still not clear. Recent searches for M subdwarf binary systems \citep{ria08,lod09} show a small binary fraction of $\sim$ 3\%, and suggest a sharp cut-off in the multiplicity fraction from G to M subdwarfs. While a more extensive survey conducted by \citet{jao09} shows a multiplicity rate of 26$\pm$6\% for K and M type cool subdwarfs. 

Forty four RSDs from our original PM selected sample are found in binary systems with projected separation of $>$ 100 au. Fainter companions are missed due to the survey depth and the incompleteness of the PM catalogue used for the companion search. Although our binary search is not complete, this binary sample does however indicate a changing trend of binary fraction by masses and metallicities. We find that the binary fraction of RSDs reduces with decreasing masses and metallicities. Table \ref{mul} shows the statistics of binary frequency of RSDs from our sample by spectra and metallicity classes.

Companions of wide binaries in Table \ref{tcsdb} without SDSS spectra are not in our original sample. We group binaries by spectral types of companions in our original sample. Both companions of two sdK + sdM systems 18AB and 30AB in Table \ref{tcsdb} are in our original sample, we count as one  in both sdK and sdM groups.

\begin{table}
\caption{Statistics of binaries of late-type K and M subdwarfs}
\begin{center}
\begin{tabular}{|r|r|r|c|}
\hline\hline
  Group & Number & Binary & Fraction\\
 \hline
sdK & 204~~~ & 11~~~ & 5.39 per cent \\
sdM  & 622~~~ & 15~~~ & 2.41 per cent \\
esdK & 175~~~ & 4~~~ & 2.29 per cent \\
esdM  & 486~~~ & 12~~~ & 2.47 per cent \\
usdK  & 84~~~ & 0~~~ & --- \\
usdM  & 255~~~ & 2~~~ & 0.78 per cent \\
\hline
K & 463~~~ & 15~~~ & 3.24 per cent \\
M & 1363~~~ & 29~~~ & 2.13 per cent \\
\hline
sd & 826~~~ & 26~~~ & 3.15 per cent \\
esd & 661~~~ & 16~~~ & 2.42 per cent \\
usd & 339~~~ & 2~~~ & 0.59 per cent \\
\hline
Total & 1826~~~ & 44~~~ & 2.41 per cent \\
\hline
\end{tabular}
\end{center}
%\begin{list}{}{}
%\item[]Note. The spectroscopically confirmed esdK7+WD binary (SDSS J1633) is not used for the binary fraction calculation.   
%\item[$^{a}$] 
%\end{list}
%\end{minipage}
\label{mul}
\end{table}%

This binary sample also allows us to put a lower limit on the binary fraction ($>$ 100 au) of RSDs.  2.41\% of our RSDs are confirmed in binary systems. As our search of binaries is not complete the binary fraction will be higher than 2.41\%. There are another seven binary candidates listed in Table \ref{close} to be confirmed with second epoch imaging. The completeness of SDSS+USNO PM catalogue is 0.7 for SDSS $i$ = 19, and 0.3 for SDSS $i$ = 20 \citep{mun04}. RSDs in our sample are  at distances of 100--500 pc. Faint UCSD companions of these RSDs would be missed due to the survey depth. Massive ($M/M_{\sun}>$ 1) companions of RSDs would have evolved (e.g. cool WDs, neutron stars and black holes) and are too faint to be detected by SDSS and UKIDSS. Thus the RSD binary fraction at $>$ 100 au should be $\ga$ 5\%. The binary fractions of K- and M subdwarfs at $ > $ 100 au and $ > $ 100 au measured by \citet{jao09} are 14 and 12 per cent, respectively. Assuming that RSD binary fractions are comparable, the total RSD binary fraction would be $ \ga $ 10\% according to our sample. More complete and deeper PM catalogues (e.g. Gaia) and deep imaging surveys are needed to find wide and cooler companions of our RSDs. High spatial resolution imaging is needed to search for close binaries ($<$ 100 au). These new searches will allow us to put a tighter constraint on the binary fraction of RSDs.

\section{Binary systems of note}
\label{snote}
In this section we discuss in more detail systems of particular interest, including those where the secondary is a relatively nearby late-type star, and also those where a companion has somewhat unusual properties.

\subsection{G 224--58 AB (esdK2 + esdM5.5)}
The spectrum of SDSS J151650.33+605305.4 (SDSS J1516) is shown in Fig. \ref{wbspec}. It has been spectroscopically classified as an esdM5.5 subdwarf, and is a wide companion to the esdK2 subdwarf G 224--58 (22A in Table \ref{tcsdb}). With a separation of 93 arcsec, this system is one of our widest binaries. An optical spectrum of G224--58 was observed with the Intermediate Dispersion Spectrograph mounted on the Isaac Newton Telescope on 2010 December 24. The data were reduced and the spectrum extracted using standard software packages, and we measured the radial velocity of G 224--58 by cross-correlation with a radial velocity standard over several wavelength ranges. We avoided regions contaminated with telluric lines, and also avoided possible emission lines, and any lines that appeared to be broadened. When we cross-correlated with the reference star HD3765, which has a radial velocity of $-$63.30 km s$^{-1}$ \citep{udr99}, we derived a radial velocity of $-$189.07$\pm$0.15 km s$^{-1}$ for G 224--58. We also measured the radial velocity using a different reference star; HD10780 ($+$2.70 km s$^{-1}$), and measured a consistent value. The radial velocity of G 224--58 is consistent with that of SDSS J1516 or G 224--58 B, $-$185.02$\pm$3.20 km s$^{-1}$ measured from the SDSS spectrum. This is consistent with expectations for a physically associated system. G 224--58 AB has the highest PM (449.77$\pm$7.10 mas$\cdot$yr$^{-1}$) amongst our binary sample. It is at a distance of 137$\pm$28 pc, estimated using the relationship between absolute magnitudes and spectral type (Fig. \ref{sptm}), and has clear halo space velocities ($U=278$ km s$^{-1}$, $V = -$202 km s$^{-1}$, $W=-$39 km s$^{-1}$). 

 Precise metallicity measurements of M dwarfs for calibration have become popular in the last few years \citep[e.g.][]{roj10,roj12,ter12,one12,nev13}. These works are based on M dwarfs in binary systems with FGK dwarfs primaries. Precise metallicities are measured from high resolution spectra of early-type primaries adapted to the M dwarf secondaries to calibrate metallicity features in their spectra. The lowest M dwarf metallicity calibrations are currently  [$M$/H] $\sim -0.5$ due to the lack of binaries of M + FGK subdwarfs.  With an early-type esdK subdwarf and a late-type esdM subdwarf, G 224--58 AB is an ideal benchmark for subsolar metallicity calibration of M subdwarfs down to $= -1.5 <$ [$M$/H]  $ < -1.2$.  

UCDs in wide binary systems have been identified \citep[e.g.][]{luh07,burm09,bur11,zha10,fah10,pin12} and used to calibrate spectral analysis techniques \citep[e.g.][]{bur06a}, and test atmospheric and evolutionary models \citep[e.g.][]{dup09,leg08}. Yet among metal-poor ultracool dwarfs, only one benchmark has been discovered, the d/sdM9 ([Fe/H] = $-$0.7) HD 114762B \citep{bow09}, the proximity to its sdF9 primary has made it a challenge to observe. No L and T subdwarf benchmark has been found so far. Although there are a few mild metal-poor T dwarf benchmarks (with [$M$/H] $\sim -0.3\pm0.1$) that have been found: SDSS J1416AB \citep{bur10}; HIP 73786B \citep{mur11}; BD+01$^{\circ}$ 2920B \citep{pin12}. G 224--58B is a very cool extreme subdwarf and could provide a precise metallicity constraint from its early-type primary, it is thus a benchmark object that could be used for testing and calibration of  atmospheric and evolutionary models of metal-poor low-mass stars.

\subsection{SDSS J210105.37$-$065633.0AB (esdM1+esdM5.5)}
The spectrum of SDSS J210105.44$-$065639.0 (SDSS J2101 B) is shown in Fig. \ref{wbspec}, and is classified as an esdM5.5 subdwarf. It is a companion to the esdM1 subdwarf SDSS J210105.37$-$065633.0 (SDSS J2101A). This binary has an angular separation of 6 arcsec, and it is at a distance of 183$\pm$37 parsec, derived using our relationship between absolute magnitudes and spectral type (Fig. \ref{sptm}). The system has halo space velocities ($U=-$90 km s$^{-1}$, $V=-$310 km s$^{-1}$, $W=-$49 km s$^{-1}$). This system is of particular use as a test for the M subdwarf classification methods which is still in debate \citep[e.g.][]{lep07,jao08}.

\subsection{Three carbon subdwarfs in binary systems}
 \label{3sdc}
 
\begin{figure} %  figure placement: here, top, bottom, or page
   \centering
   \includegraphics[width=\columnwidth, angle=0]{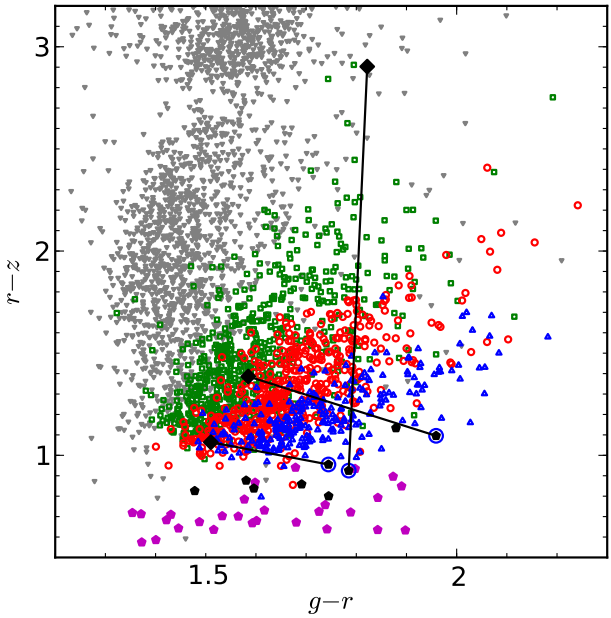}
   \caption{$g-r$ versus $r-z$ colours of the three carbon subdwarf binaries. Grey down-pointing triangles: dMs; green squares: sdMs; red circles: esdMs; blue up-pointing triangles: usdMs; filled magenta pentagons: carbon dwarfs; black pentagons: carbon subdwarfs. Blue circles with black pentagons inside are spectroscopically confirmed carbon subdwarfs in binaries (SDSS J1535B, SDSS J1457A and SDSS J1614A, from left to right), and black diamonds are their companions which do not have measured spectra. Companions in confirmed binaries are joined with black lines.}
   \label{grzbc}
\end{figure}

Three of the companion objects have features that are characteristic of RSDs and also have features that are characteristic of carbon dwarfs. We have examined these objects closely in the SDSS $i-$ and $z-$ band images (0.4 arcsec pixel$^-1$), and they show no evidence for being partially resolved multiple systems. Fig. \ref{grzbc} shows the relative location of each component with respect to the dM, sdM, esdM and usdM sequences in $g r z$ colour-space. The companions to these three objects appear to be normal RSDs occupying typical colour space in the $g-r$ versus $r-z$ diagram. This indicates that the carbon in subdwarfs did not originate in their formation environment, but has presumably comes from the progenitors of unseen WD companions.

\begin{figure} %  figure placement: here, top, bottom, or page. 
%   \centering
   \includegraphics[width= \columnwidth]{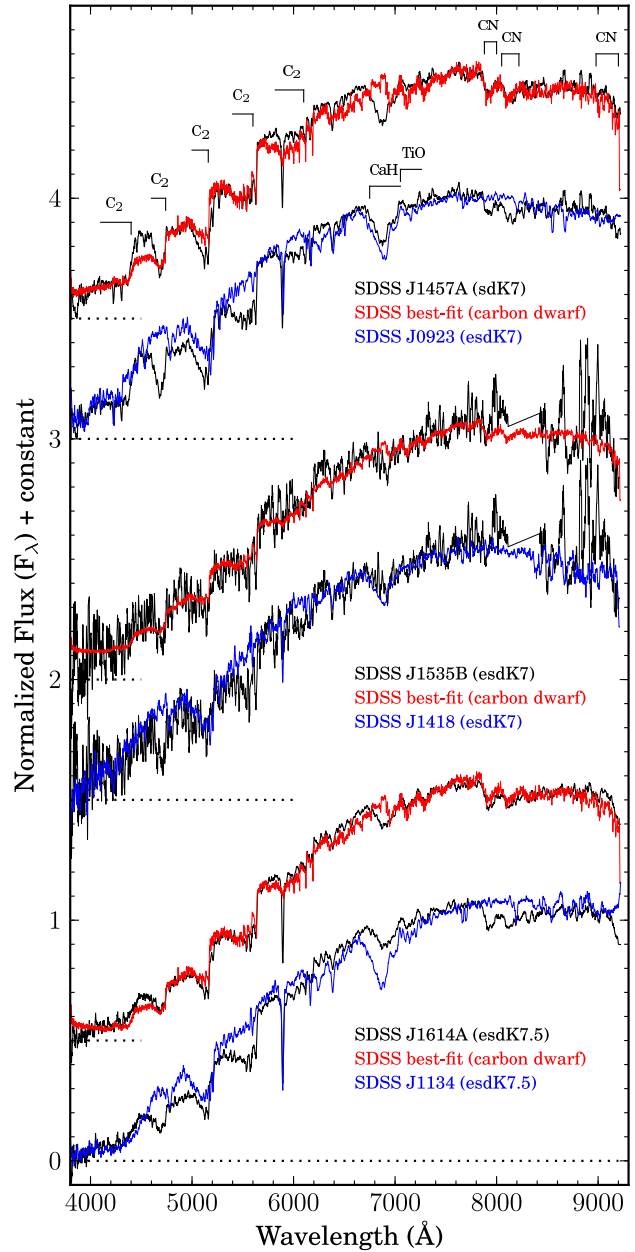}
   \caption{SDSS spectra of three carbon subdwarf SDSS J1457A, SDSS J1535B and J1614A. SDSS best-fitting spectra (red, carbon dwarf) and spectrum of SDSS J0923 (top panel, blue, esdM7), SDSS J1448 (middle panel, blue, esdM7) and SDSS J1134 (bottom panel, blue, esdM7.5) are over-plotted for comparison. Spectra are normalized at  7500 \AA~and binned by 11 pixels except SDSS best-fitting  carbon dwarf spectra.}
   \label{sdc}
\end{figure}

\textit{SDSS J145725.85+234125.4} (SDSS J1457 A) was classified as an usdK6.5 subdwarf according to the metallicity index $\zeta_{\rm TiO/CaH}$ defined by \citet{lep07}. It has a high PM of $\mu_{\rm RA}=-348.71\pm10.73$ mas$\cdot$yr$^{-1}$, $\mu_{\rm Dec}=-59.22\pm15.17$ mas$\cdot$yr$^{-1}$. SDSS J1457AB is a robust genuine binary with separation of 3.96 arcsec. The top panel of Fig. \ref{sdc} shows the spectrum of SDSS J1457A plotted along with a best-fitting carbon dwarf spectrum (also from SDSS) and an esdK7 type subdwarf spectrum. We can see that the spectrum of the carbon dwarf is very similar to that of SDSS J1457A, apart from the CaH region around 6700--7000 \AA. However, the spectrum of the esdK7 subdwarf SDSS J092302.40+301919.7 (SDSS J0923)  compares well with the spectrum of SDSS J1457A in this 6400--7900 \AA~ range. Features of both RSDs (CaH, TiO bands) as well as carbon dwarfs (e.g. the C$_{2}$ swan bands) are clear in the spectrum of SDSS J1457A. Thus, we think SDSS J1457A is a K7-type carbon subdwarf. 

Although the CaH and TiO indices match well with that of an esdK7 subdwarf, but SDSS J1457A is actually an sdK7 subdwarf. The reason why the SDSS J1457A appears like an esdK7 is because the TiO index is  sensitive to both metallicity and carbon abundance. When the C/O ratio is greater than 1, all of the oxygen is bound in CO, and none is left to bond with titanium to form TiO \citep[chapter 2,][]{rei05}. Thus it is not possible to measure the correct metallicity of carbon subdwarfs by their CaH and TiO indices without consideration of carbon abundance. In the case of SDSS J457A, we can measure the metallicity from its binary companion. SDSS J145726.02+234122.2 (SDSS J1457B)  is an M6.5-type subdwarf according to the $M_{r, i, z}$--spectral type relationships shown in Fig. \ref{sptm}.  Fig. \ref{grzbc} shows that SDSS J1457B is an sdM subdwarf, as it is located at the  sdM sequence and very close to the dM sequence. SDSS J1457A should share the same metallicity as SDSS J1457B. Thus, we conclude that SDSS J1457AB is an sdK7+sdM6.5 carbon subdwarf system.

\textit{SDSS J153554.81+105323.7} (SDSS J1535B) was classified as an usdK6.5 subdwarf according to the metallicity index $\zeta_{\rm TiO/CaH}$ defined by \citet{lep07}. The spectrum of SDSS J1535B shown in the middle of Fig. \ref{sdc} is similar to that of SDSS J1457A but with shallower CaH indices and C$_{2}$ swan bands. The best-fitting carbon dwarf spectrum does not provide a good fit for the CaH indices at 6700-7000 \AA.  SDSS J1535B compares well with an esdK7 type subdwarf SDSS J144819.31+363400.4 (SDSS J1448) in the spectral range 6400-7900 \AA. The carbon dwarf emission index at 6750-6950 \AA~explains why the CaH indices of SDSS J1535A are slightly shallower than those of the esdK7 subdwarf. We classify SDSS J1535B as a K7 type carbon subdwarf. SDSS J1535A is a K5.5 subdwarf according to Fig. \ref{sptm}. SDSS J1535AB is an esdK5.5 + esdK7 carbon subdwarf system according to its CaH and TiO indices. They may have higher metallicity than a typical esdK because the TiO strength will also be affected by a higher carbon abundance. Fig. \ref{grzbc} shows that SDSS J1535A probably has metallicity between esdK7 and sdK7.

\textit{SDSS J161454.33+145314.7} (SDSS J1614A) was classified as an usdK7 subdwarf according to the metallicity index $\zeta_{\rm TiO/CaH}$ defined by \citet{lep07}. The spectrum of SDSS J1614A is shown at the bottom of Fig. \ref{sdc}. The best-fitting SDSS spectrum of SDSS J1614A is a carbon dwarf. The spectrum of the carbon dwarf does not provide a good fit for the CaH indices at 6700-7000 \AA~which is the major spectroscopic feature of an RSD. The spectrum of SDSS J1614A compares well with that of an esdK7.5 subdwarf SDSS J113419.66+345807.8 (SDSS J1134), and we thus classify it as a K7.5 type carbon subdwarf. Its fainter companion, SDSS J161519.36+145719.9 which is 437.85 arcsec away, has a spectral type of esdM2.5 according to Fig. \ref{sptm}. We also note that the colours of SDSS J1614B are consistent with an esdM2.5 subdwarf (Fig. \ref{grzbc}).

The classification system for dwarf carbon stars has not been established. Spectral types assigned to these three carbon subdwarfs are equivalents of RSDs, do not have information of carbon abundance.  The C$_{2}$ swan bands in these carbon subdwarf spectra are somewhat weaker than in normal carbon dwarf spectra, suggesting that they have lower carbon abundance than normal carbon dwarfs. This could explain why the CaH indices of these late-type K subdwarfs are not as strong as we might expect. There are three possibilities to explain why these carbon subdwarfs have less carbon abundance than carbon dwarfs. (1) The progenitors of their WD companions had lower mass leading to a lower level of carbon accretion on to the secondary. Subdwarfs are older than dwarfs and thus have enough time for solar-mass companions to evolve through the red giant, AGB and WD stages. (2) These carbon subdwarfs have wider separation from their unseen WD companions than younger carbon dwarfs, again leading to lower levels of accretion. (3) The metal-poor WD progenitors have a lower carbon abundance.

\subsection{Six RSDs companion to WDs}
WDs provide important constraints on Galactic time-scales \citep{sch59} because their age can be well estimated from WD cooling time-scales combined with the evolutionary life-times of their progenitors. Binary systems containing old WDs and subsolar metallicity RSD components could provide a link between age and chemical abundance. Two old WD companions to early type K dwarfs with low metallicity ([$M$/H]$\sim -0.5$) have been identified \citep{jao03,jao05}. \citet{mon06} measured the age of these WDs to be 6-9 Gyr, concluding that they were not likely to be members of the halo because they are younger than the canonical halo age of 12-14 Gyr \citep{gil89}.

The ratio between the strength of TiO and CaH bands near 7000 \AA~for RSDs is a metallicity diagnostic \citep{bes82,all95}. Thus WDs with RSD companions have advantages for the study of chemical enhancement and the early formation history of the Galaxy. Four M and two late-type K subdwarfs in our sample are found to be companions to probable WDs (see Table \ref{tcsdb}). Fig. \ref{wbspec} shows spectra of two RSD companions to WDs (SDSS J124559.97+300325.2 and SDSS J141055.98+450222.6). WD companions are identified using reduced PMs (Fig. \ref{hrzb2}) and SDSS colours (Fig. \ref{grzb2}). The reduced PMs and $ r-z $ colour of these six binaries are plotted in Fig. \ref{hrzb2} and joined with cyan lines. These six WD companions are associated with confirmed WDs on the left of the plot. The RSD companions are located on the  sequence of RSDs in Fig. \ref{hrzb2}. The $ g-r $ versus $ r-z $ plot in Fig. \ref{grzb2} also suggests that these six binary systems contain WD and RSD components. WDs are located at the bottom left in the  $ g-r $ versus $ r-z $ plot (see Fig. \ref{grzbc}), while cool WDs overlap with the hot tail of the main sequence. Most of RSDs in our sample are beyond 200 pc and therefore very cool WD companions will be missed by our search. These six RSDs with WD companions are classified as sdM or sdK could be members of either the thick disc or inner halo of the Galaxy.  

%With age calibration from WD companions and metallicity calibration from cool subdwarfs, we can study the metallicity enhancement during the early formation history of the Galaxy. 

%\subsection{A spectroscopic binary candidate}
\subsection{An esdK7+WD spectroscopic binary}
\begin{figure} %  figure placement: here, top, bottom, or page
   \centering
   \includegraphics[width=\columnwidth, angle=0]{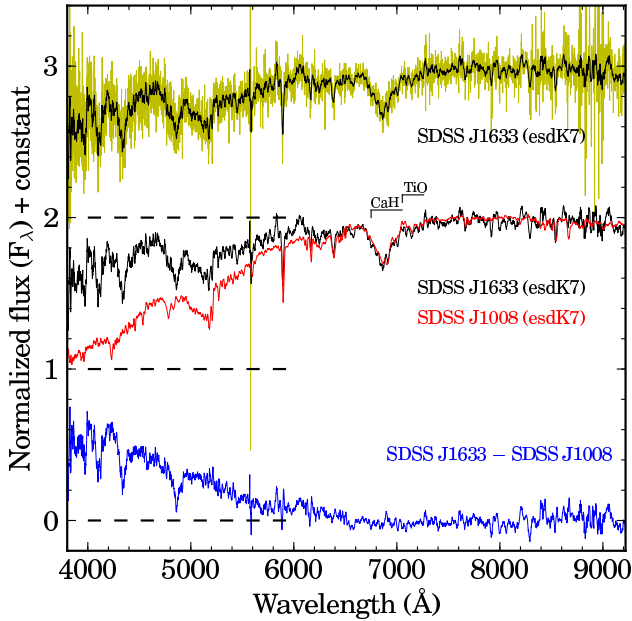}
   \caption{The spectrum of SDSS J1633. Black: spectrum of SDSS J1633 smoothed by 11 pixels. Red: SDSS J1008. Blue: the difference between SDSS J1613 and SDSS J1008. The spectra of SDSS J1633 and SDSS J1008 are normalized at  8000 \AA.}
   \label{1633}
\end{figure}

Fig. \ref{1633} shows the spectrum of SDSS J163340.83+133417.0 (SDSS J1633), which is classified as an esdK7 subdwarf. It has a significant flux excess in the blue band when compared to the normal esdK7 subdwarf SDSS J100849.85+200923.4 (SDSS J1008). When we remove the spectrum of SDSS J1008 from that of SDSS J1633, a typical WD spectrum remained. So SDSS J1633 is actually a WD+esdK7 spectroscopic binary system. It has a PM of $\mu_{\rm RA}=-107.63\pm3.61$ mas$\cdot$yr$^{-1}$; $\mu_{\rm Dec}=-68.73\pm3.61$ mas$\cdot$yr$^{-1}$ and a radial velocity of $-135.47\pm7.45$ km$\cdot$s$ ^{-1} $. SDSS $r, i$ and $z$ band absolute magnitudes of K7 subdwarfs are $9.53<{\rm M}_{r}<12.23$, $8.97<{\rm M}_{i}<10.42$ and $8.68<{\rm M} _{z}<10.08$ respectively, based on sdK7 and esdK7 subdwarfs with parallax measurements. The distance of SDSS J1633 has been estimated (by averaging the absolute magnitudes in $M_{i}$ and $M_{z}$) as 592$^{+229}_{-165}$ pc. Its tangential velocity is 356$^{+138}_{-100}$ km s$^{-1}$, and the resulting space velocity is $U=-66.1^{+22.3}_{-18.4}$ km s$^{-1}$, $V=-245.4^{+72.6}_{-78.0}$ km s$^{-1}$ and $W=14.8^{+45.3}_{-41.0}$ km s$^{-1}$.

\begin{figure} %  figure placement: here, top, bottom, or page
   \centering
   \includegraphics[width=0.332\textwidth, angle=0]{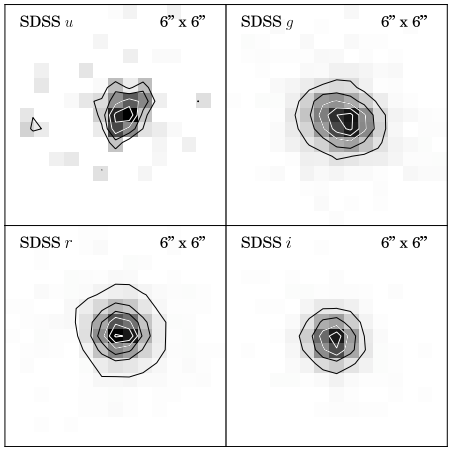}
   \caption{SDSS \emph{u,g,r,i} images of SDSS J1633. All images are observed on 2004 June 12. Each image has a  size of 6 arcsec $\times$ 6 arcsec with north up and east left. The differences between contour levels in each image are equal.}
   \label{1633m}
\end{figure}

\begin{figure} %  figure placement: here, top, bottom, or page
   \centering
   \includegraphics[width=\columnwidth, angle=0]{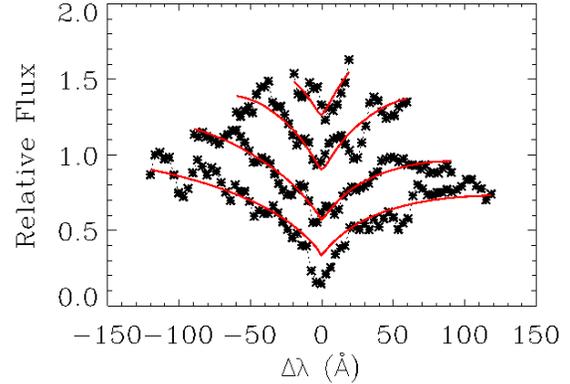}
   \caption{Fits of the observed Balmer lines for the WD companion of SDSS J1633. Dotted lines correspond to the observations and the red line corresponds to the best-fitting (WD models). Balmer lines range from H$\beta$ (bottom) to H$\epsilon$ (top).}
   \label{wdage}
\end{figure}

We checked the SDSS images of SDSS J1633, to see if any companion was resolved. Fig. \ref{1633m} shows the SDSS \emph{u-, g-, r-, i-} band images of SDSS J1633. It is less likely to be resolved in the SDSS $u, i$ and $z$ bands because WDs are too faint in $i, z$ bands and RSDs are relatively faint in the $u$ band. There is no evidence that the system is resolved in the $u r z$ bands. The SDSS $g$ band image shows a slightly elliptical profile; however, all stars around SDSS J1633 in this $g-$band image show elliptical profiles so this must be a characteristic of this particular SDSS image. Thus we conclude that the WD+esdK7 binary is not resolved in the SDSS images, and that the separation of this binary is less than $\sim$ 0.4 arcsec ($\sim $ 171--328 au).

The atmospheric parameters ($T_{\rm eff}$ and log \textit{g}) of the WD component were derived 
by performing a fit of the observed Balmer lines to hydrogen-rich WD models (Koester, 
private communication), following the procedure described in \citet{gar11}. The Balmer lines 
in such WD models were calculated with the modified Stark broadening profiles of \citet{tre09}. For the line fitting we used the code fitsb2 \citep{nap04}, which follows a procedure based on $\chi^{2}$ minimization. H$\alpha$ was not included in the fit, since it was not clearly visible in the spectrum, probably due to the contribution of the subdwarf companion. The atmospheric parameters obtained were: $T_{\rm eff}$ = 12980$\pm$770 K and log \textit{g} = 8.4$\pm$0.19 dex. As can be seen in Fig. \ref{wdage} the fit in H$\beta$ is poor, which can also be due to the contribution of the companion. Considering these parameters and using the WD cooling sequences of \citet{sal00} we determined the mass and cooling time of this WD, obtaining 0.86$\pm$0.08 $M_{\sun}$ and 0.53$\pm$0.12 Gyr, respectively. The total age of a WD is the WD cooling time plus its progenitor lifetime. When a WD is isolated we can calculate the progenitor mass by using an initial--final mass relationship \citep[e.g.][]{cat08} and then determine the progenitor lifetime using stellar tracks. In this case the WD is in a close binary, so, we cannot follow this procedure since the two stars may have interacted in the past. It is difficult then to obtain the total age for this WD since the progenitor lifetime could range from 0.2 to 6 Gyr if we consider progenitor masses from 4 to 1 $M_{\sun}$ and the stellar tracks of \citet{dom99} for $Z$=0.004. It is worth noting that the temperature obtained, 12980K, is also quite high for a halo WD with such a large mass. Typical halo WDs with $T_{\rm eff}>$10000 K have an average mass of 0.45$M_{\sun}$\citep[derived from the ESO SN Ia progenitor survey project;][]{pau06}. However, we note that the mass of the WD may have changed as a member of a close binary. \citet{jao05} discovered two WDs in systems with RSDs, LHS 193AB and LHS 300AB. These two systems have large tangential velocities and are likely members of the thick disc population of the Galaxy. \citet{mon06} estimated ages of these two WDs of 6-9 Gyr. If SDSS J1633 is a halo object, it should have an age of $>$ 9 Gyr \citep{gil89}. The age of the system estimated from its WD companion does not fit with that of a halo object.

There are however uncertainties about this system: (1) it could be a binary system formed when the Galactic disc had a lower metallicity; (2) it could be a member of a stream; (3) the spectrum of the WD companion (after subtraction of the RSD) has a low signal to noise; (4) mass transfer between the two components could make the WD look younger; and (5) the RSD could have been captured by the WD or its progenitor.

\section{Summary \& conclusions }
\label{ssummary}
We have selected $ \sim $ 1800 RSDs from SDSS with PMs greater than 100 mas$\cdot$yr$^{-1}$. 42 of these objects are late-type M subdwarfs with spectral type of $\geqslant$M6, 30 of them are new ones.  We fitted an absolute magnitude -- spectral type relationship (in the $r, i, z, J, H, K$ bands) for M and L subdwarfs, showing that subdwarfs have different sequences to M and L dwarfs. Metal-poor cool dwarfs are subdwarfs only for spectral types of $\leq$M5, and become `super' dwarfs for later types ($>$M5). We estimated distances of our M subdwarfs using the absolute magnitude -- spectral type relationships, and placed constraints on the $U, V, W$ space velocities. Our sample shows that halo and disc populations have overlaps of metallicity and kinematics.

Five M ultra subdwarfs are found to have considerably higher gravity than normal M subdwarfs. Their CaH absorption features are significantly deeper than normal M subdwarfs (whose spectra are similar in other respects). These objects provide a good tests for how surface gravity effects the spectra of cool stars. These high-gravity features are only found in M ultra subdwarfs which may reveal the role that metallicity plays in the formation and evolution history of low-mass stars. We also identify fourteen carbon rich RSDs which represent a new population of carbon subdwarfs. These objects can help us to study carbon star populations over a much greater age range.

We have presented 45 RSDs in wide binary systems ($>$ 100 au) containing sources with SDSS spectroscopy, confirming associations through common PMs.
%We measured the binary fraction of late K and M subdwarfs to be $\geq$2.41\%.
Their separations range from 0.4 arcsec to 9 arcmin, and the secondaries have spectral types ranging from late-type K to late-type M.
30 are wide and 15 are partially resolved binary systems. 
G 224--58AB is one of our widest binary systems, and contains an esdK2- and an esdM5.5-type subdwarf.  SDSS J210105.37--065633.0AB is a closer  binary system that contains an esdM1- and an esdK5.5-type subdwarf. 
We found  one spectroscopic and six wide WD + RSD binary systems. With age constraints from the WD companions in these systems, we can study the chemical evolution of the Galaxy. Three metal-poor carbon dwarfs are found in binary systems with subdwarf companions. Kinematics and radial velocity follow up would be very useful to better understand the physics of carbon subdwarfs.  
Although our binary search is not complete, our sample shows that the binary fraction of RSDs goes down with decreasing mass and metallicity, and we estimate a RSD  binary fraction of $\ga$ 5\% for separation $>$ 100 au and $\ga$ 10\% for all separation distances. 

In the future it will be possible to use UCSD binary systems (e.g. G 224--58AB) as benchmarks to test  metal-poor ultracool atmospheric models. It will also be possible to use M subdwarf binaries systems (e.g. SDSS J2101AB, SDSS J143305.04+301727.6AB) to test M subdwarf classification methods, particularly  gravity effects. We can also measure the metallicity of carbon subdwarfs most effectively if we are able to study their subdwarf companions (which do not suffer from carbon pollution). A larger sample of  carbon dwarfs/subdwarfs in wide binaries is expected in the future, from surveys/facilities such as Pan-STARRS, LAMOST, Gaia and LSST, providing the potential to identify large numbers of new nearby M, L and T subdwarf multiple systems.

\section*{Acknowledgements}

Funding for the SDSS and SDSS-II has been provided by the Alfred P. Sloan Foundation, the Participating Institutions, the National Science Foundation, the U.S. Department of Energy, the National Aeronautics and Space Administration, the Japanese Monbukagakusho, the Max Planck Society, and the Higher Education Funding Council for England. The SDSS website is http://www.sdss.org/. Funding for SDSS-III has been provided by the Alfred P. Sloan Foundation, the Participating Institutions, the National Science Foundation, and the U.S. Department of Energy Office of Science. The SDSS-III web site is http://www.sdss3.org/. This work is based in part on data obtained as part of the UKIRT Infrared Deep Sky Survey. 
This publication makes use of data products from the Two Micron All Sky Survey, the Palomar Observatory Sky Atlas (POSS-I) and The Second Palomar Observatory Sky Survey (POSS-II). The Isaac Newton Telescope is operated on the island of La Palma by the Isaac Newton Group in the Spanish Observatorio del Roque de los Muchachos of the Instituto de Astrof\'{i}sica de Canarias. This research has made use of the VizieR catalogue access tool, CDS, Strasbourg, France. IRAF is distributed by the National Optical Observatory, which is operated by the Association of Universities for Research in Astronomy, Inc., under contract with the National Science Foundation. 

ZHZ thanks Kieran Forde and Roberto Raddi for the help with data analysis, thank Paul Steele for useful discussion about ages of WDs. 
ZHZ has received support from the Royal Astronomical Society during this research. ZHZ, BB, HRAJ and RLS have received support from the Marie Curie 7th European Community Framework Programme grant n.247593  Interpretation and Parametrization of Extremely Red COOL dwarfs (IPERCOOL) International Research Staff Exchange Scheme. ZHZ, DJP, HRAJ, MCGO, SC and YVP have received support from RoPACS during this research, a Marie Curie Initial Training Network funded by the European Commission's Seventh Framework Programme. MKK and JG were supported by RoPACS. MCGO acknowledges the support of a JAE-Doc CSIC fellowship co-funded with the European Social Fund under the program {\em `Junta para la Ampliaci\'on de Estudios'}. SC acknowledges financial support from the European Commission in the form of a Marie Curie Intra European Fellowship (PIEF-GA-2009-237718). The authors thank the referee, Dr Nigel Hambly for the useful and constructive comments.

\appendix

\section{Byproduct}

\begin{table*}
 \centering
 %\begin{minipage}{140mm}
  \caption{37 carbon dwarfs and subdwarfs}
  \begin{tabular}{c c c c c r r c}
\hline\hline
    SDSS Name &  SDSS \emph{g} &  SDSS \emph{r}  & SDSS \emph{i} & SDSS \emph{z} & $\mu_{\rm RA}$(mas/yr)  & $\mu_{\rm Dec}$(mas/yr)  & SpT$^a$  \\
\hline
J012518.66$-$104448.2 & 17.90$\pm$0.02 & 16.42$\pm$0.02 & 15.89$\pm$0.02 & 15.60$\pm$0.02 & $-$28.58$\pm$2.73 & $-$127.96$\pm$2.73  & sdC\\
J093619.49+374034.6 & 19.18$\pm$0.02 & 17.60$\pm$0.03 & 17.07$\pm$0.01 & 16.72$\pm$0.03 & $-$139.67$\pm$3.08 & $-$96.52$\pm$3.08 &  sdC\\
J121341.40+451919.9 & 20.47$\pm$0.03 & 18.59$\pm$0.02 & 17.85$\pm$0.02 & 17.46$\pm$0.03 & $-$129.55$\pm$4.34 & $-$218.48$\pm$4.34 &  sdC\\
J131639.62+121529.2 & 20.71$\pm$0.03 & 18.96$\pm$0.02 & 18.42$\pm$0.02 & 18.16$\pm$0.03 & $-$106.82$\pm$4.32 & $-$27.69$\pm$4.32 &  sdC\\
J145318.82+600421.4 & 19.20$\pm$0.02 & 17.60$\pm$0.02 & 17.00$\pm$0.02 & 16.76$\pm$0.02 & $-$235.83$\pm$3.39 & $-$6.25$\pm$3.39 &  sdC\\
J145703.10+363041.9 & 20.43$\pm$0.03 & 18.74$\pm$0.01 & 18.16$\pm$0.02 & 17.88$\pm$0.02 & $-$101.29$\pm$3.62 & 60.83$\pm$3.62 &  sdC\\
J145725.85+234125.4 & 18.55$\pm$0.02 & 16.77$\pm$0.02 & 16.12$\pm$0.02 & 15.84$\pm$0.02 & $-$360.44$\pm$2.60 & $-$67.76$\pm$2.60  & sdC\\
J153554.81+105323.7 & 21.63$\pm$0.05 & 19.89$\pm$0.02 & 19.27$\pm$0.02 & 18.93$\pm$0.04 & $-$88.59$\pm$4.32 & $-$82.95$\pm$4.32  &  sdC \\
J161454.33+145314.7 & 19.56$\pm$0.02 & 17.60$\pm$0.01 & 16.92$\pm$0.01 & 16.50$\pm$0.02 & $-$16.60$\pm$2.79 & $-$151.41$\pm$2.79& sdC\\
J004853.30$-$090435.8 & 21.06$\pm$0.04 & 19.46$\pm$0.02 & 18.87$\pm$0.02 & 18.59$\pm$0.04 & $-$27.20$\pm$4.71 & $-$117.35$\pm$4.71  & dCf\\
J010717.91$-$091329.5 & 20.62$\pm$0.04 & 18.82$\pm$0.02 & 18.22$\pm$0.01 & 17.89$\pm$0.03 & 167.36$\pm$3.75 & $-$60.07$\pm$3.75 &  dCf\\
J094352.09+362545.3 & 20.14$\pm$0.02 & 18.78$\pm$0.01 & 18.32$\pm$0.02 & 18.07$\pm$0.02 & $-$72.40$\pm$3.52 & $-$79.97$\pm$3.52 &  dCf\\
J144448.40+043944.2 & 21.41$\pm$0.05 & 19.62$\pm$0.02 & 19.10$\pm$0.02 & 18.90$\pm$0.05 & $-$111.73$\pm$5.60 & 42.56$\pm$5.60 &  dCf\\
J163132.74+355328.9 & 19.09$\pm$0.02 & 17.41$\pm$0.01 & 16.89$\pm$0.01 & 16.73$\pm$0.02 & $-$142.10$\pm$3.24 & 107.02$\pm$3.24 & dCf\\
J012028.55$-$083630.8 & 18.72$\pm$0.02 & 17.00$\pm$0.02 & 16.45$\pm$0.01 & 16.27$\pm$0.02 & 145.82$\pm$3.01 & $-$49.40$\pm$3.01 & dC\\
J012150.11+011302.6 & 18.74$\pm$0.02 & 17.00$\pm$0.02 & 16.52$\pm$0.02 & 16.36$\pm$0.02 & 206.08$\pm$3.06 & $-$132.26$\pm$3.06  & dC\\
J074257.17+465918.1 & 17.64$\pm$0.02 & 15.77$\pm$0.02 & 15.15$\pm$0.02 & 14.87$\pm$0.02 & $-$58.81$\pm$11.84 & $-$142.40$\pm$11.84 &  dC\\
J074638.20+400403.4 & 21.40$\pm$0.05 & 19.50$\pm$0.02 & 19.00$\pm$0.02 & 18.86$\pm$0.05 & $-$276.50$\pm$5.79 & $-$173.34$\pm$5.79 &  dC\\
J081807.44+223427.5 & 17.15$\pm$0.01 & 15.56$\pm$0.01 & 15.09$\pm$0.01 & 14.89$\pm$0.02 & 46.63$\pm$2.70 & $-$241.94$\pm$2.70 &  dC\\
J095005.05+584124.2 & 18.47$\pm$0.01 & 17.02$\pm$0.01 & 16.57$\pm$0.02 & 16.38$\pm$0.02 & 22.66$\pm$3.11 & $-$97.64$\pm$3.11 &  dC\\
J095114.98+261207.5 & 21.24$\pm$0.04 & 19.35$\pm$0.02 & 18.80$\pm$0.02 & 18.50$\pm$0.03 & $-$77.49$\pm$4.43 & $-$79.51$\pm$4.43 &  dC\\
J100353.99+100807.2 & 18.50$\pm$0.02 & 17.13$\pm$0.02 & 16.66$\pm$0.02 & 16.42$\pm$0.03 & $-$32.54$\pm$4.40 & $-$113.36$\pm$4.40 & dC\\
J105429.38+340225.0 & 18.59$\pm$0.02 & 17.19$\pm$0.02 & 16.78$\pm$0.02 & 16.60$\pm$0.02 & $-$58.31$\pm$2.86 & $-$116.22$\pm$2.86 & dC\\
J111320.81+115234.9 & 20.38$\pm$0.03 & 18.81$\pm$0.02 & 18.34$\pm$0.02 & 18.11$\pm$0.03 & $-$21.79$\pm$4.01 & $-$112.77$\pm$4.01 &  dC\\
J111449.32+420252.9 & 19.52$\pm$0.02 & 18.09$\pm$0.02 & 17.60$\pm$0.01 & 17.38$\pm$0.02 & $-$26.57$\pm$3.30 & $-$101.59$\pm$3.30 &  dC\\
J114807.41+401008.9 & 19.91$\pm$0.02 & 18.17$\pm$0.02 & 17.64$\pm$0.01 & 17.41$\pm$0.02 & $-$50.41$\pm$3.17 & $-$101.35$\pm$3.17 &  dC\\
J115854.95+232234.6 & 21.35$\pm$0.05 & 19.51$\pm$0.03 & 18.94$\pm$0.04 & 18.72$\pm$0.04 & 19.13$\pm$4.45 & $-$124.24$\pm$4.45 &  dC\\
J115929.27+640501.9 & 19.19$\pm$0.03 & 17.66$\pm$0.01 & 17.17$\pm$0.01 & 16.95$\pm$0.02 & $-$98.94$\pm$3.17 & $-$69.79$\pm$3.17 &  dC\\
J124358.58+183042.6 & 20.04$\pm$0.02 & 18.42$\pm$0.01 & 17.89$\pm$0.03 & 17.69$\pm$0.03 & 7.24$\pm$4.41 & $-$133.08$\pm$4.41 &  dC\\
J130323.99+092543.6 & 19.24$\pm$0.02 & 17.64$\pm$0.01 & 17.15$\pm$0.02 & 16.96$\pm$0.02 & $-$119.34$\pm$3.04 & $-$3.44$\pm$3.04 &  dC\\
J133101.12+140613.1 & 21.14$\pm$0.04 & 19.30$\pm$0.02 & 18.81$\pm$0.02 & 18.66$\pm$0.04 & 4.83$\pm$5.49 & $-$110.20$\pm$5.49 &  dC\\
J140325.56+202827.1 & 18.75$\pm$0.02 & 17.33$\pm$0.02 & 16.85$\pm$0.02 & 16.65$\pm$0.03 & $-$139.35$\pm$2.58 & 5.18$\pm$2.58 &  dC\\
J151542.93+520145.9 & 18.77$\pm$0.02 & 17.39$\pm$0.01 & 16.98$\pm$0.02 & 16.82$\pm$0.02 & $-$132.57$\pm$3.10 & $-$15.71$\pm$3.10 & dC\\
J154809.21+322724.9 & 18.08$\pm$0.01 & 16.51$\pm$0.01 & 15.99$\pm$0.01 & 15.72$\pm$0.02 & $-$136.65$\pm$2.62 & 50.23$\pm$2.62  & dC\\
J155237.29+292758.5 & 17.07$\pm$0.01 & 15.56$\pm$0.01 & 15.08$\pm$0.02 & 14.92$\pm$0.02 & $-$218.51$\pm$2.45 & $-$192.91$\pm$2.45 & dC\\
J172843.18+270829.1 & 18.24$\pm$0.01 & 16.75$\pm$0.01 & 16.26$\pm$0.01 & 16.08$\pm$0.01 & 34.18$\pm$2.94 & $-$113.35$\pm$2.94 & dC\\
J223250.79$-$003437.0 & 21.07$\pm$0.04 & 19.39$\pm$0.02 & 18.79$\pm$0.02 & 18.45$\pm$0.03 & 4.03$\pm$3.32 & $-$188.01$\pm$3.32 & dC\\
\hline
\end{tabular}
\begin{list}{}{}
%\item[]Note. 
\item[$^{a}$]The sdC represents carbon subdwarfs, dC represents carbon dwarfs, dCf represents subdwarfs with weak carbon features.   
\end{list}
%\end{minipage}
\label{37dc}
\end{table*}

%\begin{landscape}
\begin{table*}
 \centering
 %\begin{minipage}{140mm}
  \caption{Nine common proper motion pairs of ultracool dwarfs.}
  \begin{tabular}{c c c c c c r r r l}
\hline\hline
 Comp & SDSS Name &  SDSS \emph{g} & SDSS \emph{r} & SDSS \emph{i} & SDSS \emph{z} &$\mu_{\rm RA}$(mas/yr)  & $\mu_{\rm Dec}$(mas/yr)  & Sep(\arcmin)  &  SpT   \\
\hline
01A & J042604.36+170714.3 & 16.52$\pm$0.03 & 14.99$\pm$0.02 & 13.33$\pm$0.00 & 12.41$\pm$0.02 & 109.24$\pm$2.52 & $-$30.43$\pm$2.52 & 329.08  & M5~~ \\
01B & J042619.09+170301.9 & 19.23$\pm$0.02 & 17.64$\pm$0.03 & 15.66$\pm$0.02 & 14.56$\pm$0.02 & 101.69$\pm$2.83 & $-$28.77$\pm$2.83 &  &   M6~~~\\
02A & J120719.63+405349.9 & 16.80$\pm$0.01 & 15.39$\pm$0.02 & 14.38$\pm$0.02 & 13.86$\pm$0.02 & $-$115.23$\pm$2.67 & $-$2.43$\pm$2.67 & 13.98 &        \\
02B & J120718.57+405357.1 & 21.70$\pm$0.06 & 20.01$\pm$0.03 & 17.82$\pm$0.02 & 16.66$\pm$0.02 & $-$122.59$\pm$4.91 & 3.76$\pm$4.91 &  &   M6~~ \\
03A & J125945.65+195659.1 & 19.67$\pm$0.02 & 18.14$\pm$0.01 & 16.11$\pm$0.02 & 15.02$\pm$0.02 & $-$480.80$\pm$3.15 & 151.47$\pm$3.15 & 39.93 &   M6~~ \\
03B & J125948.18+195641.2 & 20.54$\pm$0.03 & 19.04$\pm$0.02 & 16.66$\pm$0.02 & 15.43$\pm$0.02 & $-$477.93$\pm$4.44 & 156.55$\pm$4.44 &  &        \\
04A & J091244.25+560450.5 & 16.99$\pm$0.01 & 15.65$\pm$0.01 & 14.90$\pm$0.01 & 14.45$\pm$0.02 & $-$53.74$\pm$2.82 & $-$93.51$\pm$2.82 & 265.96 &        \\
04B & J091222.55+560136.3 & 22.19$\pm$0.09 & 20.76$\pm$0.04 & 18.62$\pm$0.03 & 17.53$\pm$0.02 & $-$52.31$\pm$5.62 & $-$88.92$\pm$5.62 &  &  M7~~ \\
05A & J090215.19+033524.8 & 18.16$\pm$0.01 & 16.65$\pm$0.01 & 15.17$\pm$0.02 & 14.37$\pm$0.02 & $-$107.63$\pm$2.95 & 160.27$\pm$2.95 & 165.29 &       \\
05B & J090226.21+033534.6 & 21.21$\pm$0.04 & 19.51$\pm$0.02 & 17.25$\pm$0.02 & 16.03$\pm$0.02 & $-$110.16$\pm$4.97 & 162.28$\pm$4.97 &  &   M7~~ \\
06A & J135208.56+502044.8 & 18.86$\pm$0.03 & 17.42$\pm$0.03 & 16.60$\pm$0.02 & 16.17$\pm$0.02 & $-$23.34$\pm$3.14 & $-$107.45$\pm$3.14 & 340.26 &        \\
06B & J135238.95+501748.7 & 20.69$\pm$0.02 & 19.06$\pm$0.02 & 16.82$\pm$0.01 & 15.61$\pm$0.02 & $-$18.55$\pm$3.56 & $-$99.93$\pm$3.56 & &  M7~~ \\
07A & J084909.03+560439.8 & 15.63$\pm$0.01 & 14.17$\pm$0.00 & 13.05$\pm$0.00 & 12.49$\pm$0.01 & 170.08$\pm$2.61 & $-$73.97$\pm$2.61 & 90.51   &      \\
07B & J084918.85+560517.7 & 20.63$\pm$0.04 & 18.84$\pm$0.01 & 16.35$\pm$0.01 & 15.03$\pm$0.01 & 168.26$\pm$4.28 & $-$74.28$\pm$4.28 &  &   M7.5 \\
08A & J133619.15+615601.4 & 22.39$\pm$0.13 & 20.81$\pm$0.05 & 18.65$\pm$0.02 & 17.47$\pm$0.03 & $-$35.93$\pm$5.61 & $-$98.37$\pm$5.61 & 282.51 &   M6~~ \\
08B & J133554.63+615944.7 & 23.60$\pm$0.32 & 21.98$\pm$0.11 & 19.39$\pm$0.02 & 18.00$\pm$0.03 & $-$39.89$\pm$6.20 & $-$105.94$\pm$6.20 & &   M7.5 \\
09A & J085147.11+413415.0 & 18.25$\pm$0.04 & 16.65$\pm$0.02 & 14.96$\pm$0.02 & 13.98$\pm$0.02 & $-$113.94$\pm$2.88 & 18.69$\pm$2.88 & 32.66 &        \\
09B & J085148.19+413445.3 & 21.68$\pm$0.06 & 19.97$\pm$0.02 & 17.30$\pm$0.01 & 15.85$\pm$0.01 & $-$109.39$\pm$4.96 & 18.78$\pm$4.96 &  &   M7.5\\
\hline
\end{tabular}
%\begin{list}{}{}
%\item[]Note. 
%\item[$^{a}$]10A and 10B are binary. 10C is CPM companion to 10AB 162\arcsec away. 10C is redder than 10B in $r-z$ colour but brighter than 10B suggesting it may not be connected to 10AB.   \\
%\item[$_{b}$]This binary system is not in the original proper motion search. \\
%\item[$_{c}$]Objects have carbon lines and classified as carbon subdwarfs.
%\end{list}
\label{tucdb}
\end{table*}

\bsp

\label{lastpage}

\end{document}